\documentclass[onecolumn,amsmath,amssymb,nofootinbib,12pt]{article}
\pdfoutput=1
\usepackage{jheppub}
\usepackage{ifpdf}
\usepackage[bb=boondox]{mathalfa}
\usepackage{graphicx,subcaption}
\graphicspath{ {figures/} }
\usepackage{amsfonts}
\usepackage{amsmath}
\usepackage{amssymb}
\usepackage{epsfig}
\usepackage{pdfpages}
\usepackage{bbm}
\usepackage{graphicx,epstopdf}
\usepackage[numbers]{natbib} 
\usepackage[makeroom]{cancel}
\usepackage{hyperref}
\hypersetup{pdftex,colorlinks=true,allcolors=blue}
\usepackage{array}
\usepackage[export]{adjustbox}

\usepackage[normalem]{ulem}
\usepackage{romannum}

\newcommand{\RN}[1]{%
  \textup{\uppercase\expandafter{\romannumeral#1}}%
}

\usepackage[thinlines]{easytable}
\usepackage{textcomp}

\newcommand{\eg}{{\it e.g.,}\ }
\newcommand{\ie}{{\it i.e.,}\ }
\newcommand{\viz}{{\it viz.,}\ }
\newcommand{\reef}[1]{(\ref{#1})}

\newcommand{\mt}[1]{\textrm{\tiny #1}}

\newcommand{\GN}{G_\mt{N}}

\newcommand{\mB}{\mathcal{B}}

\newcommand{\rmax}{r_{\rm{max}}}
\newcommand{\rmin}{r_{\rm{min}}}
\newcommand{\rturn}{r_{\rm{turn}}}
\newcommand{\tR}{t_{\mt{R}}}
\newcommand{\tL}{t_{\mt{L}}}
\newcommand{\tinf}{t_{\infty}}
\newcommand{\tauinf}{\tau_{\infty}}

\newcommand{\dS}{\text{dS}}
\newcommand{\arctanh}{\text{arctanh}}

\newcommand{\beq}{\begin{equation}}
\newcommand{\eeq}{\end{equation}}
\newcommand{\beqa}{\begin{eqnarray}}
\newcommand{\eeqa}{\end{eqnarray}}

\newcommand{\bea}{\begin{eqnarray}}
\newcommand{\eea}{\end{eqnarray}}

\renewcommand{\(}{\left(}
\renewcommand{\)}{\right)}
\renewcommand{\[}{\left[}
\renewcommand{\]}{\right]}

\renewcommand{\S}{\Sigma}

\newcommand{\CV}{\mathcal{C}_{\mt{V}}}
\newcommand{\mO}{\mathcal{O}}

\newcommand{\mV}{\mathcal{V}}
\newcommand{\mL}{\mathcal{L}}

\newcommand{\RNum}[1]{\uppercase\expandafter{\romannumeral #1\relax}}
\newcommand{\veps}{\varepsilon}

\newcommand{\cv}{{\cal C}_\mt{V}}
\newcommand{\ca}{{\cal C}_\mt{A}}
\newcommand{\cvv}{{\cal C}_\mt{SV}}

\title{Holographic Complexity in $\dS_{d+1}$}

\author[a,b]{Eivind J\o rstad}
\author[a]{Robert C. Myers}
\author[c]{and Shan-Ming Ruan}

\affiliation[a]{Perimeter Institute for Theoretical Physics, Waterloo, ON N2L 2Y5, Canada}
\affiliation[b]{Dept. of Physics \& Astronomy, University of Waterloo, Waterloo, ON N2L 3G1, Canada}
\affiliation[c]{Center for Gravitational Physics, Yukawa Institute for Theoretical Physics, \\
Kyoto University, Kitashirakawa Oiwakecho, Sakyo-ku, Kyoto 606-8502, Japan}

\emailAdd{ejorstad@perimeterinstitute.ca}
\emailAdd{rmyers@perimeterinstitute.ca}
\emailAdd{ruan.shanming@yukawa.kyoto-u.ac.jp}

\date{\today}

\abstract{We study the CV, CA, and CV2.0 approaches to holographic complexity in $(d+1)$-dimensional de Sitter spacetime. We find that holographic complexity and corresponding growth rate presents universal behaviour for all three approaches. In particular, the holographic complexity exhibits `hyperfast' growth \cite{Susskind:2021esx} and appears to diverge with a universal power law at a (finite) critical time. We introduce a cutoff surface to regulate this divergence, and the subsequent growth of the holographic complexity is linear in time.}

\begin{document}

\begin{flushright}
	\hfill{ YITP-22-15}
\end{flushright}

\maketitle

\section{Introduction}

The $\text{AdS}_{d+1}/\text{CFT}_{d}$ correspondence \cite{Maldacena:1997re} or gauge/gravity duality is the duality between quantum gravity theory in $(d+1)$-dimensional asymptotically Anti-de Sitter (AdS) spacetime and $d$-dimensional conformal field theory (CFT) living on the asymptotic  boundary of AdS spacetime. Remarkably, the AdS/CFT correspondence provides an ideal proving ground for probing many deep questions in quantum gravity, \eg understanding the microscopic origin of black hole entropy. Even though this holographic duality provides us with the most promising and precise definition of nonperturbative quantum gravity, our universe is quite different from AdS spacetime. Defining a quantum gravity theory for cosmological spacetimes is a crucial question. In the light of the success of the AdS/CFT correspondence, it is natural to investigate the holographic description of asymptotically de Sitter (dS) spacetimes.

The discussion of dS holography  begins with the observation that an inertial observer in $\dS_{d+1}$ is surrounded by a causal or cosmological horizon and experiences a non-vanishing Hawking temperature given by  \cite{Gibbons:1977mu}  
\begin{equation}\label{temp}
T=\frac{1}{2\pi\,L}\,,
\end{equation}
where $L$ is the background curvature scale.
The entropy associated to the cosmological horizon is determined by the Bekenstein-Hawking formula \cite{Gibbons:1977mu},\footnote{Quantum corrections to this expression were evaluated in \cite{Anninos:2020hfj}.}
\begin{equation}\label{eq:dSentropy}
S_{\rm{dS}} =\frac{\mathrm{Area}}{4 \, \GN}= \frac{L^{d-1}\,\Omega_{d-1}}{4\,\GN} \equiv  \ N
\end{equation}
where $\Omega_{d-1}=2\pi^{d/2}/\Gamma(d/2)$ is the volume of a unit $(d-1)$-sphere. These features indicate that the physics of $\dS_{d+1}$ is essentially different from that the AdS counterpart. The holographic description of dS spacetime is further challenging because dS spacetime does not have an asymptotic spatial boundary like AdS.

Despite the obstacles, various approaches toward dS holography have been developed. As an `analytic continued' analog of the standard AdS/CFT,  the so-called dS/CFT correspondence \cite{Strominger:2001pn,Strominger:2001gp,Maldacena:2002vr,Witten:2001kn} proposes a holographic duality between  gravity in asymptotically $\dS_{d+1}$ spacetime and a $d$-dimensional CFT living on the spacelike boundary at future timelike infinity of dS. These ideas were further developed, \eg \cite{Bousso:2001mw,Balasubramanian:2002zh,Balasubramanian:2001nb,Klemm:2001ea,Leblond:2002ns,Leblond:2002tf,Kabat:2002hj,Parikh:2002py,Anninos:2011ui,Hikida:2021ese}, revealing that the boundary CFT is unconventional. Implicitly this approach describes the physics of a metaobserver living at the future infinite. An alternative approach focuses on inertial or `static patch' observers, \eg \cite{Parikh:2004wh,Banks:2005bm,Banks:2006rx,Freivogel:2005qh,Anninos:2011af,Fischetti:2014uxa,Anninos:2017hhn,Anninos:2018svg}. In the context of static patch holography, it has been argued that the dS gravity is dual to a quantum mechanical system with a finite number of degrees of freedom \cite{WFN,Banks:2000fe,Bousso:2000nf,Balasubramanian:2001rb,Parikh:2004wh,Banks:2005bm,Banks:2006rx}. This is a reflection of the finite  entropy associated with the cosmological horizon, which we have designated $N$ in eq.~\reef{eq:dSentropy} to indicate that this counts the number of fundamental degrees of freedom in the dual theory describing de Sitter spacetime.\footnote{Let us add that yet another approach named the dS/dS correspondence \cite{Karch:2003em,Alishahiha:2004md,Dong:2010pm,Dong:2018cuv,Gorbenko:2018oov} instead considers the correspondence between a quantum gravity in $\dS_{d+1}$ and two UV-cutoff CFTs which are living on $\dS_{d}$ and coupled to each other by a $d$-dimensional gravitational system.}

Quantum information theory has produced astonishing new insights into many core questions in the AdS/CFT correspondence, \eg see reviews \cite{VanRaamsdonk:2016exw,Rangamani:2016dms,Chen:2021lnq}. This progress has motivated some interesting recent discussions of de Sitter holography \cite{Susskind:2021omt,Susskind:2021dfc,Susskind:2021esx,Shaghoulian:2021cef,Shaghoulian:2022fop}. In particular, \cite{Susskind:2021esx,Shaghoulian:2021cef,Shaghoulian:2022fop} proposed generalizations of the celebrated Ryu-Takayanagi  formula \cite{Ryu:2006bv} for holographic entanglement entropy to de Sitter spacetime. An essential ingredient is to replace the asymptotic AdS boundary with the boundary of the static patch, \ie the cosmological horizon (or rather a stretched horizon just inside the cosmological horizon). As a result, the entanglement entropy between the left- and right-static patches (see figure \ref{fig:PenrosedS}) is given by eq.~\reef{eq:dSentropy}. Motivated by this work  (in particular \cite{Susskind:2021esx}), we examine another quantum information inspired entry in the holographic dictionary, namely holographic complexity (\eg see  \cite{Chapman:2021jbh} and references therein), in de Sitter spacetimes.

Holographic complexity borrows the usual notions of computational complexity used in computer science or quantum information, \eg see \cite{johnw,AaronsonRev}.
Roughly speaking, holographic complexity measures the difficulty (\ie the resources needed) to construct a particular target state in the boundary theory from an unentangled reference state using a set of fundamental simple gates. From the viewpoint of the bulk spacetime (\ie asymptotically AdS spacetime), it appears that the holographic complexity may be associated with a broad family of codimension-one and -zero gravitational observables \cite{Belin:2021bga,longpaper}. However, the most studied of these are: complexity=volume (CV) \cite{Susskind:2014rva,Stanford:2014jda},  complexity=action (CA) \cite{Brown:2015bva,Brown:2015lvg} and complexity=spacetime volume (CV2.0) \cite{Couch:2016exn}. 

The CV conjecture \cite{Susskind:2014rva,Stanford:2014jda} states that the complexity is dual to the maximal volume of hypersurface anchored at the time slice $\S$ in the boundary on which the state is defined, \ie
\begin{equation}\label{defineCV}
\cv(\S) =\ \mathrel{\mathop {\rm
		max}_{\scriptscriptstyle{\S=\partial \mathcal{B}}} {}\!\!}\left[\frac{\mathcal{V(B)}}{\GN \, \ell_{ \rm bulk}}\right] \,,
\end{equation}
where $\GN$ denotes Newton's constant in the bulk gravitational theory and $\mathcal B$ corresponds to the bulk hypersurface of interest. The maximization here is performed over all possible spacelike surfaces in the bulk whose boundary is fixed at the time slice $\Sigma$. The definition of $\cv$ requires that we introduce an additional length scale $\ell_{\rm bulk}$  to make the holographic complexity dimensionless. For simplicity, we will set $\ell_{\rm bulk} = L$, \ie the curvature radius for bulk geometry in the following.

The CA proposal \cite{Brown:2015bva,Brown:2015lvg} states that the complexity is given by evaluating the gravitational action on a region of spacetime, known as the Wheeler-DeWitt (WDW) patch, which can be regarded as the causal development of a space-like bulk surface anchored on the boundary time slice $\S$. The CA proposal then is given by 
\begin{equation}\label{defineCA}
\ca(\S) =  \frac{I_\mt{WDW}}{\pi\, \hbar}\,. 
\end{equation}
The CV2.0 proposal is both a generalization and a simplification of the previous approach \cite{Couch:2016exn}. In this case, the holographic complexity is given by simply evaluating the spacetime volume of the WDW patch, namely
\begin{equation}\label{defineCV2}
\cvv(\S) =  \frac{V_\mt{WDW}}{\GN \, \ell^2_{ \rm bulk}}\,. 
\end{equation}
As with the CV proposal, an additional length scale  $\ell_{\rm bulk}$ enters the definition of this observable as well. As before, we set $\ell_{\rm bulk} = L$ in the following.

The goal of this paper is to study the generalization of all of these approaches to holographic complexity in $\dS_{d+1}$ spacetime, following \cite{Susskind:2021esx,Shaghoulian:2021cef,Shaghoulian:2022fop} with the boundary time slice $\Sigma$ fixed on the stretched horizon. The CV complexity  has been studied in dS previously by \cite{Susskind:2021esx,Chapman:2021eyy}. Notably, ref.~\cite{Susskind:2021esx} argued that the complexity growth is hyperfast, apparently diverging as the boundary time approaches some (finite) critical time. In this paper, we regulate this divergence by introducing a geometric cutoff, \ie a cutoff surface near future timelike infinity. Evaluating the above three proposals for holographic complexity in $\dS_{d+1}$, we find that they all exhibit similar behaviour. At early times, the complexity growth rate increases and tends to be divergent when approaching the critical time. With the cutoff surface, the holographic complexity remains finite, and the hyperfast growth ends before the critical time is reached. The subsequent growth is linear in time, but the rate is large and controlled by the cutoff.

The rest of the paper is organized as follows: In section \ref{sec:sectionA}, we begin by reviewing some essential concepts for $\dS_{d+1}$ spacetime and introducing the basic framework for subsequent calculations. In section \ref{sec:CV20}, we start from CV2.0 by focusing on the time evolution of WdW patch and the corresponding spacetime volume. Based on these results, we further study CA in detail in section \ref{sec:CA}. Furthermore, we explore the extremal surfaces and CV in $\dS_{d+1}$ in section \ref{sec:dSCV}. Finally, we discuss the implications of our results as well as some open questions in section \ref{sec:disc}. More analytical results about extremal surfaces and CV in $\dS_2$ are given in appendix \ref{app:dS2}. We also examine the maximization of multiple extremal surfaces in $\dS_{d+1}$ in appendix \ref{app:max}.

\section{Preliminaries}\label{sec:sectionA}

\begin{figure}[ht!]
	\centering
	\includegraphics[width=4.0in]{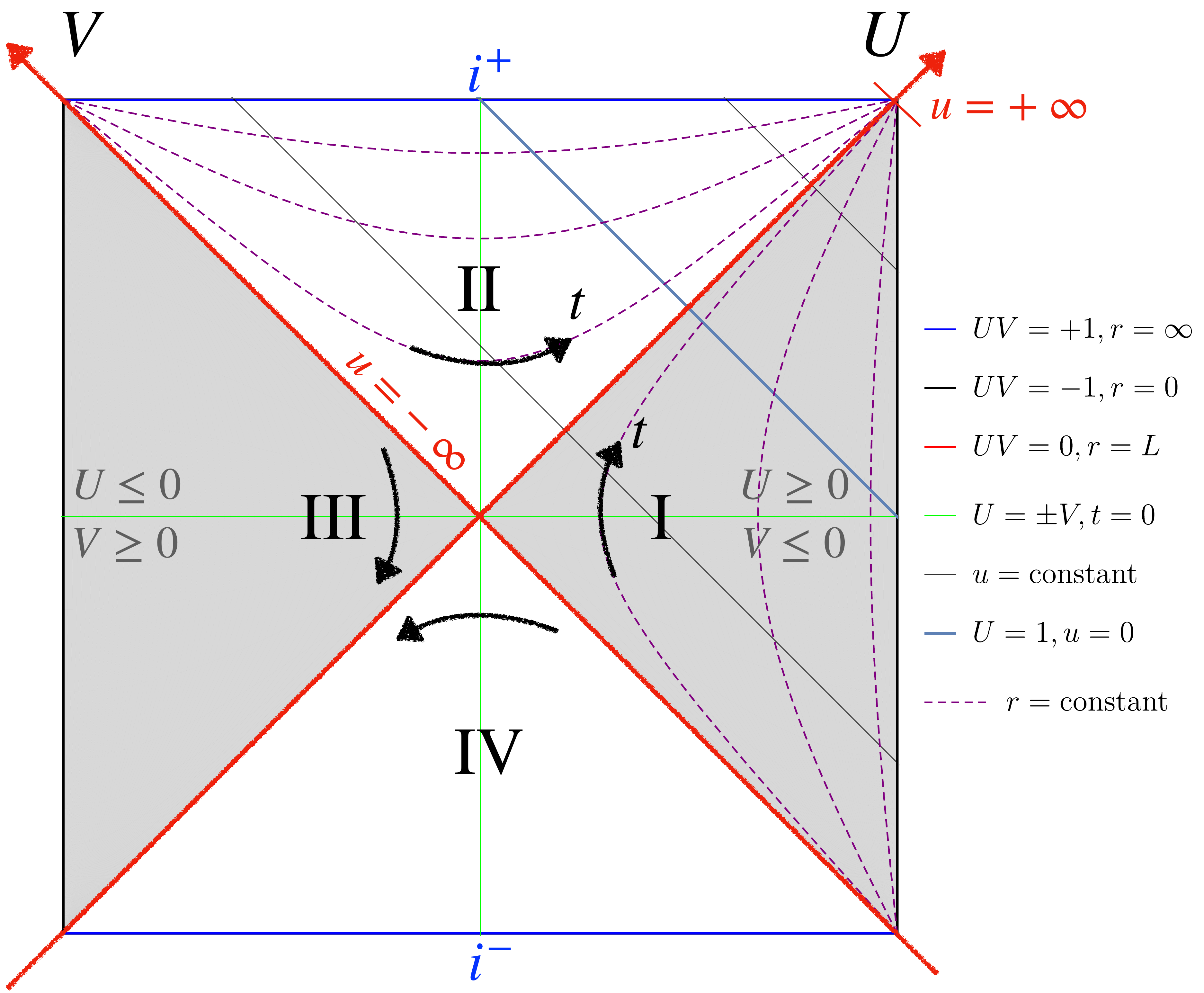}
	\caption{The Penrose diagram for dS$_{d+1}$ spacetime consists of four quadrants I, II, III, IV. The left/right gray region presents the static patch covered by $(t,r)$ coordinates. The dashed purple curves are referred to as the spacelike hypersurface with a constant radial coordinate, \ie $r=\rho L=\text{constant}$. The future and past infinity located at $r=\infty$ are denoted by blue lines and labeled by $i^+, i^-$, respectively. }
	\label{fig:PenrosedS}
\end{figure}

We will be examining the $(d+1)$-dimensional de Sitter ($\dS_{d+1}$) spacetime. This is a maximally symmetric geometry with positive curvature, illustrated by the Penrose diagram in figure \ref{fig:PenrosedS}. Each point in the diagram corresponds to a ($d-1$)-dimensional sphere. However, these spheres shrink to zero size at the right and left vertical boundaries (which we refer to as the north pole and south pole, respectively). Hence horizontal slices in the Penrose diagram correspond to $d$-dimensional spheres, and the topology of the full geometry is $R\times S^d$. Further, the $S^d$ expand to infinite size at future and past timelike infinity $i^\pm$, corresponding to the horizontal boundaries at the top and bottom of the Penrose diagram.

Much of our discussion will center on the static dS metric 
\begin{equation}\label{eq:dSmetric}
ds^2=  - \(1 - \frac{r^2}{L^2}\) dt^2 + \frac{dr^2}{1- \frac{r^2}{L^2}} + r^2 d \Omega_{d-1} \,,
\end{equation}
where $L$ is the dS curvature scale. These coordinates readily cover the `static patch' denotes quadrants I and III in figure \ref{fig:PenrosedS}. These are the regions accessible to an observer at the north or south pole, respectively, \ie $r=0$. The null boundaries of these patches correspond to the future and past cosmological horizons at $r=L$ and $t\to\pm\infty$, respectively. Quadrants II and IV correspond to the regions $r\ge L$, with $i^\pm$ corresponding to $r\to\infty$. Of course, $r$ becomes a timelike coordinate and $t$ is spacelike in these regions. The advantage of this coordinate system \ref{eq:dSmetric} is that the Killing vector $\partial_t$ is obvious. The global flow of this time coordinate is illustrated in the figure, and as usual, the time translation symmetry can be thought of as a `boost' symmetry about the bifurcation surface where the two horizons cross.

In examining the various holographic complexity proposals in $\dS_{d+1}$, we need coordinates that can extend across the cosmological horizons. Hence it is convenient to introduce Eddington-Finkelstein (EF) coordinates  
\begin{equation}\label{eq:infallv}
\begin{split}
v = t + r^*(r) \,,\qquad&\qquad u = t -  r^*(r) \,,\\
\text{where}\qquad  r^*(r) &=  \frac{L}{2} \log \left|  \frac{L+r}{L-r}  \right|\,.\\
\end{split}
\end{equation}
Using $dr^*=dr/f(r)$, the static metric  \eqref{eq:dSmetric} becomes
\begin{equation}
\label{eq:vmetric}
\begin{split}
d s^{2}&=-f(r) d u^{2}-2\,du\,dr+{r^{2}}\, d\Omega^2_{d-1}\\
&=-f(r) d v^{2}+2\,dv\,dr+r^2\, d\Omega^2_{d-1}\,.
\end{split}
\end{equation}
Surfaces of constant $u$ are illustrated in figure \ref{fig:PenrosedS}. In quadrants I and II, they correspond to null cones originating at the north pole and expanding out to future timelike infinity $i^+$.  We note that $u\to-\infty$ on the past cosmological horizon (in quadrant I) and $u\to+\infty$ in the top right corner of the Penrose diagram. Further, $u=0$ corresponds to the null cone originating at $t=0,\,r=0$ and reaching $i^+$ at $t=0,\,r=\infty$. This surface plays a special role in the following. In quadrants III and IV, surfaces of constant $u$ are null cones with their tip on the south pole and extending into the past to $i^-$. The description of surfaces of constant $v$ is similar but with the north and south poles interchanged.

It is straightforward to extend the EF coordinates above to Kruskal coordinates $(U,V)$ covering the entire $\dS_{d+1}$ geometry. In quadrant I, we write
\begin{equation}\label{eq:defineKruskal}
 U = + e^{ u /L} \,, \qquad V = - e^{ -v /L} \,,
\end{equation}
and the metric becomes
\begin{equation}\label{Kmetric}
ds^2 = \frac{L^2}{(1- UV)^2} \(  - 4 dU dV + (1+U V)^2 d\Omega^2_{d-1}  \)\,.
\end{equation}
These coordinates and the above metric extend to the entire spacetime but one chooses $ U = \pm e^{ u /L} \,, V = \pm e^{ -v /L}$, where the signs change between the different quadrants -- see figure \ref{fig:PenrosedS}. In terms of the Kruskal coordinates, the north/south pole at $r=0$ is given by $U V=-1$. The boundary of the static patch, \ie the cosmological horizon located at $r=L$ becomes to $U V=0$. The asymptotic boundaries, \ie  past and future timelike infinity $i^\pm$, at $r \to \infty$ are represented by $UV=1$. The interested reader is referred to \cite{Spradlin:2001pw,Kim:2002uz} for further discussion (and illustrations) of the $\dS_{d+1}$ spacetime.

\begin{figure}[h!]
	\centering
	\includegraphics[width=3.5in]{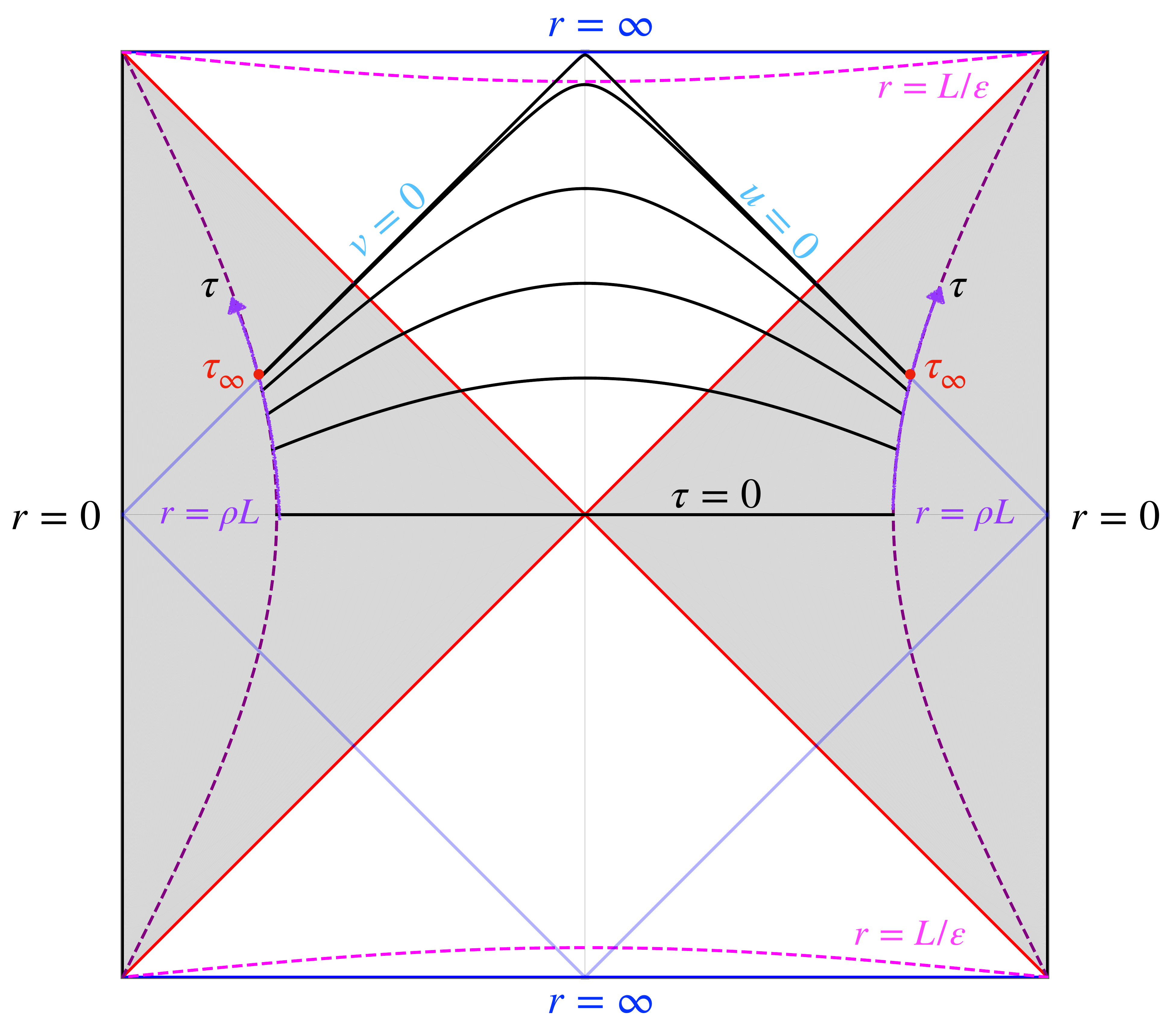}
	\caption{Extremal surfaces joining the left and right stretched horizons at the boundary time $\tau$ are denoted by the black curves.}
	\label{fig:dSExtremalSurface}
\end{figure}

We build on the recent works \cite{Susskind:2021esx,Shaghoulian:2021cef,Shaghoulian:2022fop} that examined the generalization of holographic entanglement entropy to de Sitter space. In particular, this presents an interpretation of dS entropy \reef{eq:dSentropy} as the entanglement entropy between two dual theories describing the left and right static patches -- see figure \ref{fig:dSExtremalSurface}. It is suggested that these dual theories are located on the boundary of the static patch. That is, they reside on the stretched horizon at
\begin{equation}\label{eq:defStretchedHorizon}
r= r_{\rm stretch}\equiv \rho L \,, \qquad \text{with} \qquad  0 < \rho <1\,.
\end{equation}
As in \cite{Susskind:2021esx,Shaghoulian:2021cef,Shaghoulian:2022fop}, we generally consider 
$\rho$ to be very close to 1, but our calculations will allow for any value in the range $0 < \rho <1$.
 Analogous to the Ryu-Takayanagi (RT) formula in asymptotically AdS spacetime, the holographic entanglement entropy between the left/right patch is thus given by the area of the extremal surface (minimax surface) between two stretched horizons, \ie the cosmological horizon in $\dS_{d+1}$. As a result, this dS version of the RT formula yields $S_{\rm{dS}}$ as the holographic entropy.

Following this framework, the holographic CV complexity \reef{defineCV} was also examined in \cite{Susskind:2021esx} with the extremal surfaces anchored on time slices in the two stretched horizons. More importantly, it is argued that the growth rate of holographic complexity $\cv$  in dS is hyperfast, apparently diverging when we approach a critical boundary time. This geometric result is associated with the exponential growth of the interior of dS and has striking implications for the boundary theory -- see section \ref{sec:disc}. The divergent behavior is easily explained with the Penrose diagram in figure \ref{fig:dSExtremalSurface}. We consider the stretched horizons located at $r=\rho L$ with $\rho <1$ and denote the coordinate time on the left and right stretched horizon as $\tL$ and $\tR$, respectively. Taking account of the general time evolution of the boundary times, we will focus on the symmetric case  $\tR=\tau L=-\tL$ without loss of generality (as we do throughout the following).  In particular, the extremal surfaces in $\dS$ can only be extended to a critical scale denoted by $\tauinf$, at which the extremal surface approaches the null surface 
\begin{equation}
 u = 0 \,, \qquad v=0 \,, 
\end{equation}
which are denoted by the light blue lines in figure \ref{fig:dSExtremalSurface}. It is obvious that these null surfaces extend to the future timelike infinity $i^+$ at $r=\infty$. As a result, one can find not only the divergence of holographic complexity but also the hyperfast growth, \ie 
\begin{equation}
\label{bang}
\lim_{\tau \to \tau_\infty}  \CV  \to \infty  \,, \qquad \lim_{\tau \to \tau_\infty}  \frac{d \CV}{d \tau } \to  \infty \,, 
\end{equation}
by approaching this critical time $\tau_{\infty}$. We provide a detailed analysis of this behaviour for CV, CA, and also CV2.0 (which exhibit analogous divergent growth) in the following sections. 

However, to regulate these divergences \reef{bang}, we introduce a cutoff surface near the timelike infinity $i^+$, as indicated by the pink dashed curve in figure \ref{fig:dSExtremalSurface}. For simplicity, we assume that the cutoff surface is given by 
\begin{equation}\label{eq:defrmax}
r = \rmax \equiv  \frac{L}{\varepsilon} \,, \qquad \text{with} \qquad  \varepsilon \ll 1 \,.
\end{equation}
This cutoff surface allows us to tame the divergent behaviour and examine the holographic complexity for late times, \ie $\tau\ge\tauinf$.  Interestingly, we find that the regulated holographic complexity grows linearly at late times, \viz 
\begin{equation}
\frac{d \mathcal{C} }{d\tau} \bigg|_{\tau \gtrsim  \tau_{\infty} }  \simeq  \frac{N}{ \varepsilon^d} \,,
\end{equation}
where again we find analogous behaviour for all three proposals for holographic complexity. Of course, this growth is similar to the late-time linear growth of holographic complexity in asymptotically AdS.  

We now turn to examine the CV2.0, CA, and CV proposals in de Sitter space, each in turn in the following sections.  

\section{CV2.0 in dS$_{d+1}$}\label{sec:CV20}
\begin{figure}[!]
	\centering
	\includegraphics[width=2.8in]{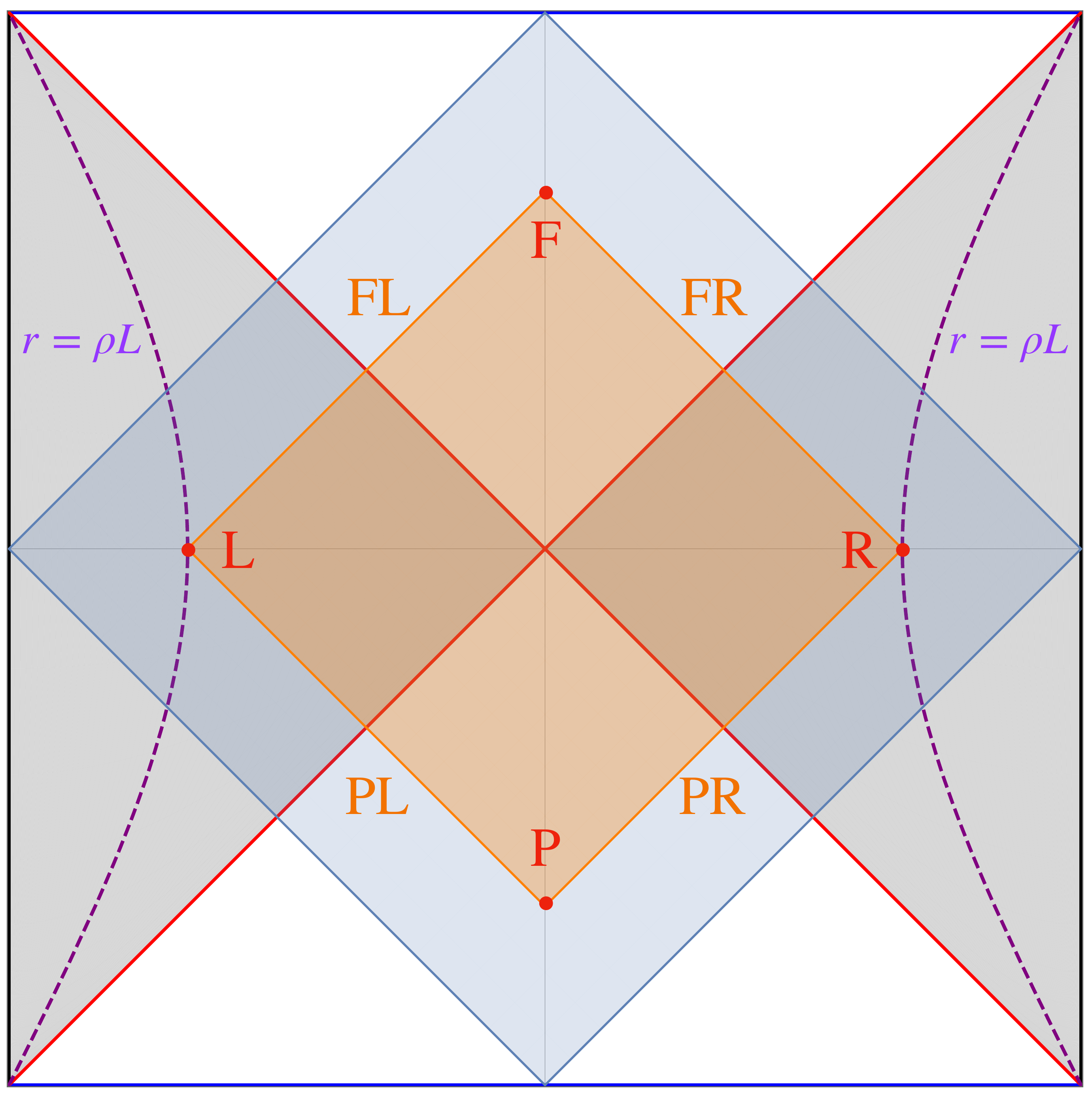}
     \qquad
		\includegraphics[width=2.8in]{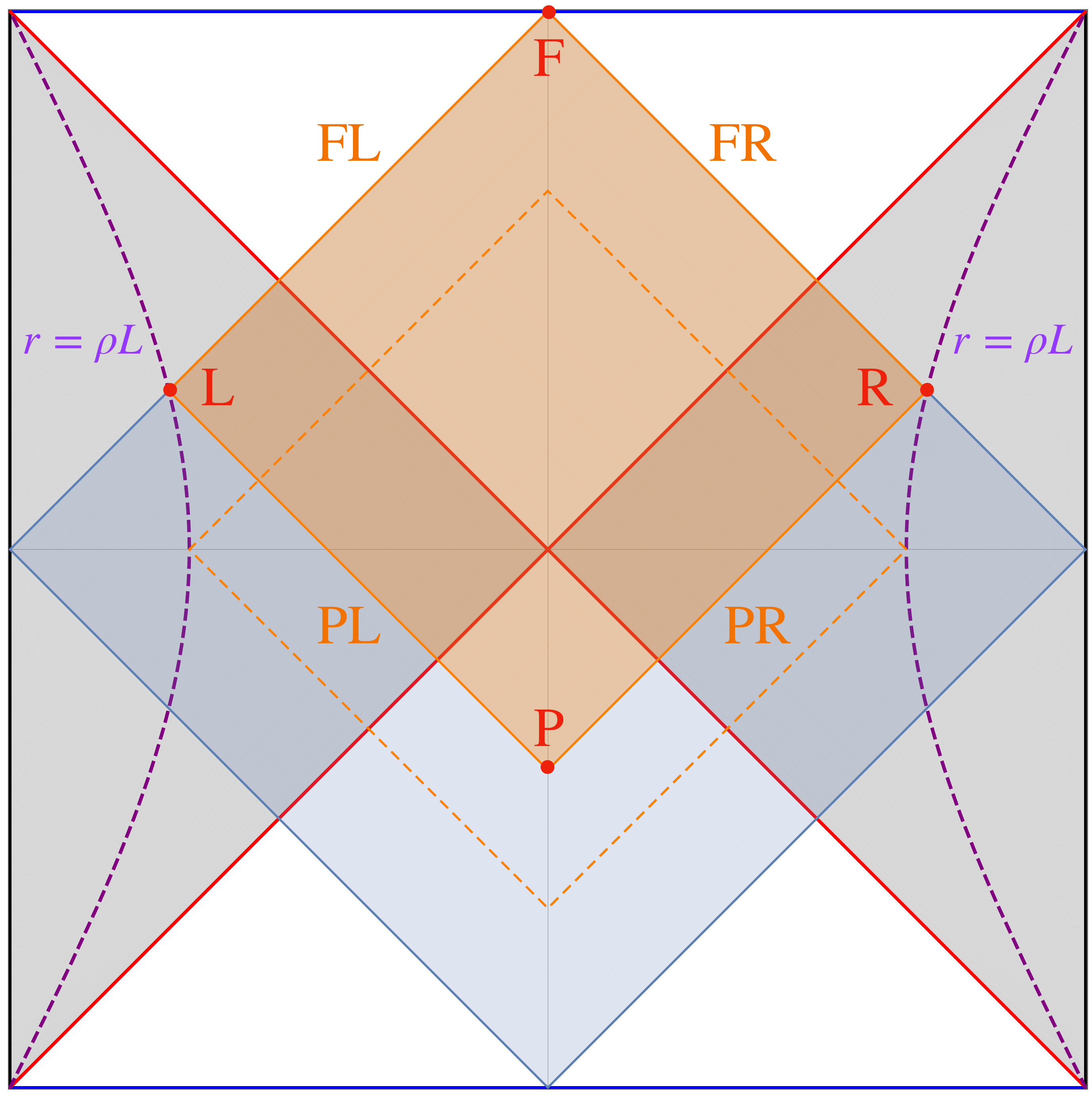}
	\caption{The orange square region denotes the WdW patch at a boundary time $\tR=\tau L =- \tL $. The purple dashed curves represent the stretch horizon at $r=\rho L$. Left: The WdW patch at $\tau=0$. Right: The WdW patch at the critical time $\tau=\tauinf$ where the future tip $F$ touches $i^+$. The shadowed blue region presents the WdW patch at $\tau=0$ with exactly anchoring the boundary at the south and north poles (\ie $\rho=0$). }
	\label{fig:WdW}
\end{figure}

In this section, we apply the CV2.0 proposal \reef{defineCV2} for holographic complexity in ($d+1$)-dimensional de Sitter space. As described above, we anchor the WdW patch to equal-time surfaces $\tR=-\tL =L\,\tau$ on the stretched horizons $r=L\,\rho$ on either side of the horizon -- see figure \ref{fig:WdW}. We show that holographic complexity and its growth rate are both divergent as the boundary time approaches the critical value  $\tau=\tauinf\equiv \arctanh \rho$. We regulate the spacetime volume by introducing a cutoff surface $r=\rmax$ near the future timelike boundary. With this regulated volume, the holographic complexity grows linearly with the boundary time for the subsequent evolution $\tau\gtrsim\tauinf$. 

\subsection*{Time evolution of WdW patch}

\begin{figure}[h]
	\centering
	\includegraphics[width=4.0in]{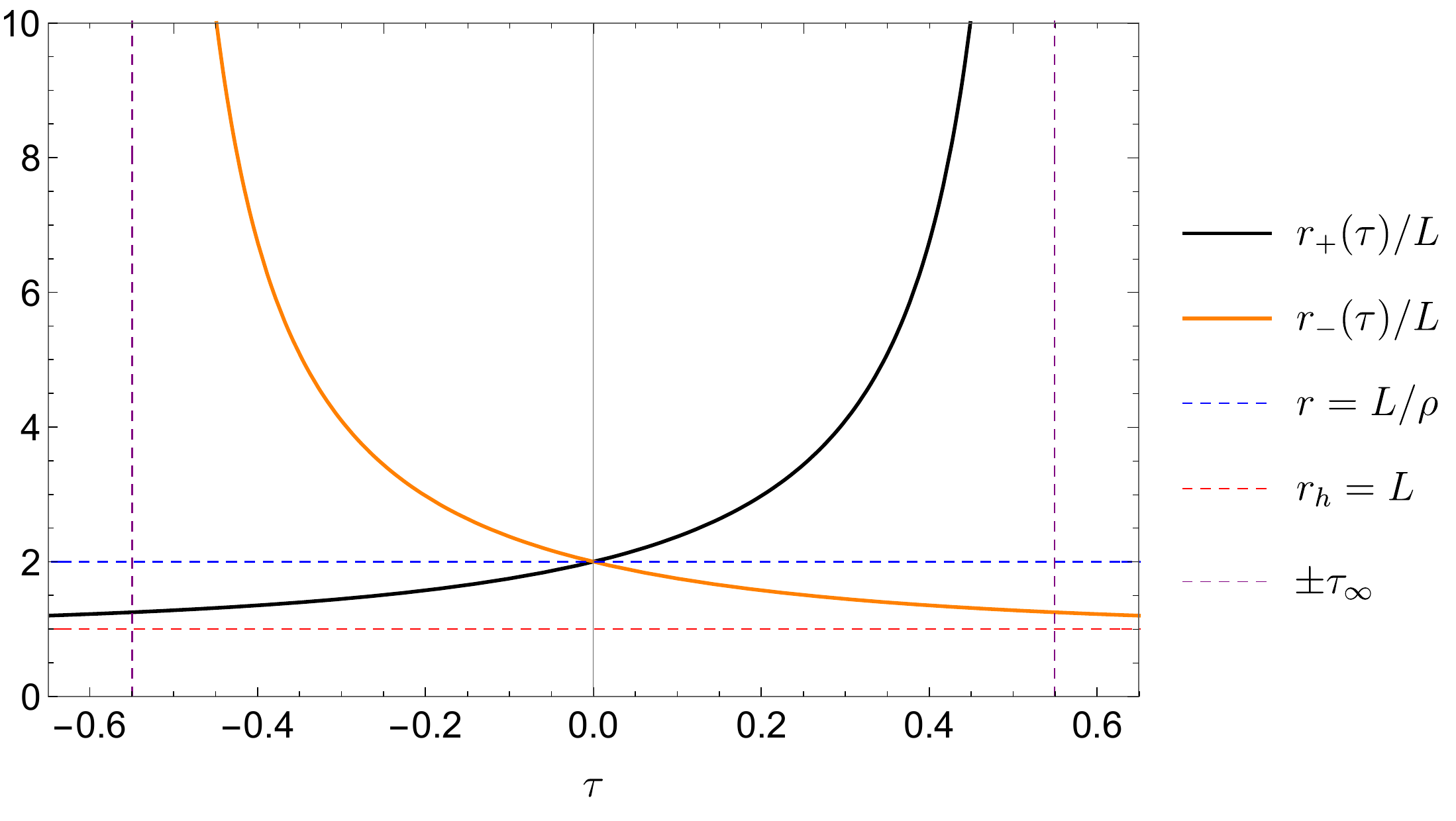}
	\caption{The time evolution of tip radii $r_\pm$ of WdW patch with $\rho =1/2$. Hence from eq.~\reef{eq:rpm}, $r_+$ diverges at $\tau=\tauinf\simeq 0.5493$, and $r_-$ diverges at $\tau=-\tauinf$.}
	\label{fig:SVrm}
\end{figure}

At early times, the boundaries of the WdW patch are four null cones, as shown in the left panel of figure \ref{fig:WdW}.  Focusing on the two boundaries on the right side of the WdW patch, they parametrized by
\begin{equation}
\begin{split}
u_{\rm max}&= L\,\tau - \frac{L}{2} \log \(  \frac{1+\rho}{1-\rho} \) =  - \frac{L}{2} \log \(  \frac{r_+ + L}{r_+-L} \)\,, \\
v_{\rm min}&= L\,\tau + \frac{L}{2} \log \(  \frac{1+\rho}{1-\rho} \) =\frac{L}{2} \log \(  \frac{r_- + L}{r_--L} \) \,, 
\end{split}
\label{square}
\end{equation}
where $r_{\pm}$ are the radii at the tips of the WdW patch, \ie the two future boundaries intersect in quadrant II at $t=0,\,r=r_+$. Using the above expressions, these positions are given by
\begin{equation}\label{eq:rpm}
 \frac{r_+}{L} =  \frac{ \cosh \tau  - \rho\, \sinh \tau  }{ \rho\,  \cosh \tau  - \sinh \tau } \,, \qquad  \frac{r_-}{L} = \frac{ \cosh \tau  + \rho\, \sinh \tau  }{ \rho\,  \cosh \tau  + \sinh \tau } \,, 
\end{equation}
whose time evolution is shown in figure \ref{fig:SVrm}. 

Evolving forward from $\tau=0$, the WdW patch retains its square shape in the Penrose diagram until the future tip reaches the asymptotic boundary $i^+$, \ie when $r_+\to\infty$. We denote this particular time as $\tauinf$ 
with\footnote{In terms of $\tauinf$, we can rewrite $r_\pm$ as $r_\pm = L\, \coth (\tauinf \mp \tau)$. \label{footy78}}
\begin{equation}
\tanh \tau_{\infty}=  \rho   \,.
\label{criticaltau}
\end{equation}
We note that this corresponds to $u_{\rm max} =0$ in eq.~\reef{square}.\footnote{We might note that in eq.~\reef{square}, $u_{\rm max}<0$ and $v_{\rm min}>0$ since the EF coordinates are defined in  \reef{eq:infallv} so that the critical null cones emerging from $t=0,\, r=0$ correspond to $u=0$ and $v=0$.} The WdW patch at this critical time is shown in the right panel of figure \ref{fig:WdW}. We will find below that the volume of the WdW patch grows rapidly as we approach $\tau=\tauinf$. The positions of the two tips  satisfy $ r_+ (\tau) = r_- (-\tau)$, which is inherited from the $t \to - t$ symmetry of the dS geometry. Hence, $r_-$ diverges at $\tau=-\tauinf$ when the lower tip of the WdW patch hits $i^-$.

With the CV2.0 proposal \reef{defineCV2}, we must evaluate the spacetime volume of the WdW patch, which is most easily done using Kruskal coordinates \reef{Kmetric}. For early times $|\tau| \le \tauinf$, the holographic complexity is given by 
\begin{equation}\label{eq:defineSpacetimeVolume}
\begin{split}
\cvv=\frac{V_{\mt{WdW}}\( |\tau| \le \tauinf \)}{\,G_N\,L^2}  &=\frac{1}{\,G_N\,L^2}  \, \int_{\mt{WdW}}\!\!\!\! \sqrt{-g}\, d\Omega_{d-1}\,dU\,dV \\
&= 
8\,N\,\int^{U_{\rm max}}_{U_{\rm min}} \!\!dU  \int^{V_{\rm max}}_{V_{\rm min}} \!\! dV \  \frac{(1+UV)^{d-1}}{\(1-U V\)^{d+1}} \,,
\end{split}
\end{equation}
where $N$ denotes the de Sitter entropy \reef{eq:dSentropy}, while $U_{\rm max},\, U_{\rm min},\, V_{\rm max},\, V_{\rm min}$ are the positions of four null boundaries of WdW patch.  With the definition of the Kruskal coordinates  in eq.~\eqref{eq:defineKruskal}, the four null boundaries of WdW patch become
\begin{equation}\label{bound77}
\begin{split}
U_{\rm max} &\equiv e^{u_{\rm max}/L}= V_{\rm max} = e^{\tau} \sqrt{\frac{1-\rho}{1+\rho}} \,,\\
V_{\rm min} &\equiv -e^{-v_{\rm min}/L}=  U_{\rm min} =- e^{-\tau} \sqrt{\frac{1-\rho}{1+\rho}}  \,.\\
\end{split}
\end{equation}

\subsubsection*{CV2.0 complexity at $\tau =0$}
As a warm-up, let us first consider the most symmetric case with $\tau=0$, which yields $U_{\rm max} = V_{\rm max} =-V_{\rm min} =  -U_{\rm min}=\sqrt{\frac{1-\rho}{1+\rho}}$ in eq.~\reef{bound77}. Correspondingly, the WdW patch is symmetric between not only the left and right halves but also the top and bottom halves. The integral \eqref{eq:defineSpacetimeVolume} for the holographic complexity yields
\begin{equation}\label{tau0}
\begin{split}
\cvv( \tau=0)  &=\frac{16\,N}{d(d+1)}\, \bigg(\rho ^{d+1} \, _2F_1\left(1,\frac{d+1}{2};\frac{d+3}{2};\rho ^2\right)
\\
&\quad-\frac{1}{\rho^{d+1} } \, _2F_1\left(1,\frac{d+1}{2};\frac{d+3}{2};\frac{1}{\rho ^2}\right)-\frac{\pi  i}{2}\,(d+1)\bigg)\,,
\end{split}
\end{equation}
where $_2F_1$ is the usual hypergeometric function. 
We note that these functions are complex for the parameters here, but subtracting the imaginary constant in the second line ensures that $\cvv( \tau=0)$ is real. To get more insight, we can explicitly write the above expression for a few dimensions:
\begin{equation}\label{eq:SpaceVolumetR0B}
 \begin{split}
\cvv( \tau=0)   = 8N\times
\begin{cases}
 -2\, \log \rho\,, 
 \qquad \qquad \qquad \ \, d= 1 \\
\ \frac{1}{\rho} - \rho \,, 
\qquad \qquad\qquad \quad \ \, d= 2 \\
\  \frac{1}{3\rho^2} -\frac{1}{3}\,\rho^2 +\frac{2}{3}\,\log \rho  \,, \quad\ \  d= 3 \\
\   \frac{1}{6 \rho ^3}+\frac{1}{2\rho }-\frac{\rho}2-\frac{\rho ^3}{6} \,, \qquad \ \,d= 4 \,.\\
\end{cases}
\end{split}
\end{equation}
As we push the stretched horizons to the cosmological horizon (\ie $\rho \to 1$), the WdW patch becomes a small diamond in the vicinity of the bifurcation surface -- see figure \ref{fig:WdW}. The corresponding complexity decays to zero as  
\begin{equation}
\cvv( \tau=0) \simeq  8N \(   2 (1-\rho )+ \(1-\rho \)^2 +  \frac{d^2+5}{9}\,  (1-\rho )^3+\mathcal{O}\( (1-\rho)^4 \)  \)\,.
\end{equation}

In contrast, if we pull the stretched horizons near the north and south poles (\ie $\rho \to 0$), the WdW patch expands with future and past tips approaching the timelike boundaries of the dS geometry. That is, the null boundaries of WdW patch at $\tau=0$ are approaching the critical null surfaces at $u_{\rm max}=0= v_{\rm min}$. For $\tau=0$, eq.~\eqref{eq:rpm} simplifies to
\begin{equation}
r_\pm |_{\( \tau=0  \)}  = \frac{L}{\rho} \,,
\end{equation}
for the positions of the future and past tips of the WdW patch. As expected, these both diverge as $\rho\to0$.
As a result, the corresponding holographic complexity \reef{tau0} is also divergent, \ie
\begin{equation}\label{eq:SpaceVolumetR0}
 \begin{split}
\cvv( \tau=0)  \simeq \frac{16 N}{d}\times
\begin{cases}
\frac{1}{(d-1)\rho^{d-1}}+ \frac{1}{(d-3)\rho^{d-3}} +\cdots - \log  \rho \,, \qquad {\rm odd}\ d \,,\\
\frac{1}{(d-1)\rho^{d-1}} + \frac{1}{(d-3)\rho^{d-3}} + \cdots \,, \qquad \qquad \quad \,{\rm even}\ d \,.\\
\end{cases}
 \end{split}
\end{equation}
Although this divergent behavior is presented here for the special case $\tau=0$ and $\rho \to 0$, we will see below that this behaviour is a universal feature associated with the limit $\tau \to \tauinf$. That is, from eq.~\reef{criticaltau}, the critical time $\tau_{\infty} \simeq \rho+ \rho^3/3+ \cdots$ decreases to zero in the  $\rho \to 0$ limit. Hence we can express the leading divergence above as 
\begin{equation}\label{leadingdiverge}
\cvv \simeq \frac{16 N}{d(d-1)}\,\frac{1}{(\tauinf-\tau)^{d-1}}+\cdots\,,
\end{equation}
which we will see corresponds (up to a factor of two\footnote{The factor two arises because both tips of the WDW patch are approaching the timelike boundaries of the de Sitter geometry here, whereas for general $\rho$ in the following, we only have the future tip approaching $i^+$.}) the leading divergence as $\tau \to \tauinf$ for general $\rho$ -- compare with eq.~\reef{eq:VWdWaround}.

\begin{figure}[ht!]
	\centering
	\includegraphics[width=2.94in]{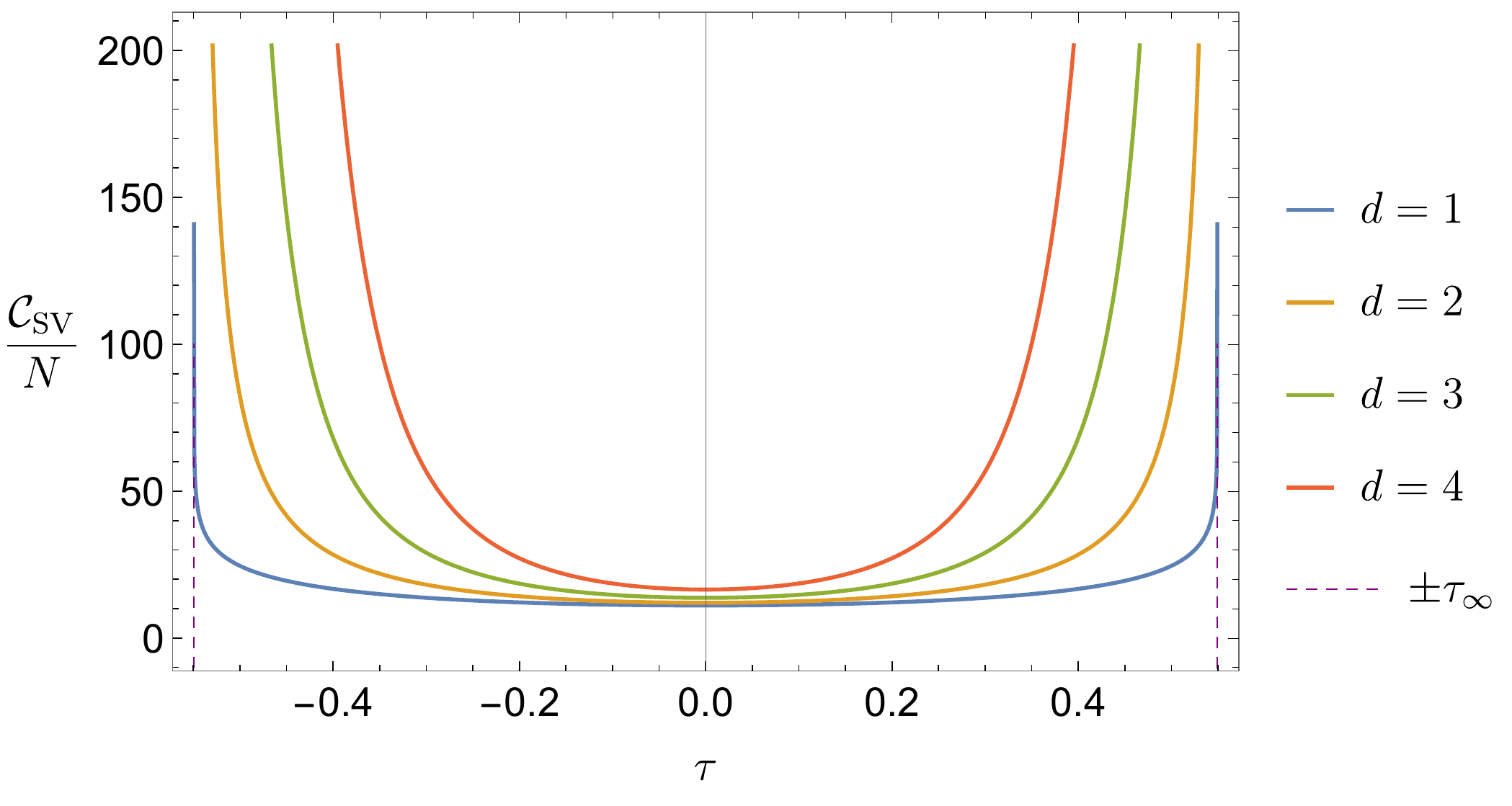}
	\includegraphics[width=3.06in]{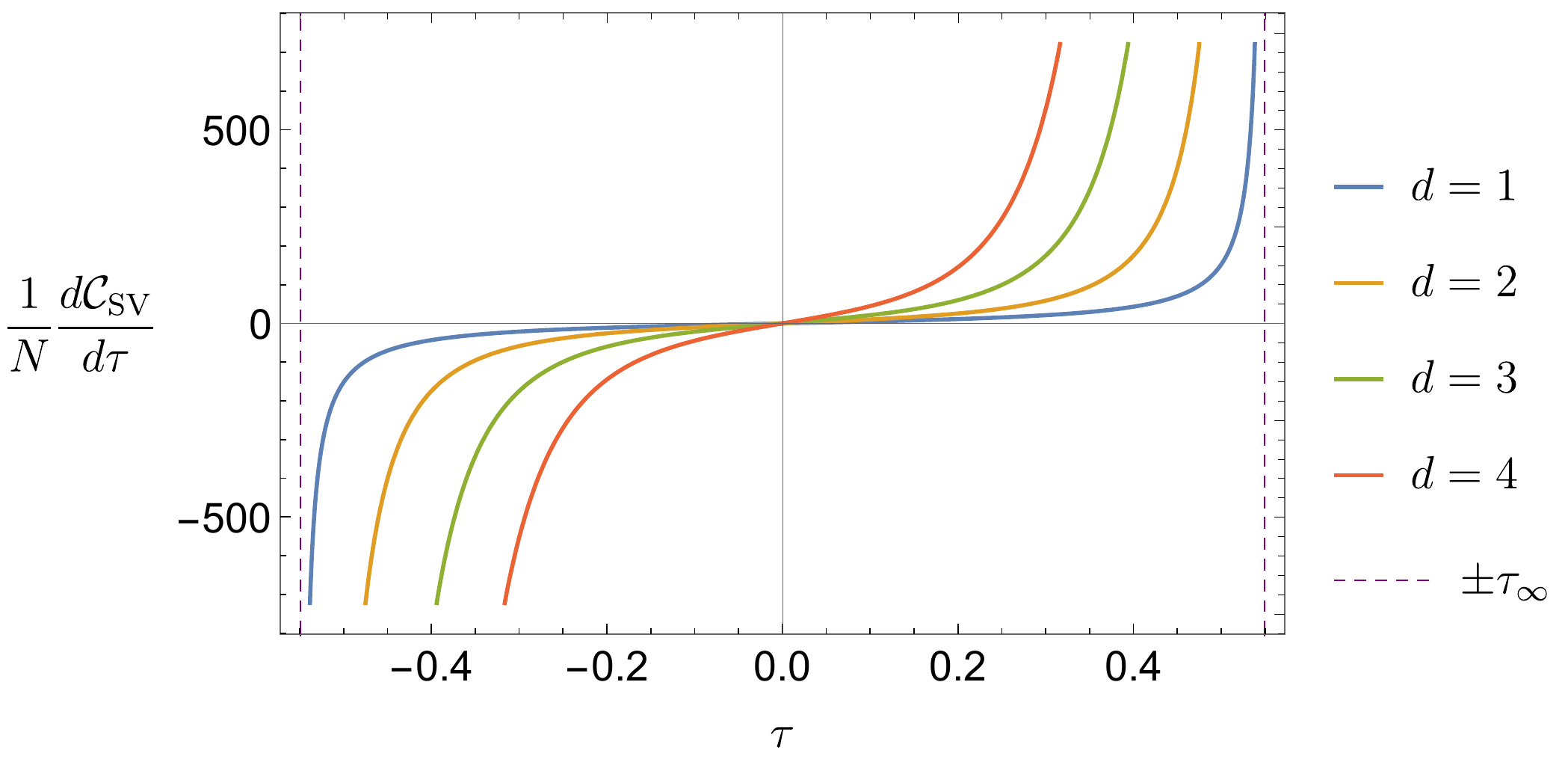}
	\caption{The time evolution of holographic complexity and its growth rate from CV2.0 with $\rho=1/2$. }
	\label{fig:SpacetimeVolumetR}
\end{figure}
\subsubsection*{Early time evolution of CV2.0 ($\tau \le \tau_{\infty}$)}

Let us move the boundary to a generic time before the critical time, \ie $|\tau| \le \tauinf $. As noted above, the WdW patch retains its square shape in this regime and we must perform the integral in eq.~\reef{eq:defineSpacetimeVolume} for the limits defined in eq.~\reef{bound77} for a specific choice of $\rho$ and $\tau$. The integral can be expressed in terms of incomplete beta functions $\mathbf{B}(z;a,b) = \int_0^z t^{a-1} (1-t)^{b-1} dt$ \cite{olver2010nist}.\footnote{Alternatively, we can write $B(z;\alpha, \beta)=\frac{z^{\alpha}}{\alpha} \, { }_{2} F_{1}(\alpha, 1-\beta ; \alpha+1 ; z)$.}
The final result is given by a sum over all four corners of the WdW patch (shown in figure \ref{fig:WdW}):
\begin{equation}\label{eq:SpaceVolumetR}
\cvv\( \tau\le \tau_{\infty} \) =\frac{4N}{d}\[  \(  \mathbf{B}\(\frac{1+W_i}{2};d, 1-d\)- \mathbf{B}\(\frac{1+W_i}{1-W_i}; d, 0\)  \)^{\rm F,P}_{\rm L,R}  - 2\pi i \] \,.
\end{equation}
Here $W\equiv U V$ and the superscripts and subscripts indicate the positions of the four corners of WdW patch. Further, we again subtract an imaginary constant, \ie $2\pi i$, to ensure that the expression is always real.\footnote{The appearance of $2\pi i$ is because the spacetime volume integral crosses the horizon and the close form involves terms like $\log \( \frac{L-r}{L+r} \)$.} 
The values $W_i$ on the corners are given by
\begin{equation}
\label{bound88}
W_i= \begin{cases}
 U_{\rm max} V_{\rm max} = e^{2\tau} \frac{1-\rho}{1+\rho}\,,\qquad\  \text{F: future tip}\\ 
  U_{\rm min} V_{\rm min} = e^{-2\tau} \frac{1-\rho}{1+\rho} \,,\qquad\,\text{P: past tip}\\ 
   U_{\rm min} V_{\rm max} = \frac{\rho-1}{\rho+1}\,,\qquad\qquad \text{L: left stretched horizon}\\ 
    U_{\rm max} V_{\rm min} = \frac{\rho-1}{\rho+1}\,,\qquad\qquad \text{R: right stretched horizon}\,.\\ 
\end{cases}
\end{equation}
We can write the results more explicitly for various dimensions, \eg
\begin{equation}\label{eq:SpaceVolumetRB}
\begin{split}
\cvv \( \tau \le \tauinf \)  =8N\times
\begin{cases}
- \log \left(1-(1-\rho^2) \cosh^2\tau\right)\,, \quad d= 1 \,,\\
\ \frac{\rho  (1-\rho ^2) \cosh ^2\tau}{1-(1-\rho^2) \cosh^2\tau} \,, \qquad
\qquad \qquad \ \  \ d= 2 \,,\\
\end{cases}
\end{split}
\end{equation}
and for $d=3$,
\begin{align*}
\cvv \( \tau \le \tauinf \)  =& \frac{8N}{3}\bigg( 1- \rho^2-  \log \left(1-(1-\rho^2) \cosh^2\tau\right)  \\
& \qquad\quad +(1-\rho^2)\,\frac{(1+\rho ^2)\cosh^2\tau-1}{\left(1-(1-\rho^2) \cosh^2\tau\right) ^2} \bigg)\,.
\end{align*}
One easily sees that the above expressions are divergent where $\cosh^2\tau\to 1/(1-\rho)^2$, which one readily verifies corresponds to $\tau\to\tauinf$ with the latter given by eq.~\reef{criticaltau}. Of course, these divergences are expected from our discussion above. Generally the divergences come from the $W_F$ contribution in eq.~\reef{eq:SpaceVolumetR} as $\tau\to\tauinf$, and we find
\begin{equation}\label{eq:VWdWaround}
\cvv \( \tau \le \tauinf \)
\simeq  8N\times
\begin{cases}
\frac{1}{d(d-1)\( \tauinf -\tau\)^{d-1}}+ \frac{1}{3(d-3)\( \tauinf -\tau \)^{d-3}} +\cdots -\frac{1}{d} \log \( \tauinf -\tau \)  \,,\ {\rm odd}\ d\\
\frac{1}{d(d-1)\( \tauinf -\tau\)^{d-1}} + \frac{1}{3(d-3)\( \tauinf -\tau \)^{d-3}} +\cdots \,, \qquad\qquad\qquad \quad\ {\rm even}\ d \,.\\
\end{cases}
\end{equation}

Equipped with the general expression \reef{eq:SpaceVolumetR} for an arbitrary $|\tau| \le \tauinf$, we easily derive the time rate of growth for the holographic complexity  before the critical time. Taking the time derivative of the holographic complexity $\cvv$, we obtain \footnote{In terms of $r_\pm$, it is rewritten as $\frac{d\cvv}{d \tau} =  \frac{8N}{d} \(  \( \frac{r_+}{L} \)^d -\( \frac{r_-}{L} \)^d     \)$.}
\begin{equation}\label{totalrate0}
\begin{split}
\frac{d\cvv}{d \tau} &=  \frac{8N}{d} \(   \(  \frac{\cosh \tau  - \rho \sinh \tau }{\rho \cosh \tau -  \sinh \tau}   \)^d -  \(  \frac{\cosh \tau  + \rho \sinh \tau }{\rho \cosh \tau +  \sinh \tau}   \)^d  \)  \\
&= \frac{8N}{d}  \(   \coth^d \( \tauinf -\tau \)  -   \coth^d \( \tauinf +\tau \)    \) \,.
\end{split}
\end{equation}
Near the critical time, the complexity exhibits hyperfast growth \cite{Susskind:2021esx}, with
\begin{equation}\label{eq:seriesdCsvdtau}
 \lim_{\tau \to \tauinf} \frac{d\cvv}{d \tau} \approx 8N\(   \frac{1}{d \( \tauinf -\tau  \)^d}+\frac{1}{3\( \tauinf -\tau  \)^{d-2}}+\cdots \)\,.
\end{equation}
Finally, we illustrate holographic complexity and its growth rate at early times in figure \ref{fig:SpacetimeVolumetR}.

\begin{figure}[ht!]
	\centering
	\includegraphics[width=2.8in]{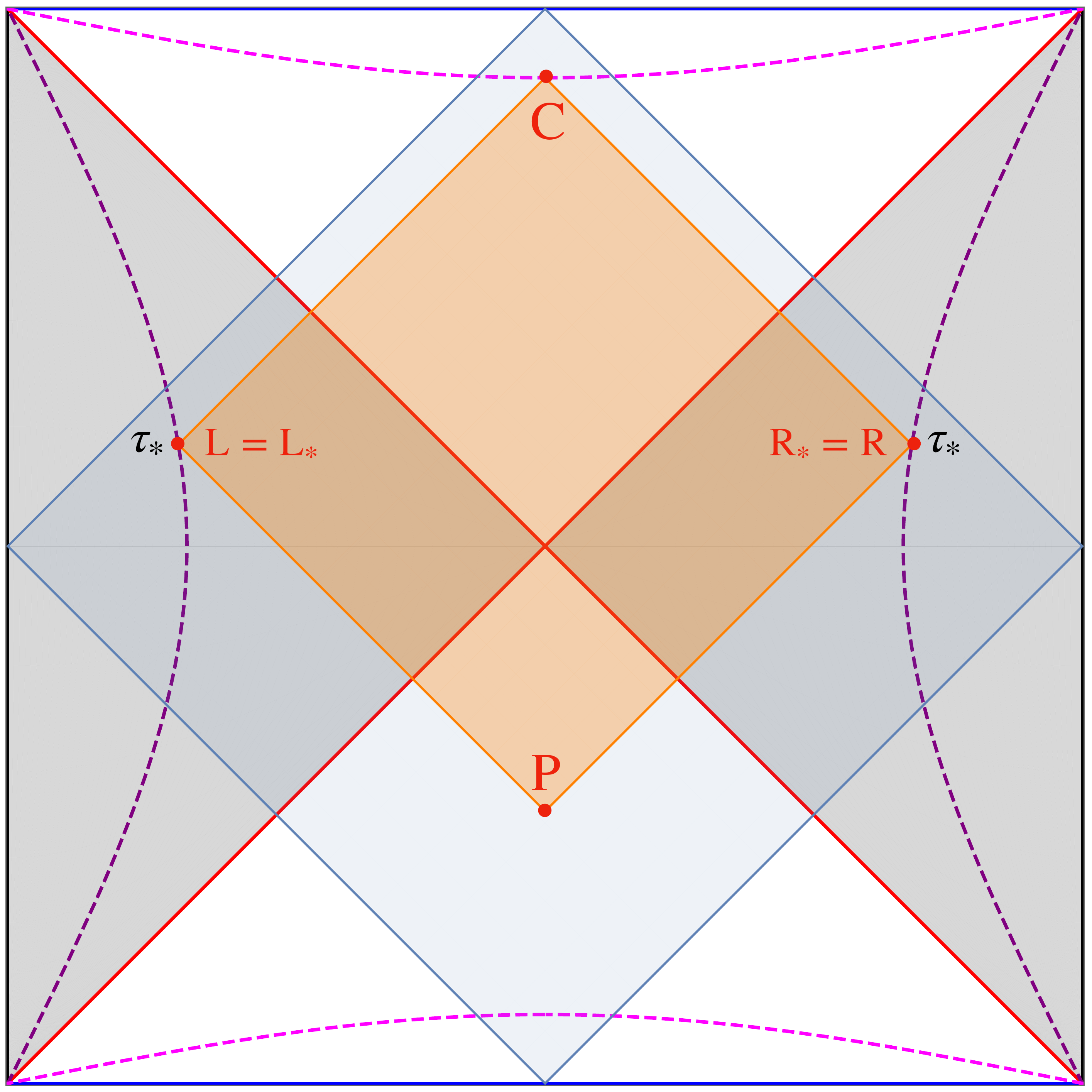}
	\qquad 
		\includegraphics[width=2.8in]{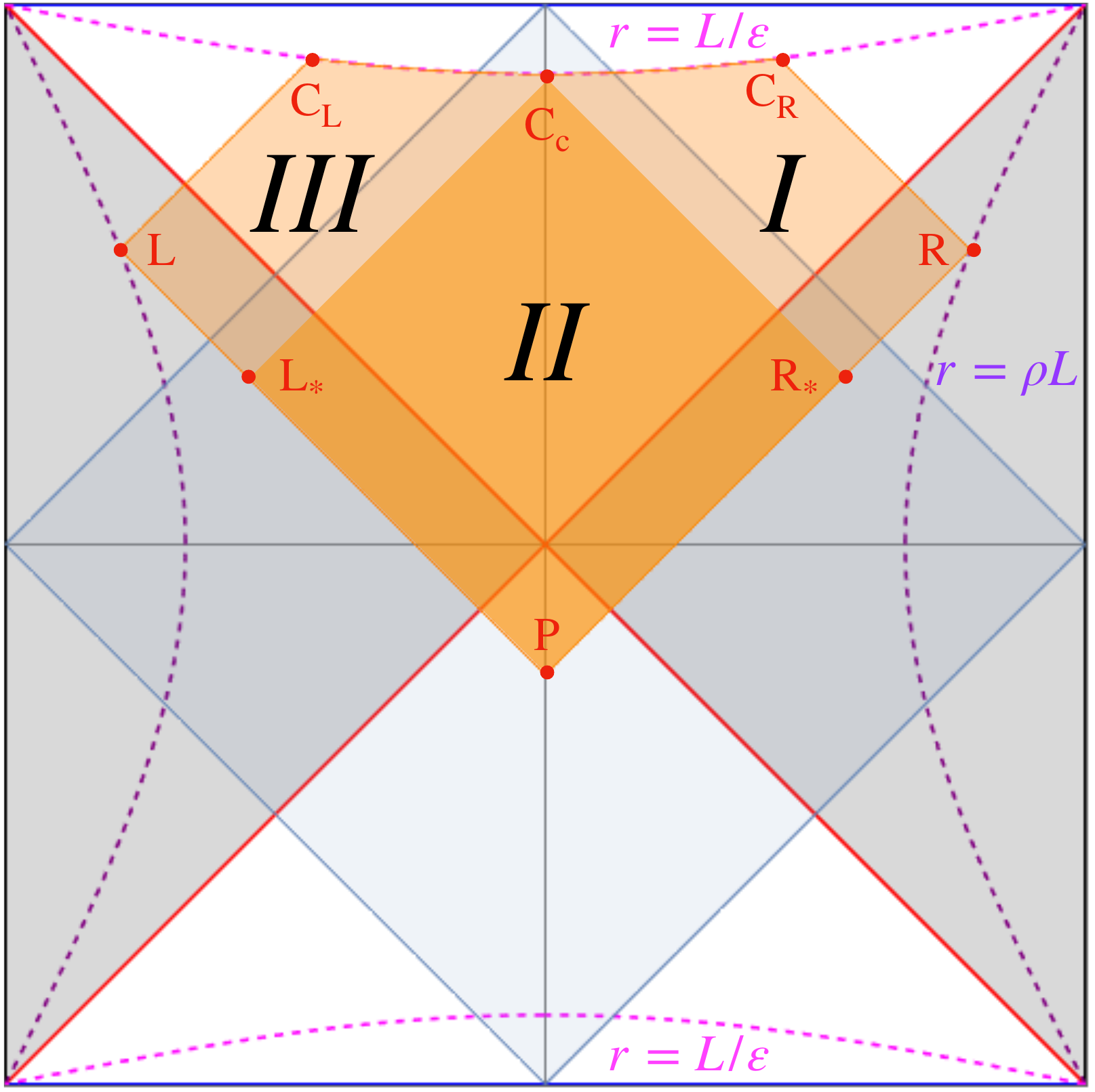}
	\caption{The orange region represents the WdW patch with a cut-off surface at $r=L/\veps$. Left: The WdW patch at the transition time, \ie $\tau= \tau_{\ast}$. Right: The WdW patch after the transition time where we divide it into three subregions as shown, with $\RNum{1}, \,\RNum{3}$ shaded light orange and $\RNum{2}$, dark orange.}\label{fig:WdWdS0034}
\end{figure}

\subsubsection*{Later time evolution of CV2.0  ($\tau \gtrsim \tau_{\infty}$)}

As observed above, the spacetime volume of the WdW patch and the corresponding holographic complexity diverges as $\tau\to\tinf$, which is only a finite time scale. To make sense of the holographic complexity at late times, we must regulate this divergence. In the present gravitational calculations, it is natural to introduce a geometric cutoff, \ie we introduce a cutoff surface at some large radius 
\begin{equation}
r= \rmax \equiv  \frac{L}{\veps} \qquad{\rm with}\quad \veps\ll1\,.
\label{regulator}
\end{equation}
For convenience, we have introduced the dimensionless parameter $\veps$ to control the position of the cutoff surface. Of course, this cutoff surface is a spacelike surface located inside the cosmological horizon. 

In evaluating the spacetime volume, we only extend our integrals up to $r=\rmax$. That is, the boundaries of the regulated WdW patch now consist of four null cones and a spacelike segment extending along the cutoff surface -- see the right panel in figure \ref{fig:WdWdS0034}. Hence, we will be able to consider the times beyond $\tauinf$. The transition between the square WdW patch considered above and this new regulated region occurs slightly before $\tinf$. That is, the WdW first touches the cutoff surface at a transition time $\tau_{\ast}$ given by
\begin{equation}
\sqrt{\frac{1-\rho}{1+\rho}}\, e^{\tau_{\ast}} = \sqrt{\frac{1-\veps}{1+\veps}} \,.
\end{equation}
 Alternatively, we can write
\begin{equation}
\label{taustar}
\begin{split}
 \tau_{\ast}&= \frac{1}{2} \log \(  \frac{(1+\rho)(1-\veps)}{(1-\rho)(1+\veps)} \)   \\
 &=\tauinf -\arctanh\,\veps
\approx   \tauinf-\veps -\frac{\veps^3}{3} +\mO\left(\veps^5\right) \,,
\end{split}
\end{equation}
where we recall $\tauinf$ is given in eq.~\reef{criticaltau}. 

Evaluating the spacetime volume is slightly more involved for $\tau \ge \tau_{\ast}$ because the corresponding WdW patch is cut off at $r=\rmax$. As shown in figure \ref{fig:WdWdS0034}, we divide the whole region into three parts. The null cones separating these subregions extend from either of the past null boundaries (at $U=U_{\rm min}$ and $V=V_{\rm min}$) up to the center of the cutoff surface (at $r=\rmax$ and $t=0$. Hence, these new null boundaries are given by
\begin{equation}
\begin{split}
U=  U_{\ast} =\sqrt{\frac{1-\veps}{1+\veps}} 
 \quad {\rm and}\quad
V=  V_{\ast} =\sqrt{\frac{1-\veps}{1+\veps}} \,.
\end{split}
\end{equation}
With this division, region $\RNum{2}$ is a square patch, and we can apply the previous calculations with appropriate boundaries. The corresponding contribution $\cvv^{\RNum{2}} $ to the holographic complexity then becomes
 \begin{equation}
\cvv^{\RNum{2}} =  \frac{4N}{d}  \( \( \mathbf{B}\(\frac{1+W_i}{2}; d, 1-d\)- \mathbf{B}\(\frac{1+W_i}{1-W_i}; d, 0\) \) \bigg|^{\rm C_c,P} _{\rm L*,R*}   - 2\pi i \) \,,\label{cvvII}
\end{equation}
with 
\begin{equation}
\begin{split}
W_{\mt{C}_\mt{c}} &= U_{\ast}  V_{\ast}  = \frac{1-\veps}{1+\veps}\,, \qquad\quad\qquad\qquad\qquad\quad
\text{C$_\mt{C}$: center of cutoff surface}\\
W_{\mt{P}} &= U_{\rm min} V_{\rm min} = e^{-2\tau} \frac{1-\rho}{1+\rho} \,, \qquad
\qquad\qquad\ \,\text{P: past tip}\\
W_{\mt{L}\ast} &= U_{\rm min} V_{\ast} = - e^{-\tau}\,\sqrt{\frac{(1-\veps)(1-\rho)}{(1+\veps)(1+\rho)}} \,,
\qquad\text{L*: left corner}\\
W_{\mt{R}\ast} &= U_{\ast} V_{\rm min} = - e^{-\tau}\,\sqrt{\frac{(1-\veps)(1-\rho)}{(1+\veps)(1+\rho)}}\,,
\qquad\text{R*: right corner} \,.\\
\end{split}
\label{bound777}
\end{equation}

Now thanks to the left-right symmetry of the configuration, the spacetime volumes of regions $I$ and $III$ are identical and we focus on the calculations for region $I$. Since the cutoff surface $r=L/\veps$ is also parametrized by $U V = \frac{1-\veps}{1+\veps}$. Accordingly, the contribution to holographic complexity from region $\RNum{1}$ is given by
\begin{equation}
\cvv^{\RNum{1}} \equiv
8N\int^{U_{\rm max}}_{U_{\ast}} dU  \int^{\frac{1-\veps}{(1+\veps)U}}_{V_{\ast}}  dV  \frac{(1+UV)^{d-1}}{\(1-U V\)^{d+1}} \,.
\end{equation}
One can perform this integral using similar methods to those above. We just note the following expression 
\begin{equation}
\int^{\frac{1-\veps}{(1+\veps)U}}_V dV \frac{2(1+UV)^{d-1}}{\(1-U V\)^{d+1}} = \frac{1}{d\, U}  \(  \frac{1}{\veps^d}- \(  \frac{1+UV}{1-UV} \)^d   \) \,.
\end{equation}
As a consequence, corresponding contribution from region \RNum{1} becomes\footnote{Note that there is no imaginary contribution, as in eqs.~\reef{eq:SpaceVolumetR}  and \reef{cvvII}. This is because both $R_\ast, R$ are outside the cosmological horizon with $W_i <0$. An imaginary contribution $-\pi i$ appears for contributions inside the horizon where $W_i >0$, \eg for F, C$_\mt{C}$, P.}
\begin{equation}
\cvv^{\RNum{1}} = \frac{4N}{d}  \(    \frac{\tau - \tau_\ast}{\veps^d}   + \(  \mathbf{B}\(\frac{1+W_i}{2}; d, 1-d\)- \mathbf{B}\(\frac{1+W_i}{1-W_i}; d, 0\)  \) \bigg|^{{\rm R}\ast}_{\rm R}   \) \,.
\label{cvvI}
\end{equation}
The corresponding $W_{\rm R}$ and $W_{\rm R*}$ are given by eqs.~\reef{bound77} and \reef{bound777}.

\begin{figure}[!]
	\centering
	\includegraphics[width=4.5in]{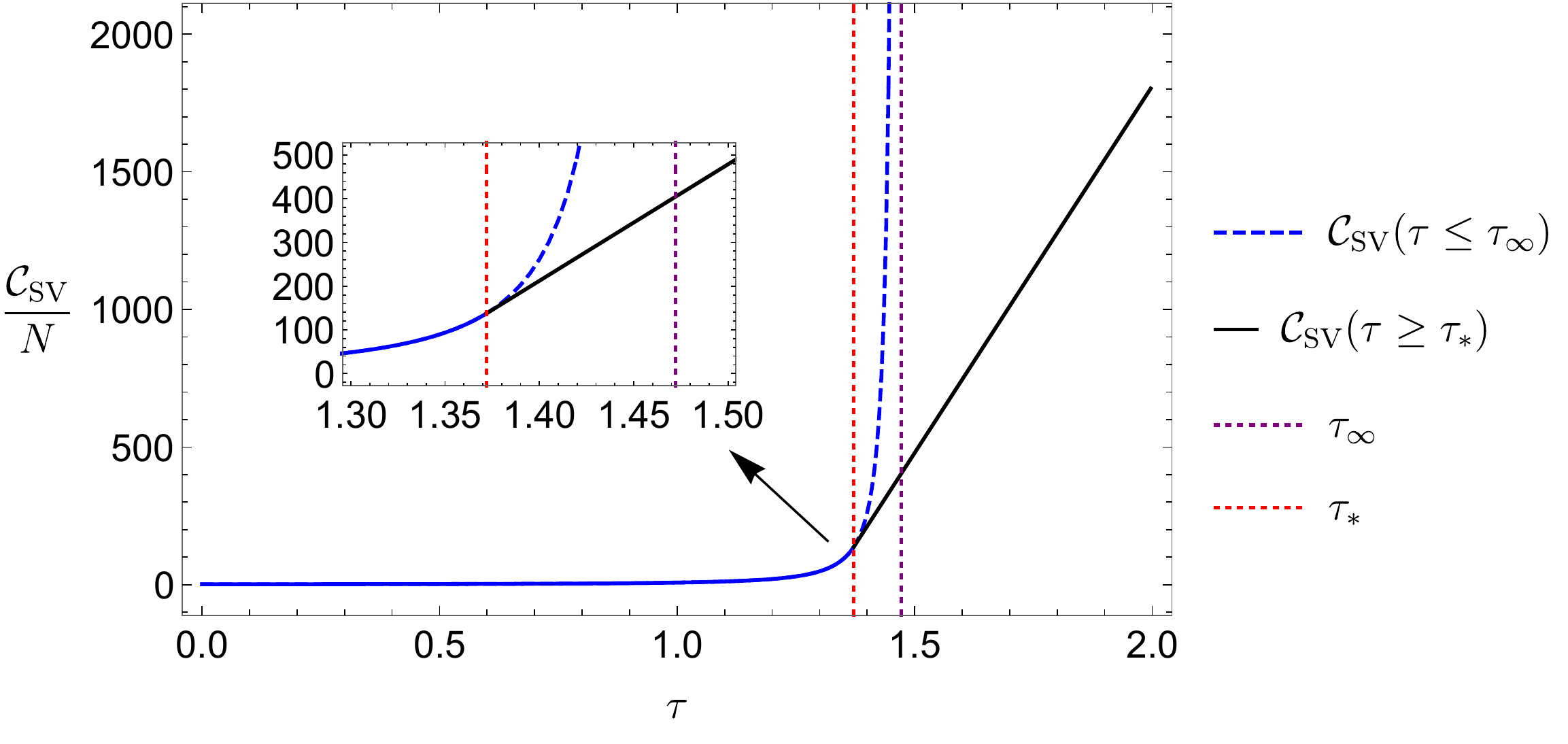}
	\caption{The time evolution of holographic complexity from spacetime volume $\cvv (\tau)$. The blue curve denotes the holographic complexity before the critical time, \ie $\cvv (\tau \le \tauinf )$. After introducing the cut-off surface, holographic complexity $\cvv (\tau \ge \tau_\ast)$ grows linearly as shown by the black line. Here we choose $d=3,\rho =\frac{9}{10}, \veps = \frac{1}{10}$.}
	\label{fig:SpacetimeVolumeFull}
\end{figure}

Combining eqs.~\reef{cvvII} and \reef{cvvI}, the regulated complexity becomes
\begin{equation}\label{eq:cvvlate}
\begin{split}
\cvv (\tau \ge \tau_\ast )&=\cvv^{\RNum{1}} + \cvv^{\RNum{2}} + \cvv^{\RNum{3}}\,,\\
&=\frac{4N}{d}  \(   \frac{2(\tau - \tau_\ast)}{\veps^d}   + \(  \mathbf{B}\(\frac{1+W_i}{2}; d, 1-d\)- \mathbf{B}\(\frac{1+W_i}{1-W_i}; d, 0\)  \) \bigg|^{{\rm C_c,P}}_{\rm L,R}  -2\pi i \) \,.
\end{split}
\end{equation}
The time evolution of holographic complexity $\cv$ from early times to late times is plotted in figure \ref{fig:SpacetimeVolumeFull}. Furthermore, the corresponding growth rate of the holographic complexity is
\begin{equation}\label{rate79}
\begin{split}
\frac{d  \cvv (\tau \ge \tau_\ast ) }{d \tau} &=  \frac{8N}{d} \(  \( \frac{\rmax}{L} \)^d -\( \frac{r_-}{L} \)^d  \) \\
&=  \frac{8N}{d} \(  \frac{1}{\veps^d}  -  \(  \frac{\cosh \tau  + \rho \sinh \tau }{\rho \cosh \tau  +  \sinh \tau}   \)^d   \)
\\
&\simeq  \frac{8N}{d} \(  \frac{1}{\veps^d}  -  1 +2d\,\frac{1-\rho}{1+\rho}\,e^{-2\tau} +{\cal O}(e^{-4\tau})    \) \,,
\end{split}
\end{equation}
where the last expression represents an expansion for large times. It is easy to show this growth rate is continuous at the transition time $\tau_*$, \ie
\begin{equation}\label{eq:CV2continuous}
\frac{d  \cvv (\tau \le \tau_\ast ) }{d \tau} \bigg|_{\tau_{\ast}}  = \frac{d  \cvv (\tau \ge \tau_\ast ) }{d \tau} \bigg|_{\tau_{\ast}} \,.
\end{equation}
Here we have evaluated eq.~\reef{totalrate0} for $\tau\to\tau_*$,
\begin{equation}
\frac{d  \cvv (\tau \le \tau_\ast ) }{d \tau} \bigg|_{\tau_{\ast}} =\frac{8N}{d}   \(  \frac{1}{\veps ^d}- \( \frac{1 + \rho^2-2\rho\veps }{2\rho -\veps(1 + \rho^2)} \)^d  \)  \simeq \frac{8N}{d \,\veps^d} \,,
\end{equation}
where the final expression is, of course, the leading  contribution for $\veps \ll 1$.

\section{CA in dS$_{d+1}$}\label{sec:CA}

In this section, we study the CA proposal \reef{defineCA} in $\dS_{d+1}$. In particular, we suppose that the holographic complexity is given by the gravitational action evaluated on the WdW patch anchored on the stretched horizon (\eg see figure \ref{fig:WdW}).  Schematically, the complexity is then defined as
\begin{equation}\label{eq:CAeq}
\ca= \frac{1}{\pi}\left(I_{\rm{bulk}}  + I_{\mt{GHY}} +  I_{\rm null} + I_{\rm{ct}} +I_{\rm{joint}}  \right)\,.
\end{equation}
Apart from the usual bulk action $I_{\rm{bulk}}$, we have included the appropriate boundary terms (\ie $I_{\mt{GHY}},\, I_{\rm null},\, I_{\rm{ct}}$) for the various codimension-one boundary segments and the joint terms $I_{\rm{joint}}$ for the codimension-two surfaces where the boundary segments intersect \eg \cite{Lehner:2016vdi,Carmi:2016wjl}. The counterterm action $I_{\rm null}$ for null boundaries are  also included to make the action invariant under reparameterization of the null boundary \cite{Lehner:2016vdi}.

The bulk contribution to the complexity is given by the usual Einstein-Hilbert action with a positive cosmological constant. Since the scalar curvature of de Sitter space is a constant, the bulk term is simply proportional to the spacetime volume of the WdW patch. Hence, we have
\begin{equation}\label{eq:Ibulk}
  \frac{I_{\rm{bulk}}}{\pi}=   \frac{1}{16\pi ^2\GN}\int d^{d+1}x \sqrt{-g}\({\cal R}-\frac{d(d-1)}{L^2}\)=\frac{d \, V_{\mt{WdW}} }{8\pi ^2\GN L^2}= \frac{d}{8 \pi^2}\, \cvv \,.
\end{equation}
Since we saw that $\cvv$ diverges as  $\tau\to \tau_{\infty}$ (see~eq.~\eqref{eq:VWdWaround}), this bulk term produces similar divergent behaviour in $\ca$. Thus, as in the previous section, we will regulate the results with a cutoff surface at $r=\rmax=L/\veps$. This is the reason that we included $I_{\mt{GHY}}$, the Gibbons-Hawking-York term \cite{GibbonsHawking,York:1972sj}, in eq.~\reef{eq:CAeq}. This term will play a role for $\tau \gtrsim \tauinf$, when the WdW patch has encountered the regulator surface and has a spacelike boundary at $r=\rmax$.

\subsubsection*{Early time evolution of CA ($\tau \le \tau_{\infty}$)}

As discussed in the previous section, at early times $\tau \le \tau_{\infty}$, the WdW patch has four null boundaries designated in eq.~\reef{bound77}.  To evaluate the corresponding boundary and joint terms, we must define the normals for these surfaces. First, we label the four null segments by the joints which they connect in the Penrose diagram (see figure \ref{fig:WdW}), \eg the null surface FR extends for the joint on the right stretched horizon to the future tip of the WdW patch. Following the prescription given in appendix A of \cite{Carmi:2016wjl}, we choose  
\begin{equation}\label{nullvec}
\begin{split}
{\rm FR\,:}\quad
   k _{\mu}dx^\mu&= \alpha\, du|_{U=U_{\rm max}} =\alpha \(dt-dr/f(r)\)|_{U=U_{\rm max}}\,,\\
   {\rm FL\,:}\quad
k'_{\mu}dx^\mu&= -\alpha'\, dv|_{V=V_{\rm max}} =- \alpha' \(dt+dr/f(r)\)|_{V=V_{\rm max}}\,,\\
{\rm PL\,:}\quad \ell_{\mu}dx^\mu&= \beta\, 
du|_{U=U_{\rm min}}=\beta \(dt-dr/f(r)\)|_{U=U_{\rm min}}\,,\\
{\rm PR\,:}\quad
\ell'_{\mu}dx^\mu&= -\beta'\, dv|_{V=V_{\rm min}} =- \beta' \(dt+dr/f(r)\)|_{V=V_{\rm min}}\,,
\end{split}
\end{equation}
where $\alpha$, $\alpha'$, $\beta$ and $\beta'$ are arbitrary (positive) constants. Note that raising the index on one of the null normals gives a vector that lies tangent to the null surface, as follows immediately from \eg $k^{\alpha}k_{\alpha}=0$. Thus, each normal defines a parametrization of the null direction along the hypersurface according to 
\begin{equation}\label{sparam}
\frac{\partial x^{\alpha}}{\partial s}=k^{\alpha}\,,
\end{equation}
with corresponding relations for each boundary segment -- see further comments below. 

Returning to eq.~\reef{eq:CAeq}, the contribution of the null boundary terms is
\begin{equation}\label{nullI}
\frac{I_{\rm null}}{\pi} = \frac{1}{8\pi^2 \GN}\sum \int ds\,d^{d-1}\Omega\,\sqrt{\gamma}\,\kappa  \,,
\end{equation}
where $\gamma_{ij}$ is the induced metric on the transverse (\ie sphere) directions and we sum over the four null segments. Now the quantity $\kappa$ is defined by $k^{\alpha}\nabla_{\alpha}k^{\beta}=\kappa k^{\beta}$ for the corresponding null normal, and measures the degree to which the parametrization fails to be affine. However, with our definition \reef{nullvec} of the normals, $s$ is an affine parameter, \ie  $k^{\alpha}\nabla_{\alpha}k^{\beta}=0$ in each case.  Hence this contribution \reef{nullI} to the holographic complexity vanishes.

Next, we have the joint contribution in eq.~\reef{eq:CAeq},
\begin{equation}\label{jointI}
\frac{I_{\rm{joint}}}{\pi} =\frac{1}{8\pi^2 \GN}\sum \int d^{d-1}\Omega\,\sqrt{\gamma}\,a\,,
\end{equation}
where we sum over the four corners of the WdW patch (F,P,L,R) in figure \ref{fig:WdW}. The integrand $a$ is defined in terms of the inner product of the null normals of the two surfaces intersecting at a given joint, and we find\footnote{The interested reader is referred to appendix A of \cite{Carmi:2016wjl} for details on the choice of signs here.}
\beqa \label{eq:WdWjoints}
{\rm F\,:}\ \ a&= & \log\frac{|k\cdot k'|}{2}=\log\frac{\alpha\alpha'}{|f\(r_+\)|}\,,
\qquad\quad\,
{\rm P\,:}\ \ a=
  \log\frac{|\ell\cdot \ell'|}{2}=\log\frac{\beta\beta'}{|f\(r_-\)|} \,,
  \\ 
{\rm L\,:}\ \ a&= &   -\log\frac{|\ell\cdot k'|}{2}=-\log\frac{\beta\alpha'}{f\(L\,\rho\)}\,,\qquad
{\rm R\,:}\ \ a=
    -\log\frac{|\ell'\cdot k|}{2}=-\log\frac{\beta'\alpha}{f\(L\,\rho\)}\,.
    \nonumber  
    \eeqa
These results are independent of the transverse coordinates and hence summing the four contributions in eq.~\reef{jointI} yields
\begin{equation}\label{jointI2}
\frac{I_{\rm{joint}}}{\pi} =\frac{N}{2\pi^2}
\[\frac{r_+^{d-1}}{L^{d-1}} \,\log\frac{\alpha\alpha' L^2}{r_+^2  - L^2}
+\frac{r_-^{d-1}}{L^{d-1}} \,\log\frac{\beta\beta'L^2}{r_-^2- L^2}-\rho^{d-1}\log\frac{\alpha\alpha'\beta\beta'}{(1-\rho^2)^2}
\]\,.
\end{equation}

Finally, we come to the null boundary counter term \cite{Lehner:2016vdi} in eq.~\reef{eq:CAeq}
\begin{equation}\label{eq:Theta}
\frac{I_{\rm{ct}} }{\pi}=\frac{1}{8\pi^2 \GN}\sum \int ds \,d^{d-1}\Omega\,\sqrt{\gamma}\ \Theta \log (\ell_{\rm ct} |\Theta|)\,,
\end{equation}
where $\Theta=\nabla_\mu k^\mu$ is the expansion on the corresponding null boundary segment. Note the appearance of an arbitrary scale $\ell_{\rm ct}$ in this expression, which introduces an ambiguity in the final value of the holographic complexity. 

In proceeding, let us focus on the FR segment for the moment. As noted in eq.~\reef{nullvec}, the null normal is simply expressed in the outgoing EF coordinate $u$. Given the metric \reef{eq:vmetric}, it then follows that $k^\mu\partial_\mu =-\alpha\partial_r$ and so from eq.~\reef{sparam}, we have the simple expression: $\partial r/\partial s=-\alpha$. Up to an overall factor, coordinate $r$ coincides with the affine parameter $s$, \ie $r=-\alpha\,s$. Similar results follow for the other null boundaries and this allows us to re-express eq.~\reef{eq:Theta} in terms of radial integrals using $ds = -dr/\alpha$. 

We now evaluate the expansion on the four boundaries with $\Theta= k^\mu \partial_\mu (\log\gamma)$ \cite{Poisson:2009pwt}, 
\begin{equation}\label{expansions}
\begin{split}
{\rm FR\,:}\quad
   \Theta&=-(d-1)\,\frac{\alpha}{r} 
   \,,\quad\quad
   {\rm FL\,:}\quad
 \Theta=-(d-1)\,\frac{\alpha'}{r} \,,\\
{\rm PR\,:}\quad 
 \Theta&=-(d-1)\,\frac{\beta'}{r} \,,\quad\quad
{\rm PL\,:}\quad
 \Theta=-(d-1)\,\frac{\beta}{r} \,.\\
\end{split}
\end{equation}
It is then straightforward to calculate the full counterterm contribution:
\begin{equation}
\begin{aligned}
\frac{I_{\rm ct}}{\pi}=&\frac{N}{2\pi^2} \bigg[ 2\(\log\frac{L}{(d-1)\ell_{\rm ct}}-\frac{1}{d-1}\)\(\frac{r_+^{d-1}}{L^{d-1}}+\frac{r_-^{d-1}}{L^{d-1}}-2\rho^{d-1}\)\\
&\qquad\qquad+\frac{r_+^{d-1}}{L^{d-1}}\log\frac{r_+^2}{\alpha\alpha' L^2}+\frac{r_-^{d-1}}{L^{d-1}}\log\frac{r_-^2}{\beta\beta' L^2}-\rho^{d-1}\log
\frac{\rho^4}{\alpha\alpha'\beta\beta'}\bigg]\,.
\end{aligned}\label{eq:fullct}
\end{equation}
Combining eqs.~\reef{eq:Ibulk}, \reef{jointI2} and \reef{eq:fullct} then yields the full expression for $\ca$, \ie 
\begin{equation}\label{eq:CAfull}
\begin{aligned}
\ca=&\frac{d}{8\pi^2}\cvv+
\frac{N}{\pi^2} \Bigg[ \(\log\frac{L}{(d-1)\ell_{\rm ct}}-\frac{1}{d-1}\)\(\frac{r_+^{d-1}}{L^{d-1}}+\frac{r_-^{d-1}}{L^{d-1}}-2\rho^{d-1}\)\\
&\qquad\qquad\quad+\frac{r_+^{d-1}}{L^{d-1}}\log\frac{r_+}{\sqrt{r_+^2-L^2}}+\frac{r_-^{d-1}}{L^{d-1}}\log\frac{r_-}{\sqrt{r_-^2-L^2}}-2\rho^{d-1}\log \frac{\rho}{\sqrt{1-\rho^2}}\Bigg]\,.
\end{aligned}
\end{equation}
where the exact expression for $\cvv$ is given in eq.~\eqref{eq:SpaceVolumetR}. We note that the normalization constants (\ie $\alpha$, $\alpha'$, $\beta$, $\beta'$) appearing in $I_{\rm{joint}}$ and $I_{\rm{ct}}$ separately have disappeared from this final expression. Of course, varying these constants corresponds to scaling the parametrization of the null boundaries, as noted in the discussion about eq.~\reef{expansions}, and the role of $I_{\rm{ct}}$ is to ensure that $\ca$ is independent of such reparametrization \cite{Lehner:2016vdi}. 

We can make the time dependence explicit using the expressions in footnote \ref{footy78} (as well as eq.~\reef{criticaltau}, which yields
\begin{equation}
\label{magnificent}
\begin{split}
\ca=\frac{d\, \cvv}{8\pi^2}&+  \frac{N}{\pi^2}  \left[   \coth^{d-1} (\tauinf -\tau)  \(   \log \frac{  L \cosh(\tauinf -\tau) }{(d-1)\ell_{\rm ct}}       -\frac{1}{d-1} \)  \right.  \\   
& \left. \qquad\quad +  (\tau \to - \tau )  - 2\tanh^{d-1}\tauinf \(   \log  \frac{ L \sinh\tauinf}{(d-1)\ell_{\rm ct}}     -\frac{1}{d-1} \)      \right] \,.
\end{split}
\end{equation}
Hence, we see that similar to the case for the CV2.0 proposal, $\ca$ diverges as the boundary time approaches $\tau_{\infty}$. In this case, divergent contributions are coming both from the spacetime volume of the WdW patch (\ie the bulk action contribution), the joint terms and the boundary counterterms. Let us also comment in passing that the boundary contribution proportional to $-\frac{1}{d-1}\,\frac{r_+^{d-1}}{L^{d-1}}$ in eq.~\reef{eq:CAfull} cancels the leading divergence in $\frac{d}{8\pi^2}\cvv$ as $\tau\to\tauinf$, or alternatively as $r_+\to\infty$. 
Expanding for small $\tauinf-\tau$, the holographic complexity $\ca$ is given by 
\begin{equation}\label{eq:CAearlylimit}
\lim_{\tau\to \tau_{\infty}}\ca \approx \frac{N}{\pi^2}\(\frac{\log\frac{L}{(d-1)\ell_{\rm ct}}}{(\tauinf-\tau)^{d-1}}+ ( d -1 ) \( \frac{1}{2(d-3)} -\frac{1}{3} \log \frac{L}{ \ell_{\rm ct} (d-1)} \) \frac{1}{(\tauinf-\tau)^{d-3}}+\cdots \)\,.
\end{equation}
Hence the strength of the leading term is controlled by $\log\frac{L}{(d-1)\ell_{\rm ct}}$ and in particular, by the counterterm length scale $\ell_{\rm ct}$ due to the cancellation noted above. There has been no constraint to fix $\ell_{\rm ct}$, but we see here that one must choose $\ell_{\rm ct} < L/(d-1)$ to make holographic complexity positive.

\begin{figure}[h!]
	\centering
	\includegraphics[width=5.5in]{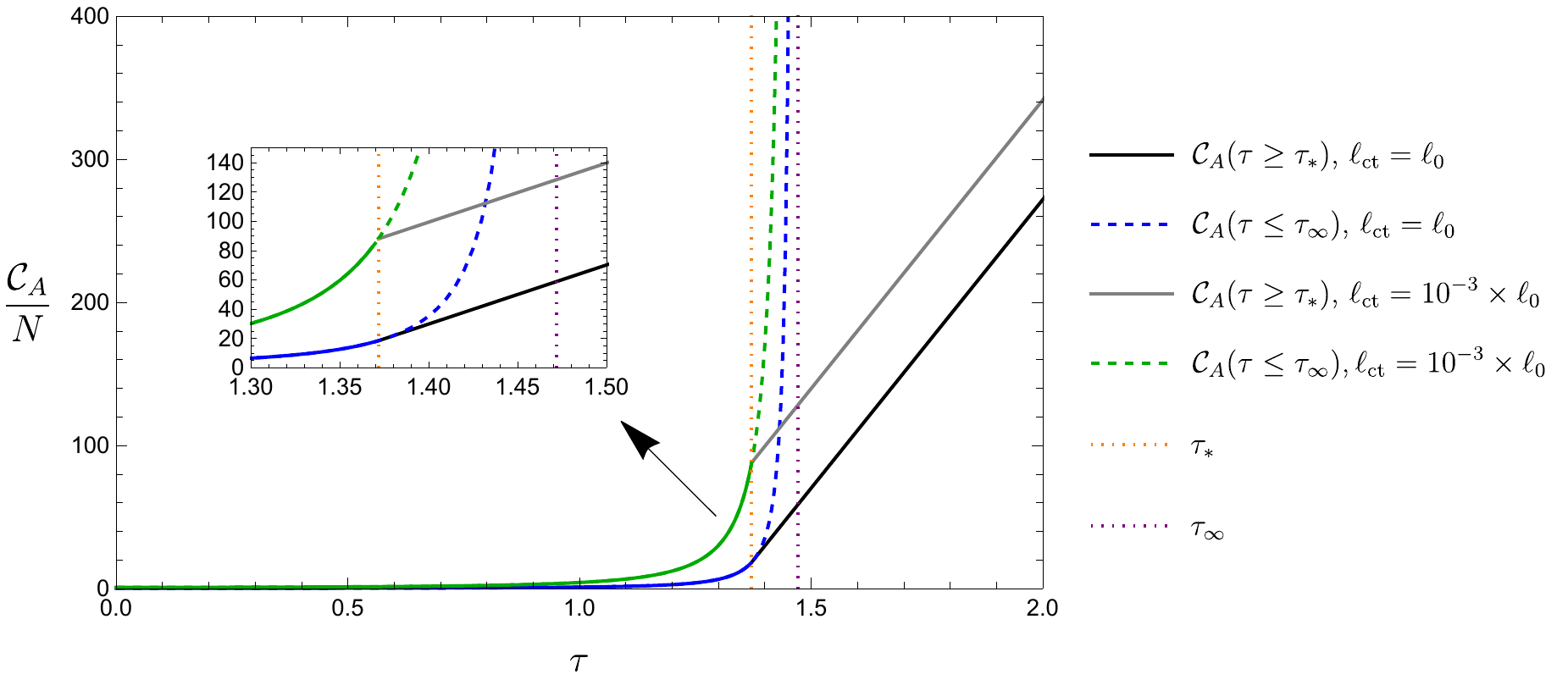}
	\caption{ The holographic complexity $\ca$ plotted against the boundary time for two choices of $\ell_{\rm ct}$. The dashed lines show the unregulated divergent behaviour for $\tau<\tauinf$. The solid black and gray curves show the corresponding evolutions after the introduction of the cutoff surface. Note the linear growth for $\tau\geq\tau_*$. For the choice $\ell_{\rm ct}=\ell_0$, the regulated curve is continuous in the first derivative at $\tau=\tau_*$ -- see eq.~\reef{eq:l0}. The typical non-smooth evolution of the regulated complexity is shown for the arbitrarily chosen value $\ell_{\rm ct}=10^{-3}\times\ell_0$. Here we choose $d=3$, $\rho=\frac{9}{10}$ and $\veps=\frac{1}{10}$. }
	\label{fig:Cafig}
\end{figure}

Taking the derivative with respect to the boundary time $\tau$, we can further obtain the growth rate of the holographic complexity \reef{magnificent} at early times, \ie 
\begin{equation}
\begin{split}
\frac{d \ca(\tau \le \tauinf)}{d\tau} &= \frac{N (d-1)}{\pi^2}  \[   \frac{\coth^d(\tauinf-\tau)}{\cosh^2(\tauinf-\tau)}\,  \log \frac{L\cosh(\tauinf-\tau)}{(d-1)\ell_{\rm ct}}    -  ( \tau \to -\tau  ) \] \,. 
\end{split}
\end{equation}
Similar to CV2.0, we obtain the hyperfast growth of $\ca$ by approaching the critical time $\tauinf$, \ie
\begin{equation} \label{eq:CAderivativeearly}
\lim_{\tau\to \tau_{\infty}} \frac{d \ca  (\tau \le \tauinf )}{d \tau}\approx \frac{N}{\pi^2} \frac{(d-1)\log\(\frac{L}{(d-1)\ell_{\rm ct}}\)}{(\tauinf -\tau)^d}   +\mathcal{O}\left(\frac{1}{(\tau_{\infty}-\tau)^{d-2}} \right) \,.
\end{equation}
This behaviour is sketched in figure \ref{fig:Cafig}.

\subsubsection*{Later time evolution of CA  ($\tau \gtrsim \tau_{\infty}$)}

Above, we have seen that the holographic complexity diverges as the boundary time approaches the critical value $\tau=\tauinf$. As in section~\ref{sec:CV20}, we regulate the divergence by introducing a cutoff surface at $r=\rmax=L/\veps$. Thus, we must modify our previous calculations to account  for the new spacelike boundary segment of the WdW patch which appears in this regime.

Recall that the WdW patch reaches the cutoff surface at the time $\tau=\tau_*$ which is slightly before the critical time $\tauinf$, as shown in eq.~\reef{taustar}.
Eq.~\reef{eq:Ibulk} still applies in this regime and hence we have the bulk contribution
\begin{equation}
\frac{I_{\rm bulk}(\tau\geq\tau_\ast)}{\pi}=\frac{d}{8\pi^2}\cvv(\tau\geq\tau_\ast)\,,
\label{Ibulk2}
\end{equation}
where $\cvv(\tau\geq\tau_\ast)$ is given by eq.~\eqref{eq:cvvlate}. From eq.~\eqref{rate79}, we see that this term contributes to the linear growth of holographic complexity $\ca$ at late times.

Turning to the joint contribution \eqref{jointI}, we note that the regulated WdW patch now has five joints labeled (C$_\mt{R}$,C$_\mt{L}$,L,R,P) in figure.~\ref{fig:WdWdS0034}. The results for the last three remain the same as in eq.~\reef{eq:WdWjoints}. 
To evaluate the joint terms on C$_\mt{R}$ and C$_\mt{L}$, where the future null boundary segments intersect the spacelike surface $r=\rmax$, we must first 
 introduce the (future-pointing) unit normal to the cutoff surface
\begin{equation}
	n_{\alpha}dx^{\alpha}=\frac{dr}{\sqrt{|f(r)|}} \bigg|_{r=\rmax}.
\end{equation}
 For these two joints, the integrand $a$ takes the form
\begin{equation}
a=\begin{cases}
	\log|k\cdot n|=\log\frac{\alpha}{\sqrt{\rmax^2/L^2-1}} \,,\,\qquad\qquad  \text{C$_\mt{R}$}\,,\\ 
  \log|k'\cdot n|=\log\frac{\alpha'}{ \sqrt{\rmax^2/L^2-1} }  \,,\qquad\qquad \text{C$_\mt{L}$}\,.\\
\end{cases}
\end{equation}
Note that these contributions from $\text{C}_\mt{R},\text{C}_\mt{L}$ on the cutoff surface do not vary with the boundary time $\tau$, since they only depend on the fixed radial coordinate $r=\rmax$. Summing over contributions from all five joints, \ie $\text{C}_\mt{R},\text{C}_\mt{L}, \text{L}, \text{R}, \text{P},$ yields
\begin{equation}
\frac{I_{\rm{joint}}(\tau\geq\tau_*)}{\pi} =\frac{N}{2\pi^2}
\[\frac{1}{\veps^{d-1}} \,\log\frac{\alpha\alpha' \veps^2}{1-\veps^2}
+\frac{r_-^{d-1}}{L^{d-1}} \,\log\frac{\beta\beta' L^2}{r_-^2-L^2}-\rho^{d-1}\log\frac{\alpha\alpha'\beta\beta'}{(1-\rho^2)^2}
\]\,.
\label{Ijoint2}
\end{equation}

The Gibbons-Hawking-York boundary term is as usual
\begin{equation}
\frac{I_{\mt{GHY}}    }{\pi}= \frac{1}{8\pi^2 \GN} \int_{\mt{cutoff}}\!\!\!\!\!   d^{d}x\,\sqrt{h}\,K,
\end{equation}
where $K$ is the trace of the extrinsic curvature $K=\nabla_{\alpha}n^{\alpha}$, in our case evaluated on the cutoff surface. It is straightforward to show that the trace of the extrinsic curvature is constant on any surface of constant $r$, \ie 
\begin{equation}
    K=  - \sqrt{-f(r)} \partial_r  \log  \sqrt{h}=-\frac{d\,r^2-(d-1)L^2}{r\sqrt{r^2-L^2}}.
\end{equation}
Noting the coordinate time at the joints is given by $t=( \tau-\tau_\ast)L$, it is then a simple matter to integrate $K$ over the cutoff surface, providing a contribution to the complexity in terms of 
\begin{equation}
 \frac{I_{\mt{GHY}}}{\pi}=\frac{N}{\pi^2} \frac{\(d-(d-1)\veps^2\)  }{\veps^{d}}\(  \tau-\tau_\ast \)\,.
 \label{IGHY2}
\end{equation}
Recall that $\tau_*$ is given in eq.~\eqref{taustar}. Clearly, this contribution to the complexity grows linearly with boundary time. 

The boundary counterterm contribution to the holographic complexity is also modified in the regime $\tau\geq\tau_*$. In particular, both the future boundary segments FR and FL are shortened, corresponding to an appropriate change of integration limits in eq.~\eqref{eq:Theta}, \ie the radial integration ends at $\rmax$ rather than $r_+$. The resulting counterterm contribution becomes
\begin{equation}
\begin{aligned}
\frac{I_{\rm ct}(\tau\geq\tau_*)}{\pi}=\frac{N}{2\pi^2} \Bigg[&2 \(\log\frac{L}{(d-1)\ell_{\rm ct}}-\frac{1}{d-1}\)\(\frac{1}{\veps^{d-1}}+\frac{r_-^{d-1}}{L^{d-1}}-2\rho^{d-1}\)\\
&+\frac{1}{\veps^{d-1}}\log\frac{1}{\veps^2 \alpha \alpha'} + \frac{r_-^{d-1}}{L^{d-1}}\log\frac{r_-^2}{\beta \beta'L^2}-\rho^{d-1}\log\frac{\rho^4}{\alpha\alpha'\beta\beta'}\Bigg].
\end{aligned}
\label{Ict2}
\end{equation}

Combining the contributions from eqs.~\reef{Ibulk2}, \reef{Ijoint2}, \reef{IGHY2} and \reef{Ict2}, the expression for the full CA complexity in the later time regime $\tau\geq\tau_*$ is\footnote{This result is written in a way that makes clear that it coincides with eq.~\reef{magnificent} when $\tau\to\tau_*$. The second line above can also be written in terms of the cutoff using $\veps=\tanh(\tauinf-\tau_*)$ from eq.~\reef{taustar}.}
%
%
%
\begin{equation}\label{eq:CAfull02}
\begin{split}
\ca(\tau\geq\tau_{\ast})=&\frac{d\, \cvv (\tau\geq\tau_*)}{8\pi^2}+ \frac{N}{\pi^2}\frac{\(d-(d-1)\veps^2\)  \(   \tau-\tau_\ast   \)}{\veps^{d}}\\
&\qquad+ \frac{N}{\pi^2}  \left[   \coth^{d-1} (\tauinf -\tau_{\ast})  \(   \log \frac{  L \cosh(\tauinf -\tau_{\ast}) }{(d-1)\ell_{\rm ct}}       -\frac{1}{d-1} \)  \right.  \\ 
& \left. \qquad\quad+  (\tau_{\ast} \to - \tau )    - 2\tanh^{d-1}\tauinf \(   \log  \frac{ L\sinh\tauinf}{(d-1)\ell_{\rm ct}}     -\frac{1}{d-1} \)     \right] \,.
\end{split}
\end{equation}
%
Further, for $\tau \ge \tau_\ast$, the complexity growth rate becomes
\begin{equation} \label{eq:CAderivativelate}
\begin{split}
\frac{d \ca  (\tau \ge \tau_{\ast} )}{d \tau} &= \frac{N}{\pi^2}   \[     \frac{d+1-(d-1)\veps^2}{\veps^{d}}   - \frac{\coth^d(\tauinf+\tau)}{\cosh^2(\tauinf+\tau)}  \,\log  \frac{L\cosh(\tauinf+\tau)}{(d-1)\ell_{\rm ct}}  \] \,.
\end{split}
\end{equation}

If we consider the limit where the cutoff is large, \ie $\veps\ll1$, we see that the holographic complexity grows linearly in this later time regime, \ie 
\begin{equation}
\frac{d \ca  (\tau \ge \tau_{\ast} )}{d \tau}\approx \,\, \frac{N}{\pi^2}\,\frac{\(d+
    1-(d-1)\veps^2\)}{\veps^d}\,,
\end{equation}
where these leading contributions are coming from both the bulk term and also the GHY boundary term.\footnote{Even with $\veps\sim{\cal O}(1)$, this linear growth appears at late times where $\tau\gg1$.}

As we have seen in section~\ref{sec:CV20}, the evolution of $\cvv$ is continuous in the first derivative when we reach the cutoff surface at $\tau=\tau_\ast$. However, our $\ca$ results are somewhat ambiguous because of the presence of the arbitrary counterterm scale $\ell_{\rm ct}$. Further, this scale controls the leading contributions to the hyperfast growth as $\tau\to\tauinf$ -- see eq.~\reef{eq:CAderivativeearly}. On the other hand, the leading  contribution (for $\veps\ll1$) to the derivative right after the critical time is independent of $\ell_{\rm ct}$.  As a result, there will generally be a discontinuous jump in the first derivative of $\ca$ when the WdW first intersects the cutoff surface. However, we note that there is a unique choice of $\ell_{\rm ct}$ which makes the complexity evolve continuously in the first derivative as the WdW patch hits the cutoff surface, \ie
\begin{equation}\label{eq:l0}
\ell_{\rm ct}=\ell_0\equiv L\frac{e^{-\frac{1+d-(d-1)\veps^2}{(d-1)(1-\veps^2)}}}{(d-1)\sqrt{1-\veps^2}}.
\end{equation}
One could then consider the continuity of the growth of the complexity as a possible matching condition which fixes $\ell_{\rm ct}$, removing the the ambiguity in the definition of $\ca$.

\section{CV in dS$_{d+1}$}\label{sec:dSCV}

In the previous two sections, we have shown that the holographic complexity, as well as the growth rate, for CV2.0 and CA in $\dS_{d+1}$ are both divergent when the boundary time approaches the finite critical time $\tauinf$. To make sense of the results beyond this time, we regulate the holographic complexity by introducing a cutoff surface near the asymptotic boundary $i^+$. With this approach, the complexity exhibits linear growth for subsequent times where the growth rate is controlled by the regulator. In the following section, we consider the dS version of complexity=volume \reef{defineCV} and show a similar story emerges. We focus on general $\dS_{d+1}$ spacetimes with $d>1$. The case of $d=1$ (\ie dS$_2$) can be solved completely analytically but is somewhat exceptional -- see comments around eq.~\reef{eq:limitCVdS2}.  Hence we reserve a complete discussion of this case to appendix \ref{app:dS2}. We note that extremal hypersurfaces and the CV proposal in $\dS_2$ were recently discussed in \cite{Chapman:2021eyy}. 


\begin{figure}[ht!]
	\centering
	\includegraphics[width=2.5in]{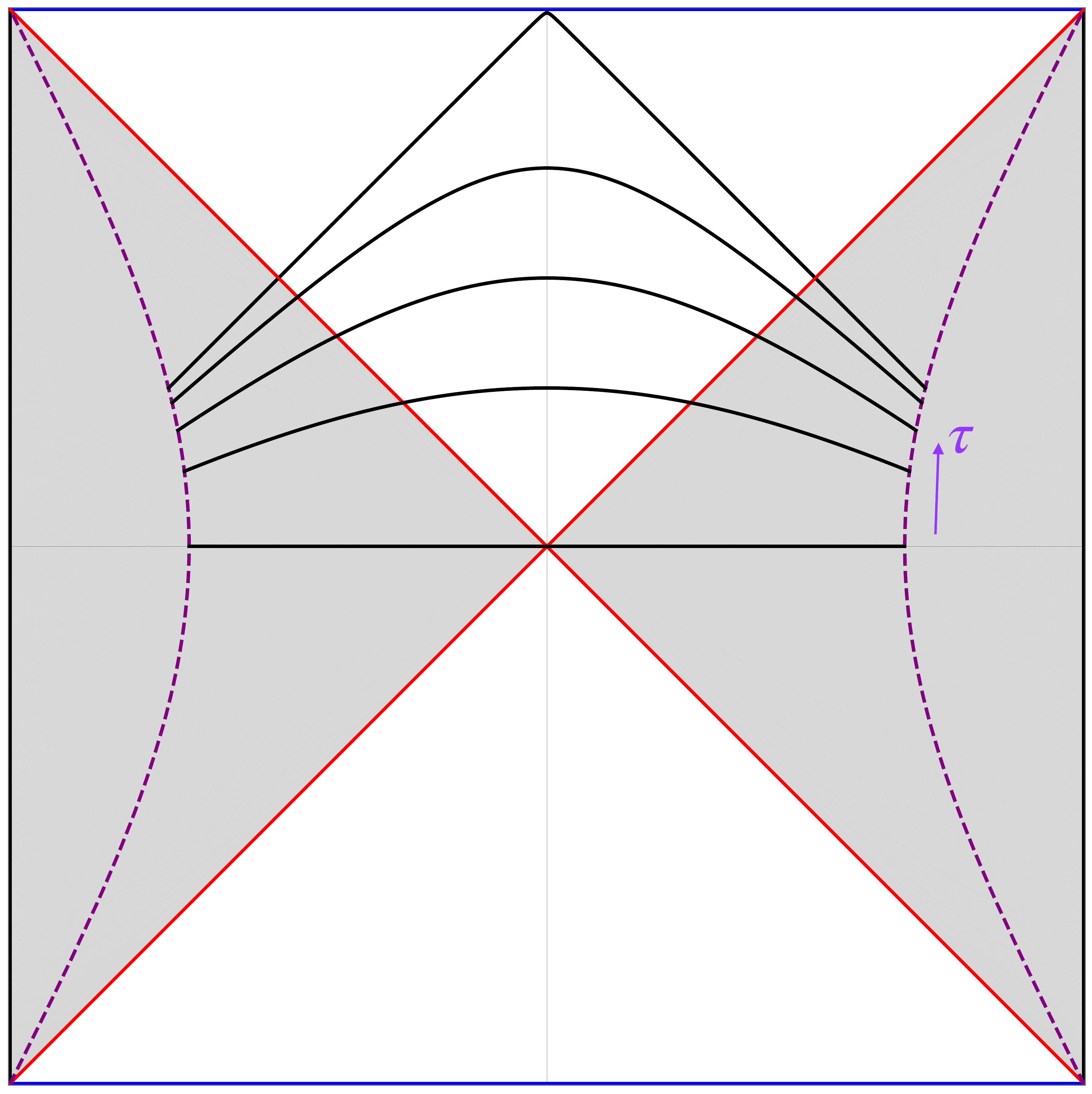}
	\caption{Extremal surfaces with anchoring on the stretched horizon $r=\rho L$ in $\dS_{d+1}$. Each black curve denotes an extremal surface associated with a conserved momentum $P_u (\tau)\ge 0$ and a boundary time $\tau \ge 0$.}
	\label{fig:dSextremal}
\end{figure}


\subsection*{Time evolution of extremal surfaces in $\dS_{d+1}$}
The CV conjecture \reef{defineCV} identifies holographic complexity as the volume of a codimension-one extremal surface, where as before we take the `boundary' time slice $\S$ to be the surfaces $\tR=-\tL=\tau L$ on the stretched horizons (\ie $r=\rho\, L$) in the two static patches. To analyze the time evolution of extremal surfaces in $\dS_{d+1}$, we follow the analysis presented for the extremal surfaces in asymptotically AdS spacetimes, \eg see \cite{Carmi:2017jqz,Chapman:2018lsv,Belin:2021bga}. The idea is that with an appropriate gauge-fixing condition, the profile of the extremal surfaces is determined by solving for the motion of a classical particle moving in an effective potential. 

Let us consider a candidate surface $\mB$ extending between the stretched horizons in $\dS_{d+1}$. Assuming the surface respects the spherical symmetry of the background geometry,  we parametrize the profile as $\(u(\lambda), r(\lambda)\)$, where $\lambda$ denotes a `radial' coordinate on $\mB$.  The corresponding  holographic complexity \reef{defineCV} would then be given by 
\begin{equation}\label{eq:dSVolume}
\cv = \frac{1}{G_{\mt{N}}L} \int_{\mB} \sqrt{h}  =\frac{4N}{L}\int_{}  \sqrt{- f(r)\dot{u}^2  - 2 \dot{u} \dot{r} } \(\frac{r(\lambda)}{L} \)^{d-1}\,d \lambda\,,
\end{equation}
where $\dot{x}=\frac{dx(\lambda)}{d\lambda}$. Finding the extremal surface is analogous to solving a one-dimensional classical mechanics problem where we  identify the Lagrangian as the integrand of the integral above: $\mL = \sqrt{- f(r)\dot{u}^2  - 2 \dot{u} \dot{r} } \(\frac{r(\lambda)}{L} \)^{d-1}$. The above integrand is invariant under reparametrizations $\lambda \to g(\lambda)$, and so we choose a convenient gauge 
\begin{equation}\label{eq:gauge}
\sqrt{- f(r)\dot{u}^2  - 2 \dot{u} \dot{r} } =  \(\frac{r}{L} \)^{d-1} \,.
\end{equation}
The holographic complexity \reef{eq:dSVolume} then reduces to 
\begin{equation}
\cv = \frac{4N}{L} \int \(\frac{r}{L} \)^{2(d-1)} \, d \lambda \,,
\label{cvv1}
\end{equation}
where the integral would be performed on the extremal surface. 

Since $\mL$ does not have any explicit dependence on $u$, the corresponding momentum is conserved\footnote{Note that we have introduced an extra minus sign in the definition of $P_u$ to simplify the following equations. Most of the later expressions are still similar to the AdS case despite this extra sign.}
\begin{equation}\label{eq:definePu}
P_u \equiv  -\frac{\partial\mathcal{L}}{\partial\dot{u}}= \( \frac{r}{L} \)^{(d-1)} \frac{f(r)\dot{u} +\dot{r}}{\sqrt{-f(r)\dot{u}^2 - 2\dot{u}\dot{r}}}= f(r)\dot{u} +\dot{r} \,,
\end{equation} 
where we have substituted the gauge-fixing condition \reef{eq:gauge} to simplify the final expression.
Combining eqs.~\reef{eq:gauge} and \reef{eq:definePu} allows us to solve for the profile of the extremal surface as follows
\begin{equation}\label{eq:dudr}
\dot{r} = \pm \sqrt{ P_u^2 + f(r) \(\frac{r}{L}\)^{2(d-1)}} \,,\qquad
\dot{u}=  \frac{ P_u- \dot{r} }{f(r)}   \,.
\end{equation} 
Without loss of the generality in the following analysis, we will only focus on the solutions with $\dot{r} \ge 0$, \ie trajectories originating at the stretched horizon and moving into the region beyond the cosmological horizon.

\begin{figure}[ht!]
	\centering
	\includegraphics[width=4in]{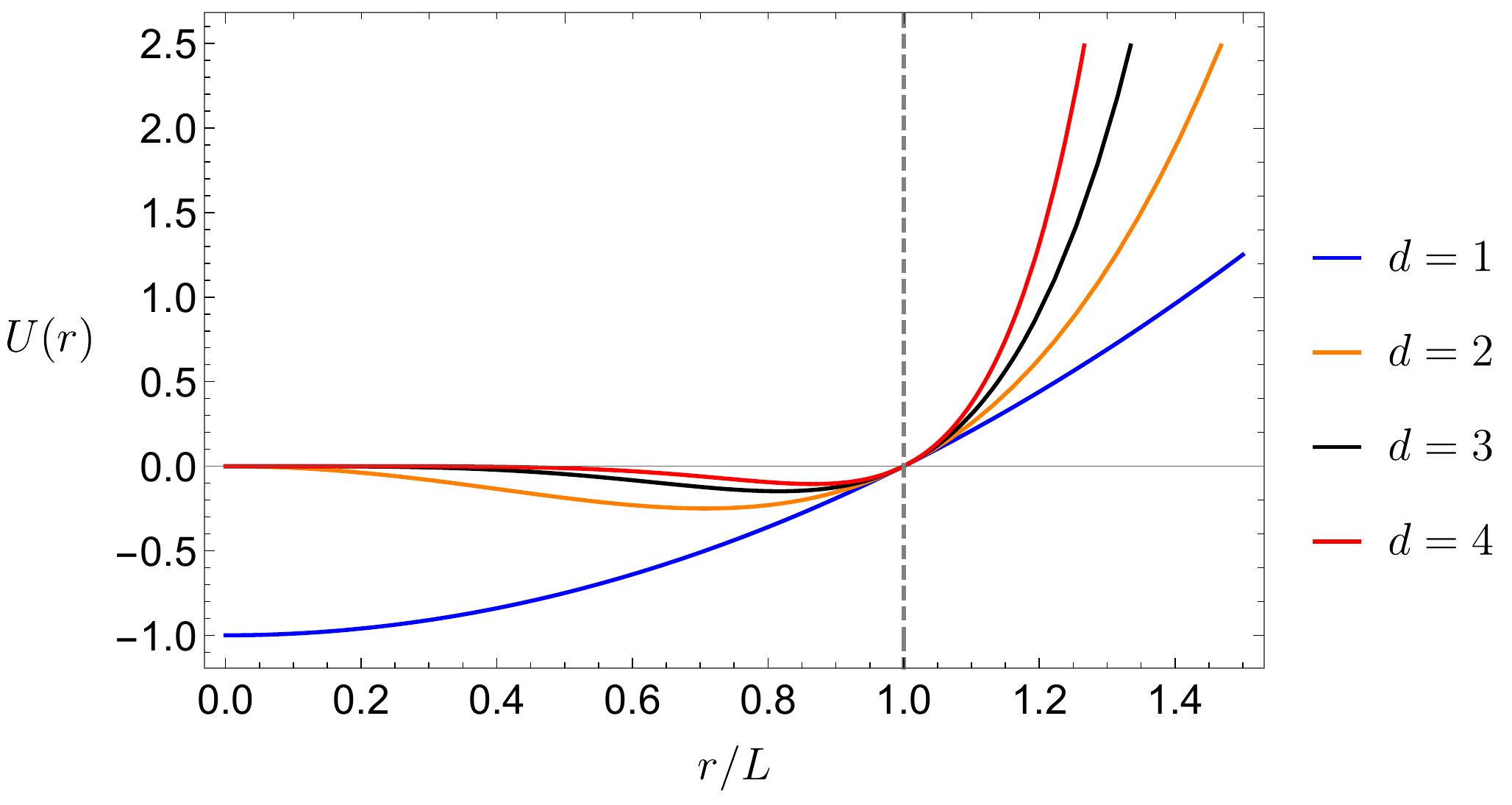}
	\caption{The effective potentials $U(r)$ for the extremal surfaces in dS$_{d+1}$.}
	\label{fig:dSpotential}
\end{figure}

Further insight comes from recasting the $\dot{r}$ equation above as
\begin{equation}\label{class}
\dot{r}^2 + U(r) = P_u^2 \qquad\text{with} \quad U(r)= -f(r) \(\frac{r}{L}\)^{2(d-1)} \,.
\end{equation}
Here we have the Hamiltonian equation for a particle moving in an effective potential $U(r)$ with an effective energy $P_u^2$. The effective potential for various dimensions is shown in figure \ref{fig:dSpotential}. A point worth stressing is the crucial difference between the potentials of $\dS_{d+1}$ and those typically studied in $\text{AdS}_{d+1}$ is the former does not contain any local maximum. In asymptotically $\text{AdS}_{d+1}$ black holes, the local maximum plays a vital role in producing the linear growth of holographic complexity, \eg \cite{Carmi:2017jqz,Chapman:2018lsv,Belin:2021bga}. For later use, we also note that the equation determining the time coordinate is given by 
\begin{equation}\label{eq:dtdr}
\dot{t}= \dot{u} + \frac{\dot{r}}{f(r)}= \frac{P_u \, \dot{r}}{
f(r)\sqrt{P_u^2+f(r)\,\,(r/L)^{2(d-1)}}}   \,.
\end{equation}

\subsection*{Time evolution of CV}
The extremal surfaces are all anchored at the stretched horizon, and hence the relevant solutions of eq.~\reef{class} begin at the minimal radius $r_{\rm min}= L\,\rho$. They then proceed to larger radii until they hit the potential at the turning point $r=r_{\rm turn} \ge L$. This turning point is determined by setting $\dot{r}=0$ in eq.~\reef{class}, which yields 
\begin{equation}
P_u^2 = \(  \frac{\rturn^2}{L^2} -1\) \(\frac{\rturn^2}{L^2} \)^{d-1} \,.
\label{return}
\end{equation}
This part of the trajectory corresponds to the first half of the extremal surface. The trajectory then `reverses' rebounding from the turning point and proceeds towards the stretched horizon on the left side with $\dot{r}<0$, as shown in figure \ref{fig:dS2extremalsurface}.   
Given  the $\dot{r}$ equation \eqref{eq:dudr}, we can rewrite the holographic complexity \reef{cvv1} as
\begin{equation}\label{eq:Vrintegral}
\begin{split}
\cv (\tau)&=\frac{8N}{L} \int^{\rturn}_{\rmin}   \frac{dr}{\dot{r}} \(\frac{r}{L}\)^{2(d-1)} 
=\frac{8N}{L} \int^{\rturn}_{\rmin} dr \, \frac{(r/L)^{2(d-1)} }{\sqrt{P_u^2+f(r)\,\,(r/L)^{2(d-1)}}}\,,\\
\end{split}
\end{equation}
where the radial integral only covers half of the extremal surface. As before, we are considering the symmetric configuration with $t_{\mt{R}}=\tau L = - \tL$ in the following. 

As noted in \cite{Belin:2021bga}, we can consider 
the time evolution of the extremal surface for an infinitesimal interval as generating a perturbation of the initial `trajectory'. Then working with the original `action' \reef{eq:dSVolume}, the standard analysis yields the conclusion that the time derivative of the holographic complexity is given by the momentum $P_u$ evaluated on the `end points', \ie
\begin{equation}\label{eq:dVdtau}
\frac{d\cv}{d\tau} = 8\, N\, P_u(\tau)\,.
\end{equation}
Comparing with the AdS case with a similar conserved momentum $P_v$ \cite{Belin:2021bga}, we note that the extra minus sign in our definition of conserved momentum \eqref{eq:definePu} is compensated by the fact that the stretched horizon at $r_{\rm min}$ is the lower limit of integration here. In contrast, the asymptotic boundary in AdS is the upper limit instead.

\begin{figure}[ht!]
	\centering
	\includegraphics[width=4in]{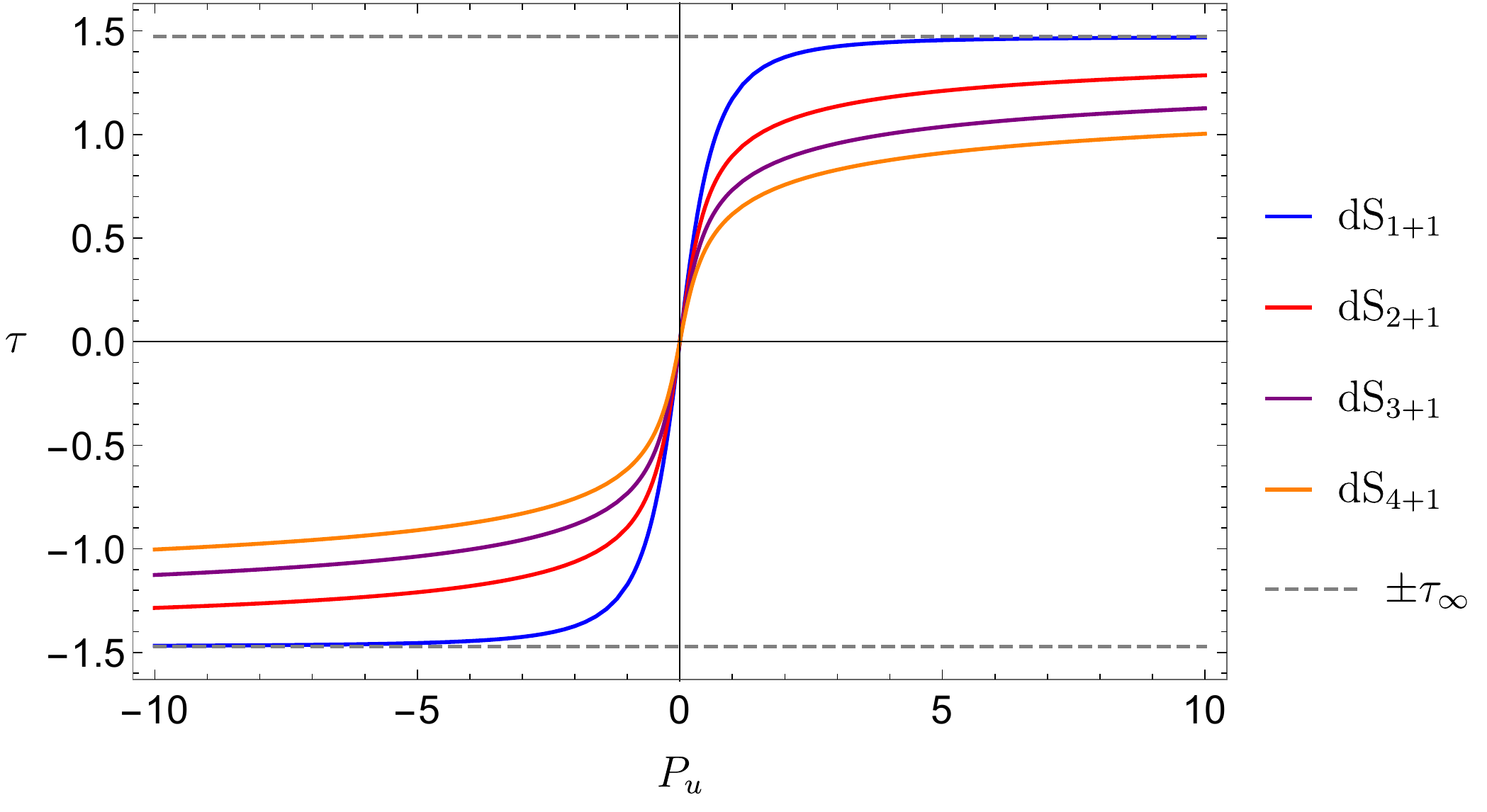}
	\caption{Boundary times $\tau (P_u)$ at various dimensional spacetime $\dS_{d+1}$ as a function of the conserved momentum $P_u$.
	We take the stretched horizon located at $\rho=9/10$ in this plot. The corresponding critical times $\pm \tauinf$ are given by the limit $|P_u| \to \infty$ and denoted by the dashed line. }
	\label{fig:dS3tR}
\end{figure}

In order to determine the evolution of holographic complexity, we still need to determine the relation between the conserved momentum $P_u$ and the time $\tau$ along the stretched horizon. Here we integrate eq.~\eqref{eq:dtdr} to find $\tau$ as a function of  $P_u$
\begin{equation}\label{eq:tauPu}
\tau=\frac{t_{\mt{R}}-t_{\rm turn}} L =-\int^{\rturn}_{r_{\rm min}} \frac{dr}{L}\,\frac{\dot{t}}{\dot{r}}=-\int^{\rturn}_{r_{\rm min}} \frac{dr}{L}\, \frac{P_u}{f(r)\sqrt{P_u^2+f(r)\,\,(r/L)^{2(d-1)}}}  \,,
\end{equation}
where we used that given our symmetric configuration, the turning point occurs at $t=t_{\rm turn} =0$. Interestingly, we should point out that the relation between $P_u$ and $\tau$ may not be a one-to-one mapping in general. That is, with a given boundary time $\tau$, one may find that several conserved momenta $P_u$ satisfy the relation \eqref{eq:tauPu}, and so there may be several extremal surfaces anchored at the same time slice on the stretched horizons. However, this feature is eliminated for $1-\rho\ll1$ (as shown in figure \ref{fig:dS3tR}) and so we defer further discussion to appendix \ref{app:max}.

From eq.~\reef{eq:tauPu}, it is obvious that $P_u=0$ corresponds to the extremal surface anchored at $\tau =0$. Increasing the conserved momentum then increases the corresponding time $\tau$. We assume that we are in the regime where $\tau$ is a monotonic function of $P_u$. Then maximum time is approached by taking the limit $P_u \to \infty$, which in turn yields $\rturn \to \infty$ in eq.~\reef{return}. In this limit, eq.~\reef{eq:tauPu} simplifies to
\begin{equation}\label{eq:deftinfity}
\tau(P_u\to\infty) = -  \int^{\infty}_{L\,\rho} \frac{dr}{L\,f(r)} = \arctanh\,\rho  =\tauinf\,. 
\end{equation}
That is, we recover precisely the critical time $\tauinf$ in eq.~\reef{criticaltau} as the maximum time that can reach. The maximum corresponds to the same critical time appearing in the CV2.0 and CA approaches. Examining eq.~\reef{eq:dudr}, we see that $\dot{r}\simeq P_u$ and $\dot{u}\simeq U(r)/P_u$ in this limit. Hence with $du/dr\to 0$, the extremal surface approaches a null surface (\ie $u=$constant) in this limit. It is straightforward to verify that in fact, it approaches $u=0$ (on the right and $v=0$ on the left after the turning point). For $\tau>\tauinf$, there are no extremal surfaces connecting these time slices on the two stretched horizons. We note that this time scale is universal for all dimensions and only depends on the position of stretch horizons $\rho$.  

Although we have explicitly shown that the limit $P_u\to \pm \infty$ yields a finite limit for the time along the stretched horizon, \ie $\tau\to \tauinf$. However, as a side point, we want to show here that the anchor time $\tau$ for our symmetric extremal surfaces is always finite, \ie the integral in \eqref{eq:tauPu} is always finite. Explicitly, this integrand is singular  at $r=\rturn$ due to 
\begin{equation}\label{eq:definermax}
P_u^2- U(\rturn) =0 \,.
\end{equation}
However, we can consider a series expansion about $r=\rturn$ as follows 
\begin{equation}
\lim_{r \to \rturn}\, \( P_u^2- U(\rturn) \) \simeq U'(\rturn) \(  \rturn-r \) + \mathcal{O}\( (\rturn-r)^2 \)\,,  
\end{equation}
where as noted above, we always have $U'(\rturn) \ne 0$. Further $\rturn>L$ and so $f(\rturn)<0$. As a result, one can find the contribution around the singular point in eq.~\eqref{eq:tauPu} is convergent with
\begin{equation}
 \int^{r \to \rturn} d{r}\, \frac{-P_u}{f({r})\sqrt{P_u^2- U({r})} } \sim \lim_{r \to \rturn}\, \frac{2P_u}{f(\rturn)\sqrt{U'(\rturn)}} \sqrt{\rturn -r} \sim 0 \,. \\
\end{equation}
So we can conclude that 
\begin{equation}
 \tau(P_u) \, \, \text{is finite} \,, \qquad  \forall  \quad  |P_u| \,.
\end{equation}
Of course, this conclusion is true even for $|P_u| \to \infty$, as we have explicitly shown above in eq.~\eqref{eq:deftinfity}. The key point in producing this finiteness is that the derivative of the effective potential was non-vanishing at the turning point. Applying the analogous analysis for the finiteness for asymptotically AdS black holes (\eg \cite{Belin:2021bga,longpaper}), one finds that $\tau \to \infty$ precisely when the trajectory approaches a local maximum in the effective potential, \ie $U' (r=\rturn)=0$.

\subsubsection*{Divergent behaviour for $\tau\to \tau_{\infty}$}

We have eq.~\reef{eq:dVdtau} relating the growth rate of the holographic complexity to the conserved momentum. Hence we know that the growth rate diverges as $\tau\to\tauinf$, since this corresponds to $P_u\to \infty$. However, we would now like to extend the analysis above to show that the CV proposal exhibits the same hyperfast growth as $\tau\to\tauinf$ that we found for the CV2.0 and CA proposals.

To make a controlled approach the critical time $\tauinf$, we consider $P_u\gg1$ and   consider a large-$P_u$ expansion of eq.~\reef{eq:tauPu},
\begin{equation}
\begin{split}
\tau 
&=\int^{\rturn}_{\rmin} \frac{dr}{L}\, \frac{-P_u}{f(r)\sqrt{P_u^2- U(r)} }\,, \\
&\approx -\int^{\rturn}_{\rmin} \frac{dr}{f(r)L}\( 1  + \frac{U(r)}{2P_u^2} + \frac{3U(r)^2}{8P_u^4}  + \mathcal{O}\(\frac{1}{P_u^6}\)    \) \,,
\end{split}
\end{equation}
where the first term matches eq.~\reef{eq:deftinfity}, which yields the critical time $\tauinf$. However, note from eq.~\reef{return} that $P_u \simeq \(\frac{\rturn}{L}\)^d$ in this regime. Hence, after integration around the turning point, all of the subleading terms yield corrections of the same order  $\frac{1}{\rturn} \sim \frac{1}{P_u^{1/d}}$. Therefore in order to derive the leading corrections to $\tauinf -\tau$ in the limit $P_u \to \infty$, we need to account for all of the $\mathcal{O}( 1/\rturn)$ contributions together, \ie
\begin{equation}\label{eq:series}
\begin{split}
 \tau (P_u) - \tauinf  &=\frac{1}{L}  \(    \int^{\infty}_{\rturn}   \frac{dr}{f(r)}  + \sum_{n=1}^{\infty} \int_{\rturn}^{\rmin} \frac{(2n-1)!!}{2^n n!}  \(\frac{U(r)}{P_u^2}\)^n  \frac{dr}{f(r)}  \)\\
 &\approx  \sum_{n=0}^{\infty} \frac{(2n-1)!!}{(2dn-1)2^n n!}\, \frac{L}{\rturn} + \mathcal{O} \( \frac{L^2}{\rturn^2} \) \,,\\
 \end{split}
\end{equation} 
where we have used the Taylor expansion for $\frac{1}{\sqrt{1-x}} = \sum_{n=0}^{\infty} \frac{(2n-1)!! x^n}{2^n n!}$. Summing this infinite series, we find  that to leading order, the time becomes\footnote{Alternatively, we can notice that the leading correction to the boundary time around $\tau_{\infty}$ is dominated by the integral around $r \approx \rturn$. Focusing on this region, the integrand is approximated by  
	\begin{equation}\nonumber
	\frac{-P_u}{f(r)\sqrt{P_u^2- U(r)} } \approx   \frac{\rturn^d}{r^2 \sqrt{\rturn^{2d} - r^{2d} }}+ \mathcal{O}\(  \frac{\rturn^d}{r^4 \sqrt{\rturn^{2d} - r^{2d} }} \) \,,
	\end{equation}
	and the first subleading term at the order $\mathcal{O}(L/\rturn)$ is given by 
	\begin{equation}\nonumber
	\int_{\rturn} dr\, \frac{L}{r^2} \(  \frac{\rturn^d}{ \sqrt{\rturn^{2d} - r^{2d} }} \) 
	\approx - \frac{\sqrt{\pi}\, \Gamma\(\frac{2d-1}{2d}\) L}{\Gamma\(\frac{d-1}{2d}\)\rturn}  + \mathcal{O}\( \frac{L^2}{\rturn^2} \) \,,
	\end{equation}
	which matches the result in eq.~\eqref{eq:series} derived from the sum of infinite series.}
\begin{equation}
\begin{split}
\tau  \simeq \tauinf - \frac{\sqrt{\pi}\, \Gamma\(\frac{2d-1}{2d}\)}{\Gamma\(\frac{d-1}{2d}\)}  \frac{L}{\rturn} \,.
\end{split}
\end{equation}
We note that this expression only applies for $d>1$, \ie $\Gamma\(\frac{d-1}{2d}\)$ diverges for $d=1$ -- we return to this special case below.

Replacing $L/\rturn\sim 1/P_u^{1/d}$ in the above expression, we can express the growth rate \reef{eq:dVdtau} as
\begin{equation}\label{eq:Pulimit}
\lim_{\tau \to \tau_\infty}  \frac{d\cv}{d\tau} \simeq 8N \(    \frac{\sqrt{\pi} \Gamma\(\frac{2d-1}{2d}\) }{\Gamma\(\frac{d-1}{2d}\)\( \tauinf -\tau \)}  \)^d   \qquad {\rm for}\ d>1\,, 
\end{equation}
which exhibits analogous divergent behaviour to that found previously for CV2.0 and CA in eqs.~\reef{eq:seriesdCsvdtau} and \reef{eq:CAderivativeearly}, respectively.
Of course, we can integrate the above expression to find the holographic CV complexity near the critical time :
\begin{equation}\label{eq:Cvlimit}
\lim_{\tau \to \tau_\infty}  \cv   \approx  \frac{8N}{d-1}  \(    \frac{\sqrt{\pi} \Gamma\(\frac{2d-1}{2d}\) }{\Gamma\(\frac{d-1}{2d}\)}  \)^d  \frac{1}{\( \tauinf -\tau \)^{d-1}} \qquad {\rm for}\ d>1\,, 
\end{equation}
which again is divergent in the limit $\tau \to \tauinf$. 
Apart from matching the powers or $\tauinf-\tau$ in eqs.~\reef{eq:Pulimit} and \reef{eq:Cvlimit}, the origin of these divergences is similar to the  CV2.0 and CA cases. The extremal surface approaches the null cones $u=0=v$ in the limit $P_u \to \infty$, and the contributions around $\rturn \to \infty$ generate the divergence.

Before we close this discussion, we return to the remark that the analysis yielding eq.~\reef{eq:series} fails for $d=1$. For this special case $\dS_2$, we need to take into account of the corrections from the next order, \ie 
\begin{equation}
\tau_{\infty} - \tau \sim \frac{L^2}{\rturn^2} \sim \frac{1}{P_u^2} \,,
\label{jump77}
\end{equation}
or equivalently, 
\begin{equation}\label{eq:limitCVdS2}
\lim_{\tau \to \tau_\infty}  P_u \( \tau  \) \approx \frac{\rho  }{\sqrt{2\( \tau_{\infty} - \tau \)  }} \,, \qquad {\rm for} \ d=1\,.
\end{equation}
Hence $d\cv/d\tau\sim \rho/\sqrt{\tau_{\infty} - \tau}$,  which does not match the $1/(\tauinf-\tau)$ behaviour found for CV2.0 and CA. We note that this result has the interesting feature that the growth rate vanishes for $\rho=0$, \ie when the extremal surfaces are anchored to the north and south pole. Further while $d\cv/d\tau$ diverges as $\tau\to \tauinf$ (with $\rho>0$), this singularity is integrable so that the complexity remains finite in this limit -- see eq.~\reef{eq:dS2CVlimit}. We refer interested readers to a complete discussion of the holographic complexity $\cv (\tau)$ for $\dS_2$ in Appendix \ref{app:dS2}.

\begin{figure}[ht!]
	\centering
	\includegraphics[width=2.5in]{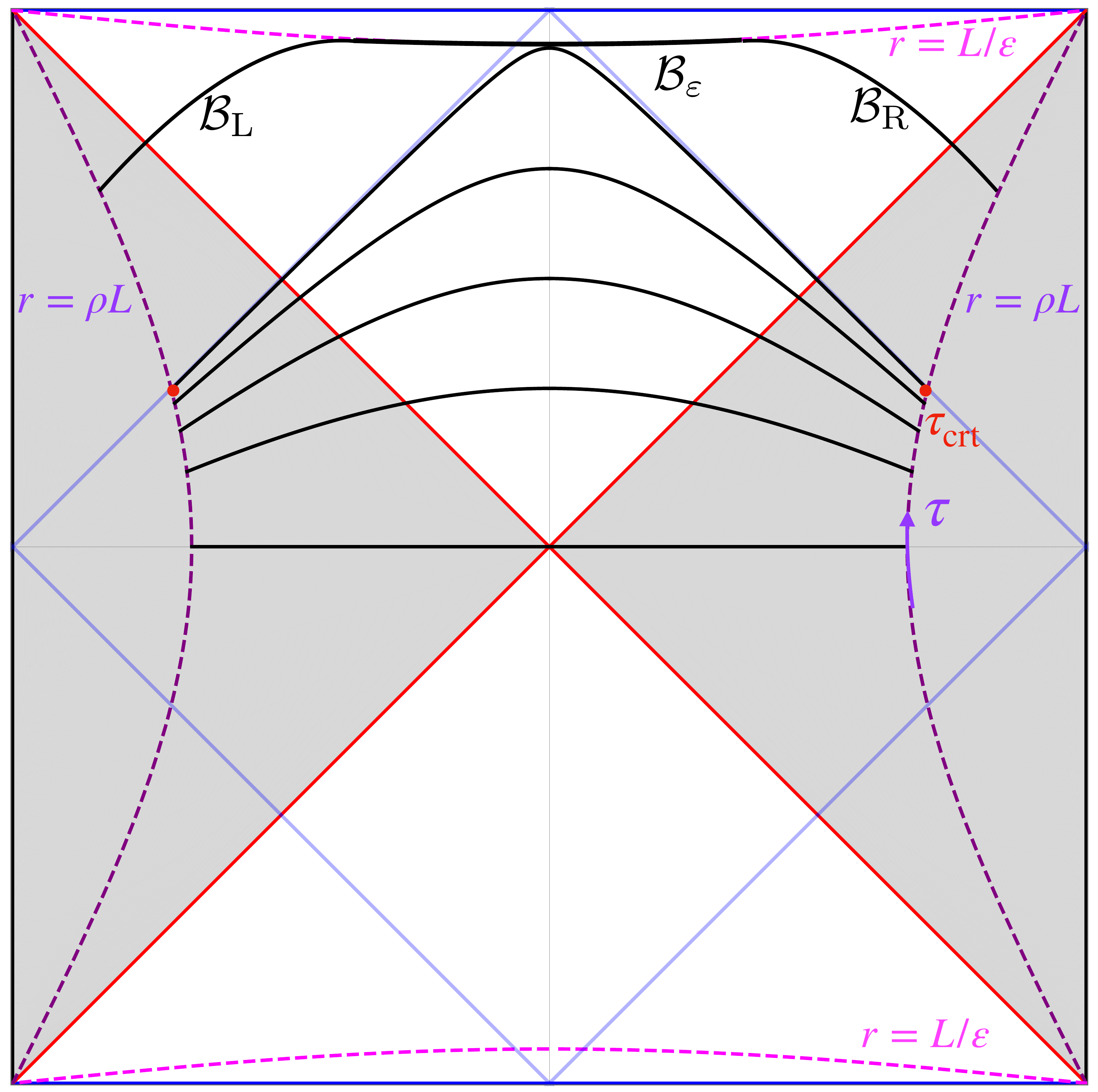}
	\qquad 
	\includegraphics[width=3in]{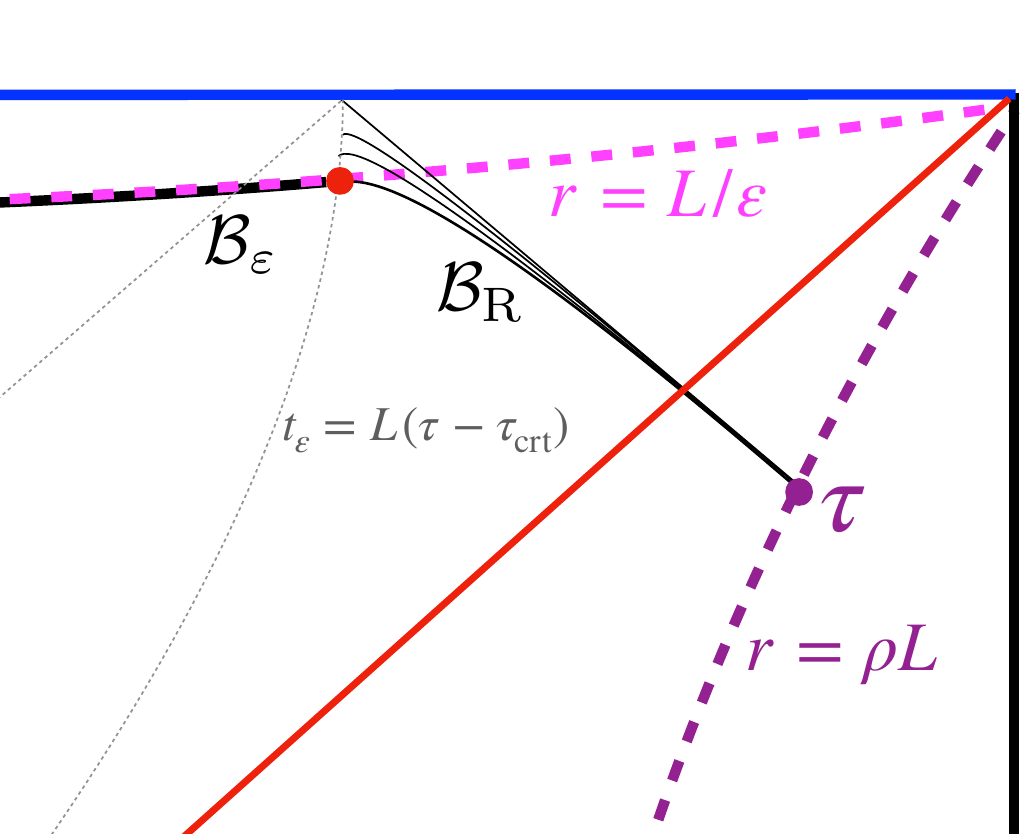}
	\caption{Left: The time evolution of the extremal surfaces for evaluating $\cv(\tau)$. After the transition time $\tau_{\rm crt}$, we consider hypersurfaces $\mB$ which are defined piecewise and do not extend beyond the cutoff surface at $r=\rmax=L/\varepsilon$. They consist of three parts, \ie $\mB=\mB_{\mt{L}} \cup \mB_\varepsilon \cup \mB_{\mt{R}}$. Right: Part of Penrose diagram with zooming into the late time regime at $\tau > \tau_{\rm crt}$. The black curves anchoring on the stretched horizon at $r=\rho L$ are various candidate surfaces ranging from $\mB_{\mt{R}}$ being null to where it connects being tangent to $\mB_\varepsilon$. The position of the intersection of $\mB_{\mt{R}}$ with $\mB_\varepsilon$ along the cutoff surface is denoted by  $t_{\varepsilon}=L\,\tau_\veps$.}
	\label{fig:dSextremalcutoff}
\end{figure}

\subsection*{Extremal surfaces joining the cutoff surface} 

The above analysis of the time evolution of the extremal surface has revealed divergent behaviour of $\cv(\tau)$
as $\tau\to\tauinf$, which reminds us of that found for CV2.0 and CA. To regulate the divergences and examine the evolution beyond $\tauinf$, we introduced a cutoff surface near future timelike infinite $i^+$ for those approaches to holographic complexity. Motivated by the results derived in previous sections, we now consider using the same geometric cutoff for the CV approach. 
Our results above (\eg see figure \ref{fig:dS3tR}) imply that the extremal surfaces connecting two boundaries at equal times on the stretched horizon simply disappear when the anchor time moves beyond a critical value $\tauinf$. In other words, as formulated initially, the CV proposal \reef{defineCV} does not work for late times (\ie $\tau\gtrsim\tauinf$) due to the absence of an extremal surface. However, when we introduce a cutoff surface at $r=\rmax$, we will demand that the surfaces yielding the holographic complexity do not extend beyond this maximal radius. Hence, we will need to modify the prescription for the CV complexity in any event. 

A natural proposal for the modified CV prescription is illustrated in figure \ref{fig:dSextremalcutoff}. In this regime, the `extremal' surface is defined piecewise with three components $\mB=\mB_{\mt{L}} \cup \mB_\varepsilon \cup \mB_{\mt{R}}$ and eq.~\reef{defineCV} is replaced by
\begin{equation}\label{eq:CVcutoff}
\cv = \mathrel{\mathop {\rm
		max}_{\scriptscriptstyle{\S=\partial \mathcal{B}}} {}\!\!}\left[\frac{\mathcal{V}(\mB_{\mt{L}}) + \mV(\mB_{\varepsilon})  + \mV( \mB_{\mt{R}}) }{G_{\mt{N}} \, L}\right] \,.
\end{equation}
Here, the segments $\mB_{\mt{L}}$ and $\mB_{\mt{R}}$ extend from the stretched horizon, across the cosmological horizon, and out to some large $r$. For early times $\tau\lesssim\tauinf$, these spacelike surfaces could not reach the cutoff surface at $r=\rmax$, and hence there will not be a segment $\mB_{\varepsilon}$, \ie this component is the empty set. With our symmetric configuration, $\mB_{\mt{L}}$ and $\mB_{\mt{R}}$ will meet at $t=0$ at some $r=r_0$. We would extremize the local profiles of these segments and also the position $r_0$. This will result in the smooth extremal surfaces found with our previous analysis for early times $\tau\lesssim\tauinf$, \eg the extremal value of the meeting point would be $r_0=\rturn$.

Now for $\tau\gtrsim\tauinf$, $\mB_{\mt{L}}$ and $\mB_{\mt{R}}$ can reach the cutoff surfaces. Hence as well as the two-component candidate surfaces considered above, we also include surfaces with a nontrivial $\mB_{\varepsilon}$ component. That is, a candidate surface $\mB_{\mt{R}}$ will extend from the stretch horizon to the cutoff surface at some $t=t_{\mt{R},\veps}>0$, while $\mB_{\mt{L}}$ intersects the cutoff surface at $t =t_{\mt{L},\veps}<0$. They are connected by $\mB_{\varepsilon}$, which simply stretches along the cutoff surface from  $t_{\mt{L},\veps}$ to $t_{\mt{R},\veps}$. In this case, the maximization in eq.~\reef{eq:CVcutoff} involves locally extremizing the profiles of $\mB_{\mt{L}}$ and $\mB_{\mt{R}}$, and also the positions of the intersection points, $t_{\mt{L},\veps}$ and $t_{\mt{R},\veps}$, on the cutoff surface.

With the anchor surfaces placed symmetrically on the stretched horizons, we can expect that even with our modified prescription \eqref{eq:CVcutoff}, the extremal surface will be left-right symmetric in the Penrose diagram. Hence, we focus our attention on $\mB_{\mt{R}}$ extending from the stretched horizon to the cutoff surface. To reduce the clutter in our equations,  we denote the position of the intersection $t=t_{\varepsilon}=L\,\tau_\veps$ and $r=\rmax= L/\varepsilon$. Because the intersection between the extremal surface and the cutoff surface is free, there are infinite extremal surfaces labeled by conserved momenta $P_u \in (0, +\infty)$ at a fixed boundary time $\tR=\tau L$. However, the extremization equations are still the same as before, but one of the boundary conditions is modified. Since the turning point should now be at $\rturn \ge \rmax$ , the conserved momenta of interest are bounded from below, \viz
\begin{equation}\label{eq:PuPcrt}
P_u \ge   \frac{\sqrt{1-\varepsilon^2} }{\varepsilon^{d}}   \equiv  P_{\rm crt}   \,.
\end{equation}
The lower bound corresponds to the critical case where the extremal surface just touches the cutoff surface with $\rturn (P_{\rm crt}) = L/\varepsilon$. As before, the infinite limit with $P_u \to \infty$ pushes the extremal surface to become a null surface located at
\begin{equation}
u = \text{constant}=\tR  - r^\ast(\rho L)= L \( \tau  -\text{arctanh}\rho   \)= L \( \tau  -\tauinf   \)\,.
\end{equation}

Following the analysis in the previous subsection, the contribution of the left and right segments is given by 
\begin{equation}\label{eq:VextdSd}
\cv^{\rm ext} =\frac{\mathcal{V}(\mB_{\mt{L}}) +  \mV( \mB_{\mt{R}}) }{G_{\mt{N}} \, L}=8N \int^{L/\veps}_{L\,\rho} \frac{(r/L)^{2(d-1)}}{\sqrt{P_u^2 +  f(r)(r/L)^{2(d-1)}}} \, \frac{dr}{L} \,.
\end{equation}
Since the cutoff surface is simply given by $r=\rmax$, it is straightforward to derive the volume for $\mB_{\varepsilon}$,
\begin{equation}\label{eq:VdeltadSd}
\cv^{\varepsilon} = \frac{  \mV( \mB_\veps) }{G_{\mt{N}} \, L}=  {8N}\,\frac{\sqrt{1-\varepsilon^2} }{\varepsilon^{d}}  \, \tau_{\varepsilon}  \,.
\end{equation}
Of course, the intersection time $\tau_{\varepsilon}$ on the cutoff surface is not totally free in that the profile of $\mB_{\mt{R}}$ connects it to the time $\tau$ on the stretched horizon for a given momentum $P_u\ge P_{\rm crt}$.  Similar to the eq.~\eqref{eq:tauPu}, the extremization equation implies the relation between $\tau$ and $\tau_{\varepsilon}$, \ie
\begin{equation}\label{eq:dSdtRtd}
\begin{split}
\tau_{\varepsilon} - \tau &  = \int^{L/\veps}_{L\rho} \frac{dr}{L}\, \frac{P_u}{f(r)\sqrt{P_u^2+f(r)\,\,(r/L)^{2(d-1)}}}  \,.\\
\end{split}
\end{equation}
Implicitly, we are considering the case where $\mB_{\mt{L}}$ and $\mB_{\mt{R}}$ do not intersect at $r=r_0<\rmax$. Hence, the intersection time (for the right part) is constrained by $\tau_{\varepsilon} \ge 0$. This implies the transition time from smooth extremal surfaces to the piecewise extremal surfaces is located at 
\begin{equation}
\tau_{\rm crt} = \int^{L/\veps=\rturn}_{L\rho} \frac{dr}{L}\, \frac{P_{\rm crt}}{f(r)\sqrt{P_{\rm crt}^2+f(r)\,\,(r/L)^{2(d-1)}}} \simeq \tau_{\infty} - \frac{\sqrt{\pi} \Gamma\(\frac{2d-1}{2d}\) }{\Gamma\(\frac{d-1}{2d}\)} \,\varepsilon \,, 
\label{taucrt}
\end{equation}
where the final approximation was derived by considering $\varepsilon \ll1$. We note that this result is similar to eq.~\reef{taustar} where it was found that the effect of the cutoff surface was first felt for CV2.0 (and CA) at a time $\tau_*=\tauinf -\mathcal{O}(\veps)$. 

For any boundary time beyond this critical time $\tau \ge \tau_{\rm crt}$, we can find a  (continuous) family of piecewise surfaces where $\mB_{\mt{L}}$ and $\mB_{\mt{R}}$ are locally extremized away from the cutoff surface. As described above, it remains to find the surface that maximizes the holographic complexity in eq.~\reef{eq:CVcutoff} by extremizing over the intersection time $\tau_{\varepsilon}$. That is, we determine the surface with the maximal volume by performing the maximization:
\begin{equation}\label{eq:definemax}
\cv \(\tau \ge \tau_{\rm crt}\) = \max_{P_u \ge P_{\rm crt}} \( \cv^{\rm ext}\(P_u\) + \cv^{\varepsilon} (\tau_{\varepsilon})  \) \,,
\end{equation}
where the time $\tau$ is fixed and so eq.~\reef{eq:dSdtRtd} determines $\tau_{\varepsilon}$ in terms of  the conserved momentum $P_u$.

First, it is easy to find that the leading contributions in  $\cv^{\rm ext}\(\tau\)$, and $ \cv^{\varepsilon} (\tau_{\varepsilon})$ in the regime $\varepsilon \ll1$ are 
\begin{equation}
\cv^{\rm ext}  \approx    \frac{8N}{\varepsilon^{d-1}}     \frac{\sqrt{\pi} \Gamma\(\frac{2d-1}{2d}\) }{(d-1)\Gamma\(\frac{d-1}{2d}\)}    \,, \qquad \cv^{\varepsilon}  \approx \frac{8 N \tau_{\varepsilon}}{\varepsilon^d} \,,
\end{equation}
respectively. One may naively expect that the maximization should identify the maximal-complexity surfaces in eq.~\reef{eq:definemax} are those which maximize the extent of $\mB_{\veps}$ along the cutoff surface because its contribution $\cv^{\varepsilon}$ dominates above. However, we note that the naive expectation is incorrect because the variations in 
$\tau_\veps$ are only $\mO(\veps)$ and so $\cv^{\rm ext}$  and $\cv^\veps$ compete on an equal footing in the maximization \reef{eq:definemax}. More precisely, we obtain the maximal and minimal values for $\tau_{\varepsilon}$ as follows  
\begin{equation}
\begin{split}
\tau_{\varepsilon}|_{\rm max} &= \tau-\tau_{\infty} + \arctanh \( \varepsilon \) \,, \quad \text{with} \qquad P_u \to \infty\,,\\
\tau_{\varepsilon}|_{\rm min} &= \tau - \tau_{\rm crt} \,,\qquad\qquad\qquad\quad \text{with} \qquad P_u= P_{\rm crt}\,,\\
\end{split}
\end{equation}
and as noted above, we will find $\tau_{\varepsilon}|_{\rm max} - \tau_{\varepsilon}|_{\rm min}  \sim \mathcal{O}(\varepsilon)$.

\begin{figure}[h!]
	\centering
	\includegraphics[width=4.5in]{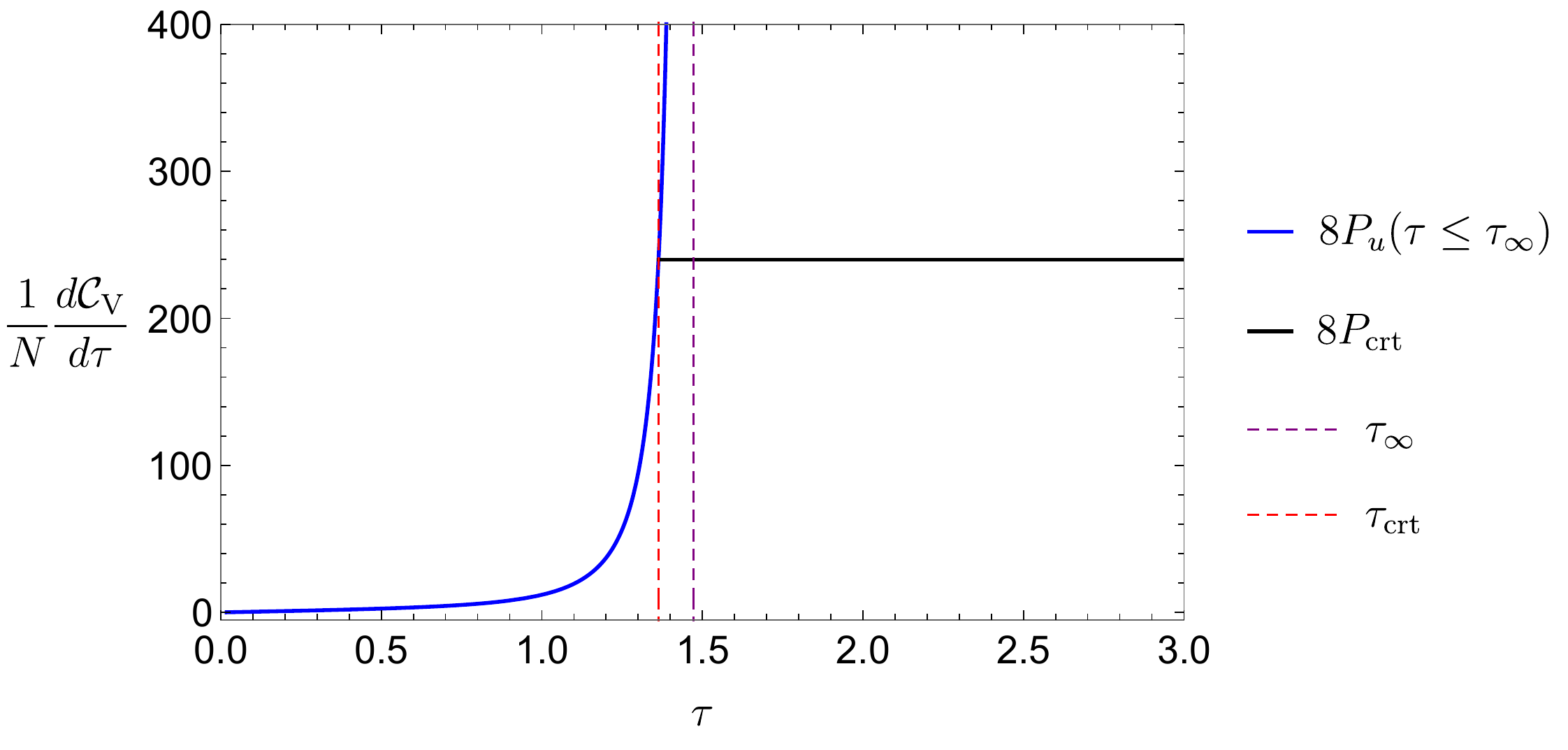}
	\caption{The time derivative of holographic complexity $\frac{d \cv(\tau)}{d\tau}$. The blue curve is referred to as the growth rate before the critical time, \ie $P_u (\tau \le \tauinf)$. After introducing the cut-off surface, holographic complexity $\cv (\tau \ge \tau_{\rm crt})$ grows linearly with a growth rate $8N P_{\rm crt}$ as indicated by the black line. We choose $d=2,\rho =\frac{9}{10}, \veps= \frac{1}{10}$ for this plot.}
	\label{fig:timederivativeCVFull}
\end{figure}

Although the full analytical results for the above integrals for higher dimensions are much more complicated, we can still show that the surface which maximizes \eqref{eq:definemax} (for  $\tau \ge \tau_{\rm crt}$) is the surface intersecting the cutoff at $\tau_{\varepsilon}|_{\rm min}$. That is, the surface is constructed with the smallest momentum $P_u=P_{\rm crt}$. We only need to focus on the derivative of the two contributions with respect to the conserved momentum, namely 
\begin{equation}
\begin{split}
\frac{\partial \cv^{\rm ext}}{\partial P_u} = -8N\, P_u \int^{L/\veps}_{L\rho}  \frac{(r/L)^{2(d-1)}}{\(P_u^2 +  f(r)(r/L)^{2(d-1)}\)^{3/2}} \, \frac{dr}L \,,
\end{split}
\end{equation}
and 
\begin{equation}
\begin{split}
\frac{\partial \cv^{\varepsilon} }{\partial P_u}  &=  8N\, \frac{\sqrt{1-\varepsilon^2} }{\varepsilon^{d}}  \, \frac{\partial \tau_{\varepsilon}}{\partial P_u}   \\
&=8N\, P_{\rm crt} \int^{L/\varepsilon}_{L \rho}  \frac{(r/L)^{2(d-1)}}{\(P_u^2 +  f(r)(r/L)^{2(d-1)}\)^{3/2}} \, \frac{dr}L \,,
\end{split}
\end{equation}
where we substituted the definition of $P_{\rm crt}$ from eq.~\reef{eq:PuPcrt} into the final expression. Combining these two expressions, we thus obtain 
\begin{equation}\label{eq:dSddVdPu}
\frac{\partial \( \cv^{\rm ext} + \cv^{\varepsilon} \)}{\partial P_u} \propto (P_{\rm crt}-P_u) \le 0 \,,
\end{equation}
due to our constraint that $P_u\ge P_{\rm crt}$. In the above analysis, we consider an arbitrary time, stretched horizon, and cutoff surface, which means that this conclusion holds for any $\tau,\,\rho,\, \varepsilon$. As a result, we conclude that the CV complexity of the piecewise surfaces $\mB$ in the late-time regime ($\tau \ge \tau_{\rm crt}$) is always associated with the extremal surfaces with a conserved momentum $P_{\rm crt}$. These are the surfaces where $\mB_{\mt{R}}$ and $\mB_{\mt{L}}$ are just tangent to the cutoff surface when they meet $\mB_\veps$, \ie the piecewise extremal surface remains smooth.  

After the transition time, the growth of holographic complexity is exactly linear. It is easy to understand this linear growth because $ \cv^{\rm ext} $ remains as a constant and $ \cv^{\varepsilon}$ grows linearly. The linear growth at late times simply reads 
\begin{equation}
\frac{d\cv }{d\tau} \bigg|_{\tau \ge \tau_{\rm crt}} = 8 N \,P_{\rm crt}=8N \,\frac{\sqrt{1-\varepsilon^2} }{\varepsilon^{d}}\,.
\label{rate98}
\end{equation}
The growth rate of holographic complexity from early to late times is shown in figure \ref{fig:timederivativeCVFull}. Finally, we remark that the transition from hyperfast to linear growth, $d \cv/d \tau$ is continuous. This is similar to the result for CV2.0 shown in eq.~\eqref{eq:CV2continuous}.

\section{Discussion}\label{sec:disc}
In this paper, we investigated three different proposals for holographic complexity applied to ($d+1$)-dimensional de Sitter space. We found that the hyperfast growth regime first discussed in \cite{Susskind:2021esx} for the CV proposal appears quite generally. That is, the complexity grows much faster than the linear growth observed for black holes, \eg \cite{Susskind:2014moa,Brown:2015lvg,Carmi:2017jqz}. In fact, all three proposals exhibit a pole of the form
\beq
\frac{d{\cal C}}{d\tau}\approx \frac{N}{(\tau_\infty-\tau)^{d}} \,,
\label{pole8}
\eeq
as we approach the critical time $\tau_\infty$ -- see eqs.~\reef{eq:seriesdCsvdtau}, \reef{eq:CAderivativeearly} and \reef{eq:Pulimit}. An exception to this behaviour was the CV complexity for $d=1$, which yields $d{\cal C}/d\tau \approx  N/\sqrt{\tau_\infty-\tau}$, as shown in eq.~\reef{rate22}.  This critical time has a clear geometric explanation in the gravity calculations \cite{Susskind:2021esx}. When the null sheets are emitted to the future from time slices on the left and right stretched horizons, they will intersect in the region behind the cosmological horizon (at early times, $0\le\tau\le\tau_\infty$). At $\tau=\tau_\infty$, the position of this intersection just reaches timelike infinity $i^+$ in the Penrose diagram, and of course, for $\tau>\tau_\infty$, these null sheets no longer intersect (\ie they reach $i^+$ before intersecting). 

As indicated in eq.~\reef{pole8}, all three approaches to holographic complexity yield a factor of $N$, the de Sitter entropy \reef{eq:dSentropy}. As this entropy is thought to measure the number of holographic degrees of freedom, \eg \cite{WFN,Banks:2000fe,Bousso:2000nf}, it is natural that such a factor should appear in the holographic complexity. Eq.~\reef{pole8} is also written in terms of a dimensionless time coordinate on the stretched horizons \cite{Susskind:2021esx}, \ie we chose $\tR=\tau L=-\tL$. Expressing the result in terms of the coordinate time $t$ would introduce factors of $L$, the dS curvature scale, or alternatively of the Hawking temperature \reef{temp} measured by an observer at the center of the static patch. Perhaps a more natural approach is to use the proper time measured along the stretched horizon, $t_p=\sqrt{1-\rho^2} L\,\tau$. Expressing  eq.~\reef{pole8} in terms of $t_p$ would then introduce factors of the blue-shifted temperature measured by observers traveling along the stretched horizon, \ie
\beq
T_{\rm{stretch}}=\frac{1}{2\pi \,\sqrt{1-\rho^2} L}  \,.
\label{blue}
\eeq
Of course, this is the natural temperature to associate with the holographic theory on the stretched horizons, and then eq.~\reef{pole8}  becomes
\begin{equation}\label{pole8a}
\frac{d{\cal C}}{dt_p}\approx \frac{N\,T_{\rm{stretch}}}{[T_{\rm{stretch}}(t_{p,\infty}-t_p)]^{d}}\,.
\end{equation}

It is interesting to compare the results for holographic complexity here with the analogous results in AdS spacetime. One immediate difference is that the present complexities are finite at early times $\tau\lesssim\tauinf$. In contrast, in AdS spacetime, one finds universal UV divergences due to the contributions near the asymptotic boundary, \eg see \cite{Carmi:2016wjl,Chapman:2016hwi}. These divergences are associated with introducing entanglement down to small UV distance scales in the boundary theory. Hence the lack of divergences at early times in de Sitter space is not surprising because it is expected that the holographic dual only has a finite number of degrees of freedom. As noted above, this number of degrees of freedom appears as the overall factor of $N$ in the dS complexity. This is analogous to the leading AdS contribution with CV or CV2.0, which is also proportional to the number of degrees of freedom in the boundary CFT \cite{Carmi:2016wjl}. That is, one finds the leading contribution to the complexity is proportional to $c_T\, {\rm vol.}/\delta^{d-1}$ where $c_T$ is a central charge characterizing the boundary CFT and  ${\rm vol.}/\delta^{d-1}$ counts the number of cutoff-sized cells in the corresponding boundary time slice. This general feature also extends to CA, although additional logarithmic factors exist.\footnote{Let us also note that if we had not included the null boundary counterterm \reef{eq:Theta} in our complexity=action calculations in section \ref{sec:CA}, the leading divergence would have been $d\ca/d\tau \approx  N \log \( \tauinf - \tau \)/\(\tauinf -\tau\) $. This raises a number of interesting questions, but it also means that the results would depend on the parametrization of the null boundaries.} For example, the factor $\log ( L/(d-1)\ell_{\rm ct})$ appearing in eq.~\eqref{eq:CAearlylimit} has similar counterparts in AdS calculations. It is interesting to note that the signs are such that we require $\ell_{\rm ct} < L/(d-1)$ for $\ca$ to be positive in dS,\footnote{See discussion under eq.\reef{eq:CAearlylimit}.} while we should choose $\ell_{\rm ct} > L/(d-1)$ in AdS \cite{Agon:2018zso,Caceres:2019pgf}. Of course, divergences appear as $\tau\to\tauinf$ (\eg see eq.~\reef{eq:VWdWaround}), and we observe that the structure of the divergences in the dS complexity is similar to the structure of UV divergences appearing in the AdS complexity. That is, in both cases, we have a series of power-law divergences beginning with $1/(\tauinf-\tau)^{d-1}$ in dS and $1/\delta^{d-1}$ in AdS \cite{Carmi:2016wjl,Chapman:2016hwi}. Further, the series includes only odd or even powers for $d$ even or odd, respectively, with an additional logarithmic term appearing for even-dimensional dS or AdS spacetime, \ie an odd-dimensional dual theory.\footnote{Of course, this logarithmic contribution is the leading term for $d=1$, \ie dS$_2$.} 

Susskind \cite{Susskind:2021esx} argued that the hyperfast growth of the complexity shown in eq.~\reef{pole8} signals that the Hamiltonian governing the time evolution of the holographic degrees of freedom is not
of the usual $k$-local type. Instead, Hamiltonian is comprised of `complex' operators where $k$ grows with $N$, the total number of degrees of freedom, \ie these operators act on a significant fraction of the degrees of freedom simultaneously. These arguments were made explicit by considering the SYK model in an unusual limit, where the temperature is large, and the number of fermions in the interactions scales as a power of $N$. This new regime allows the time evolution of the system to quickly explore the Hilbert space, in contrast to the circuits  which are used to construct the corresponding state when measuring its complexity. These `complexity' circuits are built with simple gates that only act on a finite number of degrees of freedom. This difference in the nature of the operators appearing in the Hamiltonian and the complexity circuits leads to the rapid growth of the complexity. The behaviour in eq.~\reef{pole8} would be a target in extending the discussion of the underlying microscopic degrees of freedom to higher dimensions. 

While the holographic complexity diverges in a finite time, as shown in eq.~\reef{pole8}, it was natural to regulate the geometric calculations with a cutoff surface at some large radius $\rmax$. In this case, the hyperscaling behaviour saturates at $C\sim N/\veps^d$, where $\veps=L/\rmax\ll 1$ is a dimensionless parameter  characterizing the cutoff. Subsequently, the complexity grows linearly with time
\beq
\frac{dC}{d\tau}\simeq \frac{N}{\veps^d}\,,
\label{linear8}
\eeq
as shown in eqs.~\reef{rate79}, \reef{eq:CAderivativelate}, and \reef{rate98}. Of course, this rate is somewhat ambiguous since it depends so strongly on the cutoff.  

Further, this linear growth at later times is again in agreement with the discussion of \cite{Susskind:2021esx}. However, a discrepancy between the present and earlier discussions is that in \cite{Susskind:2021esx}, it was argued that the finiteness of the Hilbert space must tame the hyperfast growth and the corresponding prefactor in eq.~\reef{linear8} would be exponential in the number of degrees of freedom. As is evident, with the cutoff which we introduced by hand, the growth rate remains linear in $N$ unless we allow the cutoff to be controlled by the number of degrees of freedom, \eg $\veps\sim e^{-aN}$. Of course, the finiteness of the Hilbert space also comes into play in discussing the long-time behaviour of holographic complexity in asymptotically AdS black holes, \ie after a time exponential in the entropy, the complexity saturates, \eg \cite{Susskind:2014moa,Brown:2016wib,Brown:2017jil}.
We note that recent calculations in JT gravity involving summing over topologies revealed the expected late-time saturation of the complexity in this context \cite{Iliesiu:2021ari}. Hence, it would be interesting to see if these calculations could be adapted to a dS version of JT gravity (as examined in \eg \cite{Maldacena:2019cbz,Cotler:2019nbi,Moitra:2022glw}) and then if they would reveal linear growth for the late-time dS complexity in line with the predictions of \cite{Susskind:2021esx}. Further, let us add that the dS complexity must also eventually saturate as in the AdS case, and it would be interesting to understand the relevant time scale for saturation either in the present regulated framework or in that considered in \cite{Susskind:2021esx}. 

Of course, setting aside the above considerations, one might ask how our geometric regulator should be interpreted in the dual theory. Here we can find guidance from the complexity=volume calculations in section \ref{sec:dSCV}. Recall that the extremal surfaces have an intuitive relation to the unitary circuits measuring the complexity of the dual state, \eg \cite{Hartman:2013qma,Susskind:2014moa,Stanford:2014jda}. While at early times, the cutoff surface does not play a role, and as usual, we imagine that the entangled state between the stretched horizons is constructed by `complexity' circuits using elementary gates acting on only a few degrees of freedom, \ie the resources available in their construction are $k$-local operators.\footnote{This assumption follows from the usual intuition developed in the conventional AdS setting.}  However, beyond $\tau\gtrsim\tauinf$, the nature of the surfaces changes, and a segment $\mB_\veps$ of the maximum volume surface lies along the cutoff surface. Hence the nature of the underlying circuit must change, which we can interpret as new resources becoming available. These resources tame the growth of the complexity, \ie the growth is linear in time albeit with a very large coefficient. Hence one might imagine that the new resource involves gates that are not $k$-local  so that the complexity circuits can keep up with the nonlocal Hamiltonian evolution of the dual theory suggested by \cite{Susskind:2021esx}.

Further progress may come from taking a `conventional' holographic interpretation of the cutoff surface. Taking over our experience from AdS/CFT, one would interpret $\rmax$ in terms of a short distance UV cutoff $\delta$ in a boundary theory, \ie
\beq
\rmax=\frac{L\,R}{\delta}\qquad{\rm or}\qquad
\veps=\delta/R\,.
\label{UV9}
\eeq
Our parameter $\veps$ becomes the ratio of this UV cutoff $\delta$ to a macroscopic scale $R$, characterizing the boundary geometry. In particular, following the standard holographic prescription, the boundary geometry here becomes
\beq
ds^2_{\rm boundary}=R^2\left(d\tau^2+ d\Omega^2_{d-1}\right)\,,
\label{bound8}
\eeq
where $\tau = t/L$, as is consistent with our previous notation.
Hence it may be natural to interpret the portion of the extremal surface that hugs $r=\rmax$ as a Euclidean path integral involving a boundary CFT in this background geometry \reef{bound8}. 

Of course, this interpretation connects with attempts to construct dS/CFT holography, \eg \cite{Strominger:2001pn,Strominger:2001gp,Maldacena:2002vr,Witten:2001kn,Bousso:2001mw,Balasubramanian:2002zh,Balasubramanian:2001nb,Klemm:2001ea,Leblond:2002ns,Leblond:2002tf,Kabat:2002hj,Parikh:2002py,Anninos:2011ui,Hikida:2021ese}, where gravity in asymptotically de Sitter space was conjectured to be dual to a boundary CFT living on timelike infinity. Of course, these studies showed that this must be an unconventional CFT (\eg with complex conformal weights). This CFT path integral certainly seems to increase the available resources  to construct the unitary circuit. In the CFT interpretation, $N$ becomes the central charge of the dual CFT, a local measure of degrees of freedom. That is, there are $N$ degrees of freedom for each cutoff-sized plaquette on a slice through the Euclidean manifold, \ie the total number of degrees of freedom is rough $N'=N (R/\delta)^{d-1}$. Then we can rewrite the rate \reef{linear8} as
\beq
\frac{dC}{d\tau}\simeq \frac{N'}{\veps}=N'\,\frac{R}\delta\,.
\label{linear8a}
\eeq
Recall that in the discussion around eq.~\reef{eq:dSddVdPu}, we showed that the segment on the cutoff surface extends from $-\tau_\veps$ to $\tau_\veps=\tau-\tau_{\rm crt}$ with the constant shift $\tau_{\rm crt}$ given in eq.~\reef{taucrt}. That is, as the holographic theory evolves for an interval $\Delta \tau$, the interval over which the path integral is performed expands by $R\, \Delta \tau$ (up to a factor of two) according to the boundary metric \reef{bound8}. Alternatively, we can say that in this interval, the Euclidean path integral expands by $R\, \Delta \tau/\delta$ cutoff-sized layers. Hence, we see that the linear growth rate is proportional to the product of the total number of degrees of freedom in the boundary CFT and the rate at which cutoff-sized layers are added to the Euclidean path integral.

In considering the new resources, we note that Euclidean path integral would not be constructed by unitary gates alone, but rather it would include `euclideons', new tensors derived directly for the Euclidean time evolution by the Hamiltonian  \cite{Milsted:2018yur,Milsted:2018san}.\footnote{See \cite{Caputa:2017urj,Caputa:2017yrh,Czech:2017ryf,Takayanagi:2018pml,Camargo:2019isp,Boruch:2021hqs}  for other approaches to explaining holographic complexity outside of the standard approach of unitary circuits.} Overall, we expect that the boundary CFT is an auxiliary system. We note that the degrees of freedom in this auxiliary theory are organized and operated on with some sense of locality in the boundary geometry \reef{bound8}.  However, embedding the total of $N$ degrees of freedom on the stretch horizons is unlikely to respect this locality and so is very much in line with the idea that we are introducing nonlocal operators in this portion of the complexity circuit.

The above speculation produces a rather pleasing description (at least to the present authors) where both perspectives on dS/CFT holography have a role to play, \ie the fundamental theory lives on the stretched horizon while a boundary CFT plays the role of an auxiliary system. However, one must ask what the nature of the underlying complexity model really is. In particular, one may wonder if the new resources are introduced (by hand) after the complexity evolves beyond a certain threshold, \ie the state crosses some distance in the Hilbert space, or if these resources are available but simply not efficient in describing the state at early times $\tau\lesssim\tauinf$. We argue that our modified CV prescription \reef{eq:CVcutoff} favors the latter approach. In finding the extremal surface, we optimize between  segments of the piecewise surface that are locally extremal and those that hug the cutoff surface, \eg we optimize over the intersection points, $t_{\mt{L},\veps}$ and $t_{\mt{R},\veps}$. This suggests a microscopic picture where one is optimizing between the standard resources, \ie simple $k$-local gates, and the new resources involving the boundary CFT path integral. In particular, our detailed analysis showed that the maximal volume surface connects as a tangent to the cutoff surface, rather than just falling into $r=\rmax$ along a null sheet. This means that the maximal surface remains away from the cutoff as long as it can and contributes $\cv^{\rm ext} \sim N/\veps^d$ before connecting to the cutoff surface.

There are a variety of directions in which the present work could be extended.
As noted above, one interesting future direction would be to adapt the JT gravity calculations of \cite{Iliesiu:2021ari}to a positive cosmological constant. This may reveal that the expected linear growth beyond $\tau\gtrsim\tauinf$ emerges naturally from a sum over spacetime topologies. Our approach of introducing a geometric regulator, \ie a cutoff surface, is a complementary approach, but it readily allows for a broader examination of dS holography including black holes, \eg \cite{Susskind:2021dfc,Shaghoulian:2021cef}
or shock waves, \eg \cite{Sfetsos:1994xa,Hotta:1992qy}. 

Indeed, extending the present discussion of holographic complexity to more general cosmological backgrounds would be interesting. The idea of using cosmological horizons as a holographic screen has been considered in \eg \cite{Sanches:2016sxy,Nomura:2016ikr,Nomura:2017fyh}. One interesting context to examine would be asymptotically dS geometries with matter excitations \cite{Balasubramanian:2001nb,Leblond:2002ns}. In general, the corresponding Penrose diagram is no longer square, and instead, the diagram would be taller than it is wide. Hence two observers fixed on antipodal points in the spacetime would then observe overlapping regions on the $t=0$ slice. It would be interesting to understand the holographic description of such a scenario.
In some cases, these matter-filled spacetimes collapse to form singularities in the future or emerge from a singularity in the past (\eg see \cite{Borde:1996pt,Balasubramanian:2001nb}). However, holographic entanglement and complexity may still prove to be exciting probes of these cosmological singularities. It would be interesting to compare their behaviour to that in the context of black hole singularities, \eg \cite{Barbon:2018mxk,Barbon:2019yrr,Barbon:2019xwc}. 

Following \cite{Belin:2021bga,longpaper}, it would be interesting to consider the present dS setting to explore the behaviour of generalized gravitational observables which have an interpretation in terms of holographic complexity. 
In this direction, one might reconsider our revised CV proposal \reef{eq:CVcutoff}. A natural question is whether the volume of the segment running along the cutoff surface could be weighted differently from the locally extremal segments? It would be interesting to investigate if such a choice changes the behaviour of the complexity in any essential way. It appears that one consequence would be that the growth rate would jump discontinuously when the cutoff surface begins to contribute. In this context, another interesting question would be to understand whether or not it is possible to construct observables that do not exhibit hyperfast growth (\ie which stay away from timelike infinity). 

Finally, we note that the discussion of holographic entanglement entropy can be framed in terms of `bit threads' 
\cite{Freedman:2016zud}. In the de Sitter setting, this leads to two distinct proposals:  the monolayer \cite{Susskind:2021esx}, and bilayer \cite{Shaghoulian:2021cef} approaches for entanglement entropy -- see   \cite{Shaghoulian:2022fop} for a discussion of the differences between the two approaches. However, we note that there is an analogous `gate line' description of complexity=volume \cite{Headrick:2017ucz,Pedraza:2021mkh,Pedraza:2021fgp}. Hence it would be helpful to examine this description of holographic complexity in the context of de Sitter space and explore if analogous subtleties arise as were found for the holographic entanglement entropy.

\begin{acknowledgments}
We are happy to thank Shira Chapman, Damian Galante, Dominik Neuenfeld, Zixia Wei, and Beni Yoshida for fruitful discussions and useful comments.  Research at Perimeter Institute is supported in part by the Government of Canada through the Department of Innovation, Science and Economic Development Canada and by the Province of Ontario through the Ministry of Colleges and Universities. RCM is supported in part by a Discovery Grant from the Natural Sciences and Engineering Research Council of Canada, and by funding from the BMO Financial Group.  RCM and SMR are supported by the Simons Foundation through the ``It from Qubit'' collaboration. SMR is also supported by MEXT-JSPS Grant-in-Aid for Transformative Research Areas (A) ``Extreme Universe'', No.\,21H05187.  

\end{acknowledgments}

\appendix

\section{CV Complexity in dS$_{2}$}\label{app:dS2}
\begin{figure}[ht!]
	\centering
	\includegraphics[width=5in]{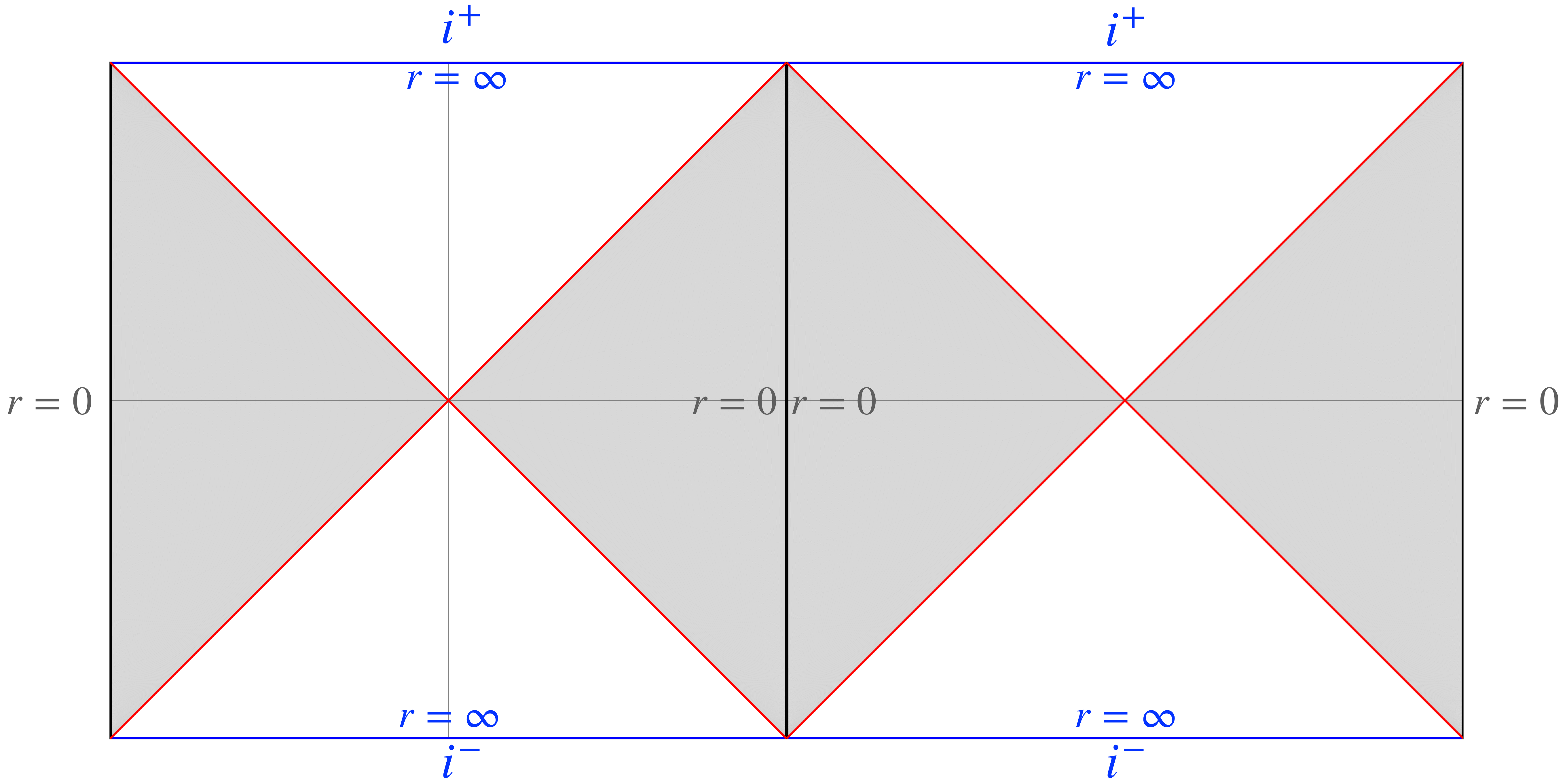}
	\caption{Penrose diagram of $\dS_{2}$ which contains two copies. In this diagram, we identify the two north poles located at $r=0$ to glue the two copies together.}
	\label{fig:dSPenrosedS2}
\end{figure}

In this appendix, we examine the extremal surfaces and holographic complexity $\CV(\tau)$ for $\dS_2$ in detail. Thanks to the simplicity of $\dS_2$, we will be able to derive all results analytically. We should also point out that the most results derived in the main section for $\dS_{d+1}$ are also valid for $\dS_2$, except the divergent power law shown in eqs.~\eqref{eq:Pulimit} and \eqref{eq:Cvlimit}. Before we move to the extremal surface in $\dS_{2}$, we also note that the crucial difference of spacetime structure of $\dS_{2}$ with its cousins the higher dimensional spacetime. In the following discussion, we will focus on the geometry represented by the Penrose diagram in figure \ref{fig:PenrosedS}. For $\dS_{d+1}$ each point in the diagram represents an $S^{d-1}$ and hence for $\dS_{2}$ \ie $d=1$, we have $S^0$. The latter actually denotes two separate points. That is, for $d=1$, a horizontal cross-section of the Penrose diagram is a circle $S^1$ with the north and south poles being two antipodal points dividing this circle into two halves. Further then, each point in the diagram represents two points equidistant from either pole, one on each of these halves. As a result, the complete Penrose diagram for $\dS_{2}$ be redrawn as shown in figure \ref{fig:dSPenrosedS2}, where each of the two halves is explicitly shown. This means that the cosmological horizon surrounding the observer at the center of the diagram actually has two separate components, one to the left of the observer and one to the right. One might imagine that there are distinct holographic degrees of freedom associated with each of these,\footnote{There is also no reason why the circumference of the $t=0$ should necessarily be $2\pi L$, which is implicitly the case in figure \ref{fig:dSPenrosedS2}. So it would also be interesting to explore the case where this circumference is larger or smaller than this canonical value. In particular, if the circumference is less than $\pi L$, a single observer would eventually be able to see the entire $t=0$ slice and realize that her universe is compact. However, we do not explore this possibility here either.}, but we do not examine this possibility here. In direct analogy with the higher dimensional discussion in the main text, we will consider extremal surfaces with two symmetric left and right components, which are always anchored on the same time slices on the stretched horizons at the same radii just outside the two horizons shown in figure \ref{fig:dSPenrosedS2}.

We also remark that the extremal surfaces in $\dS_{2}$ were recently explored in \cite{Chapman:2021eyy}. Their investigation focused on studying holographic complexity for locally $\dS_{2}$ geometries within an asymptotically AdS$_2$ spacetime \cite{Anninos:2017hhn,Anninos:2018svg}. For their discussion of pure $\dS_{2}$, the extremal surfaces were anchored  on the north and south poles at $r=0$ (and hence $t=0$). In the following, we consider the general cases with extremal surfaces anchored at $r=\rho L$, which allows the surfaces and $\cv$ to evolve along the stretched horizon. Further,  we analyze the CV complexity with a cutoff surface for late times $\tau \gtrsim \tau_{\infty}$.

\subsection*{Extremal surfaces in $\dS_2$}\label{sec:dS02}

In the following, we explore the holographic complexity $\cv$ for $\dS_{2}$, \ie 
\begin{equation}
\cv=\frac{1}{G_{\mt{N}}L} \int_{\mB} \sqrt{h}  =4N \int_{}  \sqrt{- f(r)\dot{u}^2  - 2 \dot{u} \dot{r} }\,\frac{d \lambda}{L}\,,
\end{equation}
with $N = \frac{\Omega_0}{4 \GN}$. As described above, this describes extremal surfaces with two symmetric but separate components on the left and right halves of figure \ref{fig:dSPenrosedS2}. Although our discussion only refers to a single surface, \ie a half of the entire extremal surface, the factor $\Omega_0=2$ in $N$ takes account of the contributions from the two separate components. 

\begin{figure}[ht!]
	\centering
	\includegraphics[width=3in]{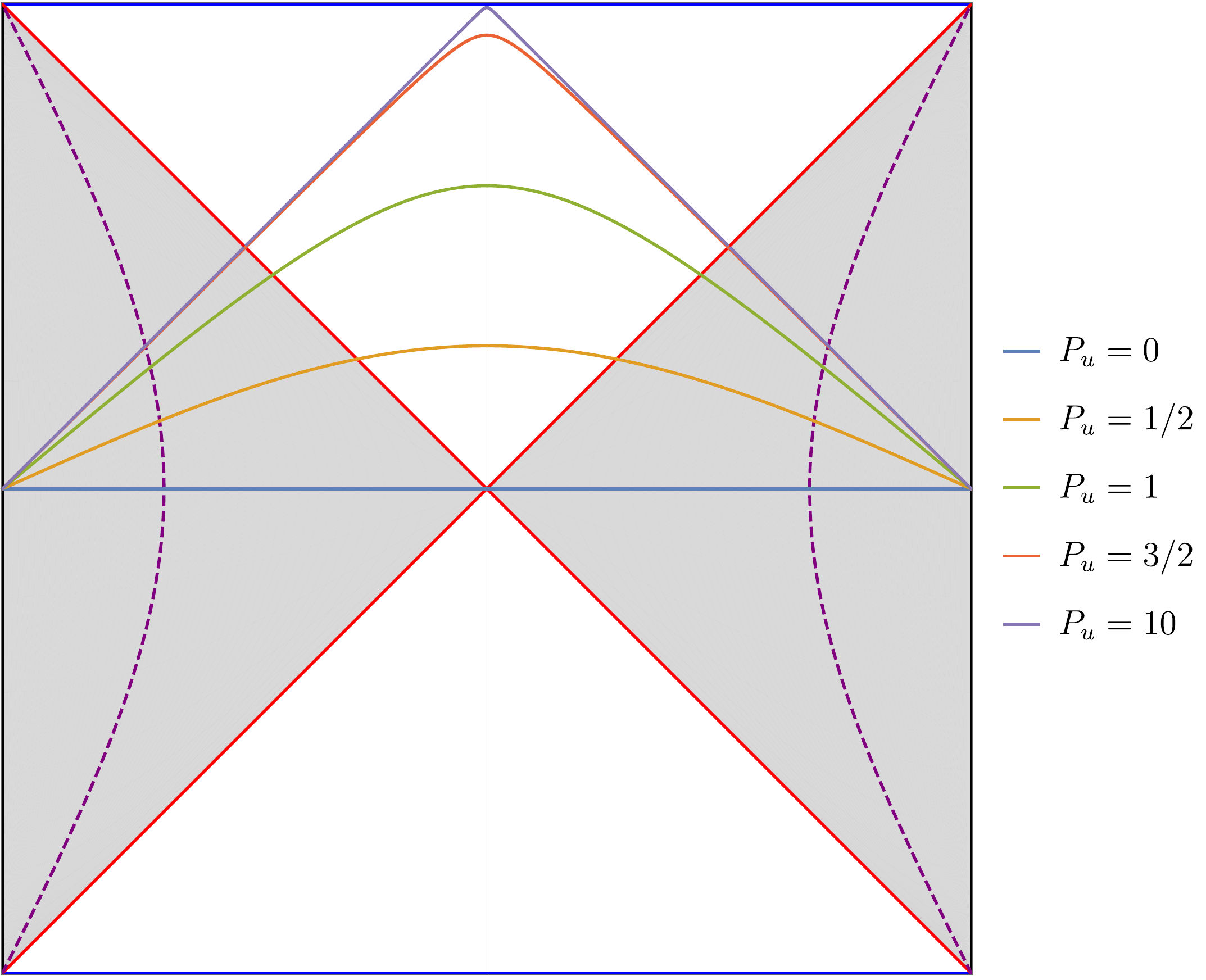}
	\caption{Extremal Surfaces with various conserved momentum in $\dS_{2}$. }
	\label{fig:dS2extremalsurface}
\end{figure}

We start from the extremizing equations in $\dS_2$, namely 
\begin{equation}
\begin{split}
\dot{r} &= \sqrt{ P_u^2 + 1 - \frac{r^2}{L^2} } \,,\quad \dot{u}=  \frac{  P_u-  \dot{r}  }{f(r)}   \,.
\end{split}
\end{equation} 
Obviously, the turning point between the left/right parts is derived as $\rturn =L  \sqrt{1+ P_u^2}$ with $\dot{r}|_{\rturn}=0$. 
First of all, we consider the simplest case with extremal surfaces starting from the north pole. The corresponding boundary conditions are given by 
\begin{equation}
r(0) = 0 \,, \quad  u\(\frac{\pi}{2}\)=t_{\rm turn} - r^\ast (\rturn)  = - \frac{L}{2}  \log  \frac{\sqrt{1+P_u^2}+1}{\sqrt{1+P_u^2}-1} \,, 
\end{equation}
in which the second condition is derived by requiring the turning point located at $t_{\rm turn}=0$ for the symmetric configuration with boundary time $\tR= \tau = -\tL$.  One can thus solve the extremizing equations and get 
\begin{equation}\label{eq:extremaldS2}
\begin{split}
 \frac{r(\lambda)}{L} &=  \sqrt{1+ P_u^2}\, \sin \( \frac{\lambda}{L} \) \,,\\
 \frac{ u(\lambda)}{L}&= -\frac{1}{2} \log \left(\frac{\left(1+\sqrt{1+P_u^2} \sin (\lambda/L)\right) (1-P_u \tan (\lambda/L))}{\left(1-\sqrt{1+P_u^2} \sin (\lambda/L)\right) (1+P_u \tan (\lambda/L))}\right) \,,
\end{split}
\end{equation}
with taking $\frac{\lambda}{L} \in [0, \frac{\pi}{2}]$. From the evolution along the radial direction, we can find the extremal hypersurface cross the cosmological horizon at the `time' $\lambda_h=\arctan \( \frac{1}{|P_u|} \)$. Obviously, we should have two symmetric branches in terms of $t,r$ coordinates, corresponding to $P_u \ge 0, P_u \le 0$, respectively. However, the extremal surface with the negative conserved momentum $P_u$ would cross the past horizon and move into the region not covered by the infalling coordinate $u$. This fact is shown by the the singularity at $\lambda=\lambda_h$ in the solution $u(\lambda)$ with $P_u <0$. In the following, we will just focus on the branch with non-negative momentum for simplicity. The corresponding extremal surfaces are shown in the Penrose diagram of $\dS_{2}$ in in figure \ref{fig:dS2extremalsurface}. At the infinite momentum limit $P_u \to \infty$, the extremal hypersurface approaches the null surface $u=0$ and the turning point happens at the infinite future $i^+$ with $\rturn  \to +\infty$.

A unique feature associated with $\dS_{2}$ is the infinite redundancy of the extremal surfaces. From the solution in eq.~\eqref{eq:extremaldS2}, we can find that $u(0)=0$ for all values of $P_u$. It means that all extremal surfaces would collapse on the north pole at $t=0, r=0$, which is faithfully shown in figure \ref{fig:dS2extremalsurface}. From this point of view, we can conclude that the extremal surfaces are infinitely degenerate, \ie there is a continuous family of extremal hypersurfaces connecting the north pole and south pole. 

An alternative way to visualize the extremal surface in $\dS_{2}$ is embedding de Sitter spacetime as the hyperboloid (with radius $L$) in a three-dimensional Minkowski spacetime $\mathrm{R}^{2,1}$. In two dimension spacetime, the extremal surfaces are nothing but geodesics. With this embedding, the spacelike and timelike geodesic (extremal surface) in $\dS_{2}$ can be shown to be the intersections with a plane through the origin of the embedding space 
 (\eg see \cite{Schrodinger:1956jnw}). Correspondingly, one can find that all spacelike geodesics are periodic, \ie starting from the north pole and ending at the south pole, as shown in the Penrose diagram in figure \ref{fig:dS2extremalsurface}. Moreover, one can notice that all these spacelike geodesics on $\dS_{2}$ are degenerate because they are related to each other by the boost in the embedding Minkowski spacetime. Consequently, we immediately conclude that the length (volume) of all geodesics in $\dS_{2}$ is the same, which we can also see in the following subsection.\footnote{Yet another way to understand this degeneracy is to imagine defining new static patch coordinates with the origin centered at the bifurcation surface shown in the Penrose diagram. Then the extremal surfaces are just constant $t$ slices in the new coordinate patch and the symmetry is time translations. The fact that the extremal surfaces are related by an isometry is why the complexity is constant when $\rho\to0$ -- see also \cite{Chapman:2021eyy}. \label{footnotedS2}} 

\subsection*{Time evolution of CV complexity}

We are more interested in exploring the time evolution of the extremal surfaces with endpoints located on a fixed stretched horizon at $r=\rho L$. 
The corresponding extremal surface is nothing but the same one derived in eq.~\eqref{eq:extremaldS2} with non-zero $\lambda$ as the start point. The boundary time $\tau = \tR/L$ defined in eq.~\eqref{eq:tauPu} then reduces to 
\begin{equation}\label{eq:dS2tau}
\begin{split}
\tau \( \rho ; P_u \) &=\int^{r_{\rm{turn}}}_{\rmin} \frac{dr}{L}\, \frac{-P_u}{f(r)\sqrt{P_u^2+1 - \frac{r^2}{L^2}}} = \text{arctanh} \( \frac{P_u \rho}{ \sqrt{P_u^2+1-\rho^2} } \) \,.
\end{split}
\end{equation}
The relation $\tau (P_u)$ is also shown in figure \ref{fig:dS2extremalsurface}. Since the stretched horizon is outside the horizon, we can find the boundary time $\tau$ is bounded from above by $\tauinf$, \ie 
\begin{equation}
\tau \( \rho ; P_u \) \le    \text{arctanh}\, \rho  = \tauinf\,,
\label{notable}
\end{equation}
where the equality is saturated when $P_u \to  \infty$. We can further obtain the series expansions in different limits, \viz 
\begin{equation}
\tau \( \rho ; P_u \) \approx \begin{cases}
\tauinf- \frac{\rho}{ 2 P_u^2}  + \mathcal{O} \( \frac{1}{P_u^3} \) \,,
\qquad \qquad \qquad \ \,  P_u  \to \infty \\
\frac{ P_u }{\sqrt{1+P_u^2}}\, \rho + \mathcal{O}(\rho^3)\,,
\qquad \qquad\qquad \quad \ \, \rho \to 0  \\
\frac{1}{2} \log \(  \frac{P_u^2}{(1-\rho)(1+P_u^2)} \)  +  \( \frac{1}{4} - \frac{1}{P_u^2} \) \( \rho -1 \) +\mathcal{O}\((\rho-1)^2\) \,,  \,  \rho \to 1 \,.\\
\end{cases}
\end{equation}
As a comparison to the higher dimensional case, let us also remark here that the relation between $\tau$ and $P_u$ is monotonic, \ie 
\begin{equation}
\frac{d \tau}{d P_u}  = \frac{\rho }{(1+P_u^2)\sqrt{P_u^2+1- \rho^2}} \ge 0\,.
\end{equation}
We note here that the non-negativity for $\dS_2$ with any nonzero $\rho$ implies that there is one and only one extremal surface anchoring on the stretched horizon $\rho$ at a specific boundary time.

Let us turn to the holography complexity $\CV$ in $\dS_{2}$ by evaluating the radial integral 
\begin{equation}\label{eq:dS2CV}
\begin{split}
\CV (\rho; P_u)&= \frac{8 N}{L} \int^{\rturn}_{\rmin}  \frac{dr}{\sqrt{P_u^2 + 1 - r^2/L^2}}= 8N \(  \frac{\pi}{2} -  \arctan \(   \frac{\rho}{\sqrt{P_u^2 + 1 - \rho^2}} \)  \)\,.
\end{split}
\end{equation} 
Here, we note that with $\rho=0$, \ie the extremal surface is anchored at the north and south poles, 
the holographic complexity reduces to a constant, 
\begin{equation}\label{eq:CVdS2rho0}
\CV (\rho=0)=4\pi\,N  \,, 
\end{equation}
which is independent of the conserved momentum $P_u$. It indicates that all extremal surfaces with different conserved momenta have the same volume. This particular result was also  discussed in \cite{Chapman:2021eyy} from the viewpoint of holographic complexity.\footnote{The constant derived in \cite{Chapman:2021eyy} differs from eq.~\eqref{eq:CVdS2rho0} by a factor $\Omega_0 =2$ since as discussed above, our extremal surfaces contain two components in the complete Penrose diagram for $\dS_2$ shown in figure \ref{fig:dSPenrosedS2}.\label{footnotefactor}} We may also note that with the choice, $\tauinf=0$ all of the surfaces are anchored at $\rho=0$ and $\tau=0$. As we discussed above, the fact that all surfaces have the same volume is related to the boost symmetry of embedding spacetime $\mathrm{R}^{2,1}$ --  see footnote \ref{footnotedS2}.

\begin{figure}[h!]
	\centering
	\includegraphics[width=3in]{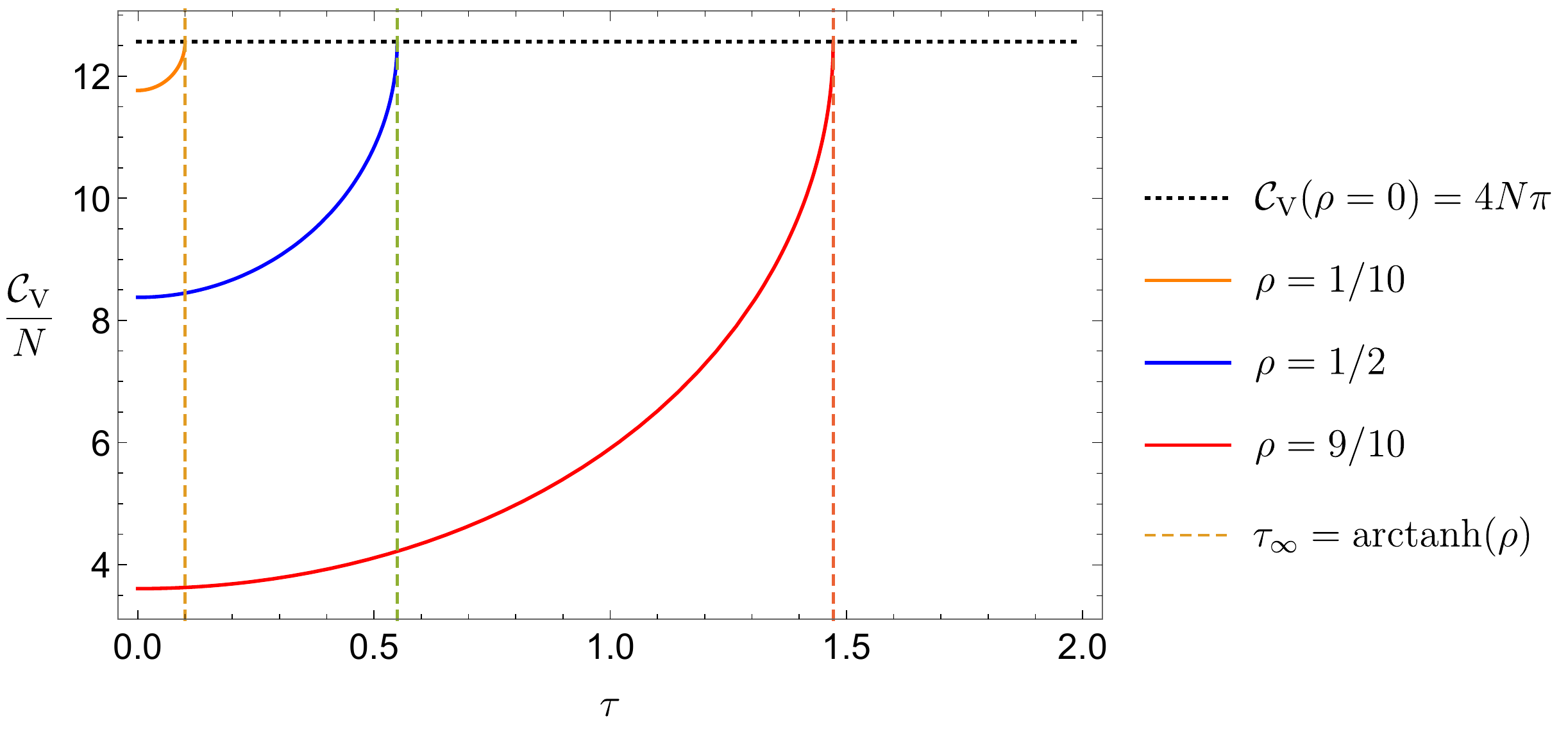}
		\includegraphics[width=3in]{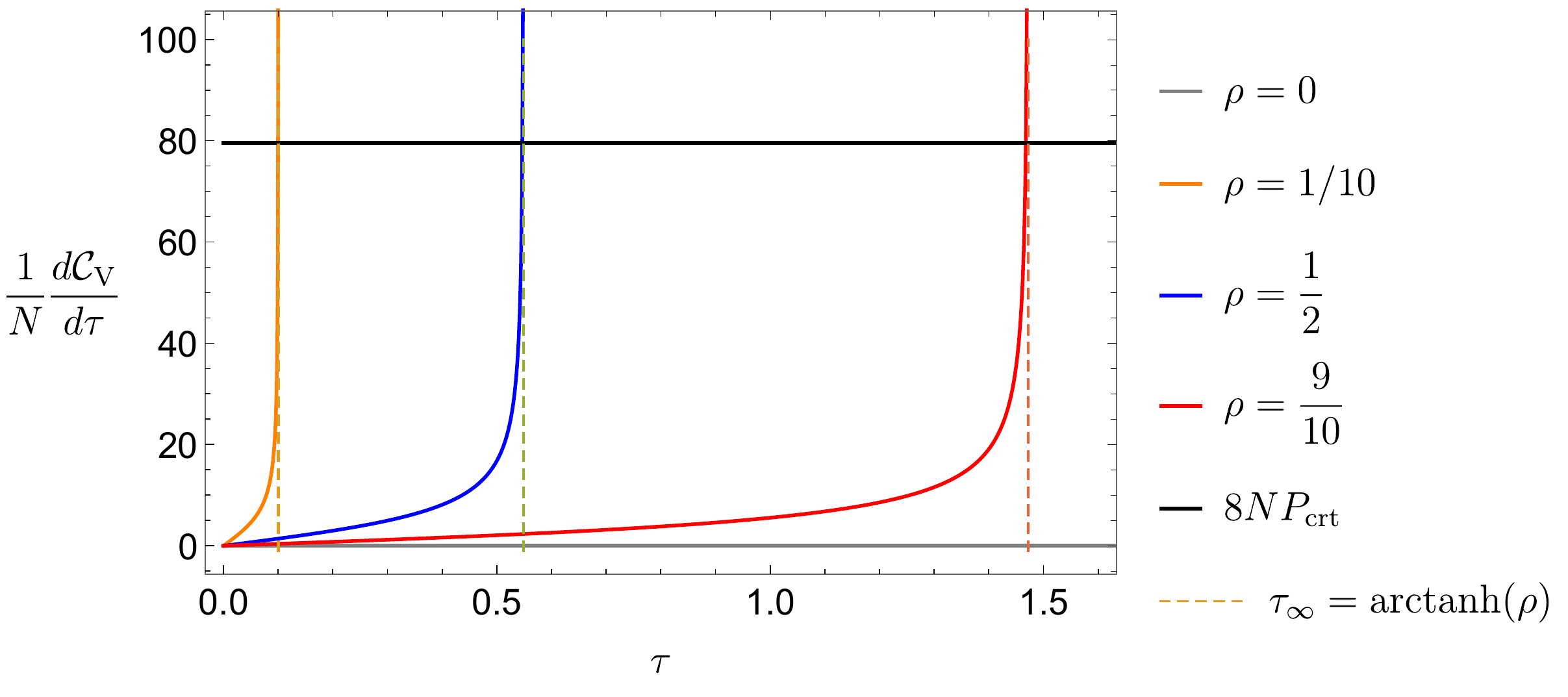}
	\caption{Left: Time evolution of holographic complexity $\CV$ in $\dS_2$. Right:Time evolution of growth rate of complexity in $\dS_2$. The black line indicates the growth rate for the linear growth after the transition time $\tau_{\rm crt}$.}
	\label{fig:dS2Volume}
\end{figure}

Unlike the particular case above,  the volume of the extremal surfaces anchored at  $\rho>0$ is sensitive to the choice of the conserved momentum, \ie depends on the boundary time $\tau$. To show the time dependence explicitly, we can also rewrite eq.~\eqref{eq:dS2tau} as 
\begin{equation}\label{eq:dS2PvtR}
P_u = \frac{e^{\tau} - e^{-\tau}}{\sqrt{2 \( \frac{1+\rho^2}{1-\rho^2}  - \cosh \(2 \tau \) \)}} \,.
\end{equation}
Substituting eq.~\eqref{eq:dS2PvtR} to eq.~\eqref{eq:dS2CV}, we finally obtain the time dependence of holographic complexity $\CV$:
\begin{equation}
\CV (\rho; P_u)=  4N  \(  \pi   - 2 \arctan \(  \frac{1}{\cosh\(\tau\)}  \sqrt{\frac{1+\rho^2 - (1-\rho^2)\cosh(2\tau)}{2(1-\rho^2)}}     \) \) \,.
\end{equation}
It is straightforward to check that its time derivative is given by 
\begin{equation}
\begin{split}
 \frac{d \CV }{d \tau} = 8N \sinh ({\tau}) \sqrt{\frac{2(1- \rho ^2)}{\rho ^2+1-\left(1-\rho ^2\right) \cosh (2 {\tau})}} = 8 N P_u \,,
\end{split}
\end{equation}
as we claimed before in eq.~\eqref{eq:dVdtau}.
Taking the limit $\tau \to \tauinf$, one can find 
\begin{equation}\label{eq:dS2CVlimit}
 \lim_{\tau \to \tauinf} \, \CV \simeq  4 \pi\,N - 8N \sqrt{2\rho (\tauinf-\tau)} + \mathcal{O}((\tauinf-\tau)^{3/2}) \,,
\end{equation}
which is approaching a constant $4\pi N$ rather than being divergent. However, we can still find the hyperfast growth of complexity, \ie 
\begin{equation}
\label{rate22}
\lim_{\tau \to \tauinf} \,  \frac{d \CV }{d \tau}   \approx  8N\sqrt{\frac{\rho }{2\(\tauinf-\tau\)}} + \mathcal{O}\( \sqrt{\tauinf-\tau}  \)\,,
\end{equation}
which is different from the power law in eq.~\eqref{eq:Pulimit} for $\dS_{d+1}$. See figure.~\ref{fig:dS2Volume} for the time dependence of complexity and its growth rate in $\dS_2$. 

\subsection*{Introducing a cutoff surface}

In contrast to the CV complexity in higher dimensional $\dS_{d+1}$, $\cv$ in $\dS_{2}$ remains finite at the critical time $\tau=\tauinf$, as shown in eq.~\eqref{eq:dS2CVlimit}. However, we still need a new description for $\cv$ to understand the behaviour after the critical time due to the absence of the extremal surfaces connecting the two stretched horizons for $\tau>\tauinf$. Following the approach in section \ref{sec:dSCV}, we introduce a cut-off surface at $r=L/\delta$ and use the modified proposal in eq.~\eqref{eq:CVcutoff}  where the extremal surfaces are defined in a piecewise manner beyond the critical time $\tauinf$. The conclusions for $\cv$ at later times in $\dS_{2}$ are then essentially the same as in higher dimensional dS. This subsection will show more analytical results to support the general analysis in section \ref{sec:dSCV}.

From the extremality equation derived in eq.~\eqref{eq:extremaldS2}, the conserved momentum of the extremal surfaces that can touch the cut-off surface should satisfy the following constrain, namely
\begin{equation}\label{eq:dS2Puconstrain}
 |P_u|  \ge  \sqrt{\frac{1}{\varepsilon^2} -1 } \,.
\end{equation}
Correspondingly, the contribution of the extremal surfaces $\mB_{\mt{L}}, \mB_{\mt{R}}$ to complexity is defined in eq.~\eqref{eq:VextdSd} and derived as
\begin{equation}
\begin{split}
\cv^{\rm ext}&=   8N \(   \arcsin \( \frac{1}{\varepsilon\sqrt{1+ P_u^2}}  \)-   \arcsin \( \frac{\rho}{\sqrt{1+ P_u^2}}  \) \)  \,.
\end{split}
\end{equation}
The complexity from the cut-off surface part $\mB_{\varepsilon}$ then reads  
\begin{equation}
\cv^{\varepsilon}= 8N \tau_{\varepsilon} \sqrt{\frac{1}{\varepsilon^2} -1}  \,,
\end{equation}
where the intersection time $\tau_{\varepsilon} $ on the cut-off surface is related to the boundary time $\tau$ by 
\begin{equation}
\tau - \tau_{\varepsilon}  = \text{arctanh} \( \frac{P_u \rho}{\sqrt{1+P_u^2 -\rho^2}} \) +  \frac{1}{2} \log \left| \frac{  \sqrt{P_u^2+1-\frac{1}{\varepsilon^2}} - P_u/\varepsilon}{ \sqrt{P_u^2+1-\frac{1}{\varepsilon^2}} + P_u /\varepsilon}  \right|\,. 
\end{equation}
The transition time is derived as 
\begin{equation}
\tau_{\rm crt} =  \text{arctanh} \( \frac{P_{\rm crt} \rho}{\sqrt{1+P_{\rm crt}^2 -\rho^2}} \) = \text{arctanh}  \(  \rho  \sqrt{\frac{\frac{1}{\varepsilon^2}-1}{\frac{1}{\varepsilon^2}-\rho^2} } \) < \tauinf \,.
\end{equation}
After the transition time $\tau_{\rm crt} $, we pick up the one with the maximal volume among those infinite discontinuous surfaces by performing the maximization:
\begin{equation}
\cv \(\tau \ge \tau_{\rm crt}\) = \max_{|P_u| \ge P_{\rm crt}} \( \cv^{\rm ext}\(P_u\) + \cv^{\varepsilon} (\tau_{\varepsilon})  \) \,.
\end{equation}
Explicitly, one can find the monotonicity with respect to the momentum $P_u$, \ie 
\begin{equation}
\frac{\partial \( \cv^{\rm ext} + \cv^{\varepsilon} \)}{\partial P_u} =\frac{ 8 N\varepsilon}{1+P_u^2} \( \sqrt{\frac{1}{\varepsilon^2}- 1}  -  P_u    \) \(    \frac{\rho}{\sqrt{1+P_u^2 - \rho^2}}   - \frac{1/\varepsilon}{\sqrt{1+ P_u^2 - \frac{1}{\varepsilon^2}}} \)  \le  0 \,, 
\end{equation}
where the non-positivity is guaranteed by the constrain in eq.~\eqref{eq:dS2Puconstrain} as well as $ \rho < 1 <1/\delta$. As a result, we conclude that the maximal volume for those surfaces anchoring at boundary time $\tau$ on the stretched horizon is derived as 
\begin{equation}
\cv \(\tau \ge \tau_{\rm crt}\) =  8N \(  \frac{\pi}{2} - \arcsin \( \rho \varepsilon \)  + \sqrt{\frac{1}{\varepsilon^2}-1} \(  \tau- \text{arctanh}  \(  \frac{\rho\sqrt{\frac{1}{\varepsilon^2}-1} }{\sqrt{\frac{1}{\varepsilon^2}-\rho^2}} \) \)  \) \,, 
\end{equation} 
with a linear growth, \ie 
\begin{equation}\label{eq:Pcrt}
\frac{d\cv }{d\tau} \bigg|_{\tau \ge \tau_{\rm crt}} = 8  N \sqrt{\frac{1}{\varepsilon^2}-1}\,.
\end{equation}

\section{Multiple Extremal Surfaces and Maximization}\label{app:max}

\begin{figure}[!]
	\centering
	\includegraphics[width=4in]{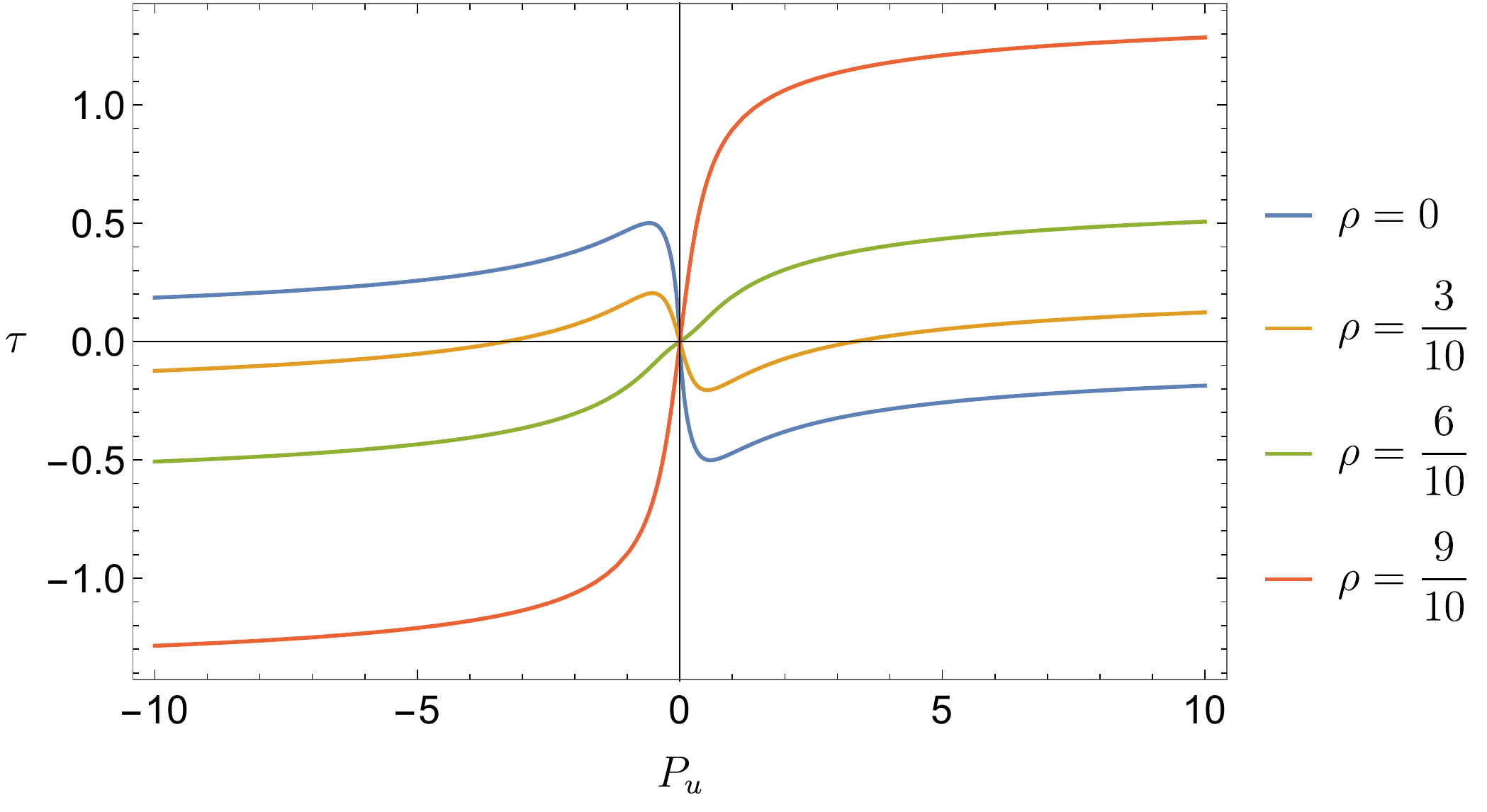}
	\caption{The boundary time $\tau$ as a function of the conserved momentum $P_u$ as derived in \eqref{eq:tauPu}. We consider three-dimensional de Sitter spacetime $\dS_{2+1}$ and choose different stretched horizons at $r=\rho L$ in this plot. We note that it is not a monotonic function for smaller values of $\rho$, \eg $\rho=0$ and 3/10.}
	\label{fig:dS3tR02}
\end{figure}

Except for the similarities with $\dS_2$, a new feature associated with the higher dimensional dS spacetime is that the relation between $\tau$ and momentum $P_u$ is not monotonic in general. In other words, the sign of the derivative $\frac{d \tau}{d P_u}$ is not fixed for $\dS_{d+1}$ for an arbitrary stretched horizon, \ie for small values of $\rho$. For example, we show the relation in figure~\ref{fig:dS3tR02} for $\dS_3$. This feature implies that there is more than one extremal surface associated with a specific boundary time $\tau$.

Taking the boundary time $\tau$ related to the extremal surfaces with a momentum $P_u$, \ie eq.~\eqref{eq:tauPu}, we first examine the derivative of $\tau$ with respective to $P_u$, \viz 
\begin{equation}\label{eq:dtaudPu}
\begin{split}
\frac{d \tau}{d P_u} =- \frac{d \rturn}{d P_u}  \frac{P_u/L}{f(r)\sqrt{P_u^2- U(r)}} \Bigg|_{r \to \rturn}+ \int^{\rturn}_{\rmin} \frac{dr}{L} \frac{U(r)}{f(r)\(P_u^2- U(r)\)^{3/2}}  \,,
\end{split}
\end{equation}
where $d\rturn/dP_u$ is given by 
\begin{equation}
\frac{dP_u}{d \rturn} = \frac{U'(\rturn)}{2P_u} \,,
\end{equation}
with using the definition of the turning point shown in eq.~\eqref{eq:definermax}. It is obvious that both terms in eq.~\eqref{eq:dtaudPu} are divergent due to the same singular point at $r=\rturn$. However, we can explicitly find that these two divergences are exactly canceled and $\frac{d \tau}{d P_u} $ is always finite. Focusing only on the divergent terms, we can get 
\begin{equation}\label{eq:purmax}
\frac{d \rturn}{d P_u}  \frac{P_u}{f(r)\sqrt{P_u^2- U(r)}} \Bigg|_{r \to \rturn} \sim  \frac{2U(\rturn)}{f(\rturn) \( U'(\rturn) \)^{3/2}} \frac{1}{\sqrt{\rturn -r }} + \mathcal{O}(\sqrt{\rturn -r }) \,,
\end{equation}
and also 
\begin{equation}
\begin{split}
\int^{r_{\rm{max}}}_{\rho} \frac{U(r)\,dr}{f(r)\(P_u^2- U(r)\)^{3/2}}  &\sim  \int^{r\to r_{\rm{max}}}\frac{U(r)\,dr}{f(r)\(U'(\rturn) (\rturn -r)\)^{3/2}}   \, d\tilde{r} \\
&\sim \frac{2U(\rturn)}{f(\rturn) \( U'(\rturn) \)^{3/2}} \frac{1}{\sqrt{\rturn -r }} + \mathcal{O}(\sqrt{\rturn -r }) \,.
\end{split}
\end{equation}
with substituting eq.~\eqref{eq:purmax} and performing the integral around the maximal radius. 
As a result, the potential divergences appearing $\frac{d \tR}{d P_u} $ are canceled. Correspondingly, we arrive at the second conclusion about the finiteness, \ie 
\begin{equation}
\frac{d \tau}{d P_u}  \,\, \text{is finite} \,, \qquad  \forall  \quad  |P_u| \,,
\end{equation}
for the extremal surfaces in $\dS_{d+1}$. This also indicates the difference with the linear growth for the extremal surfaces in asymptotically AdS whose potential contains a maximum at the turning point.

\begin{figure}[ht!]
	\centering
	\includegraphics[width=3.5in]{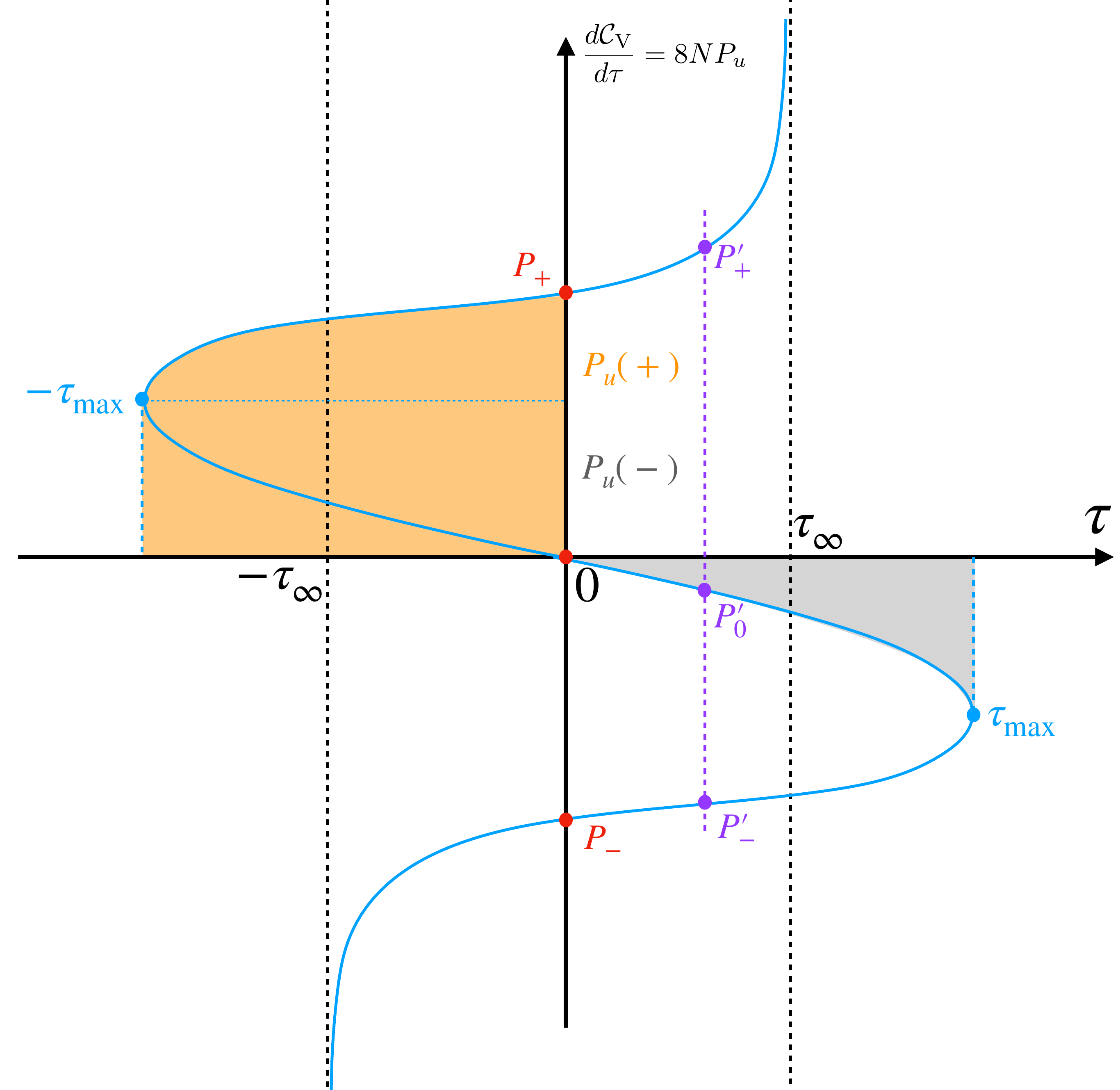}
	\caption{Multiple extremal surfaces at a fixed boundary time $\tau$.}
	\label{fig:max}
\end{figure}

Although we have shown the finiteness of $\frac{d \tau}{d P_u}$, its sign is still undetermined because it depends on the choice of the stretched horizon. Due to $f(\rturn)<0$, the two terms shown in eq.~\eqref{eq:dtaudPu} are positive and negative, respectively. However, it is still straightforward to show $\frac{d}{d \rmin} \frac{d \tau}{d P_u} > 0$. 
When the minimal radius, \ie $\rmin = \rho L$ is too small, $\frac{d \tau}{d P_u}$ could be negative since the second term in eq.~\eqref{eq:dtaudPu}. A characteristic behavior of $P_u (\tau)$ is illustrated in figure.~\ref{fig:max}. Taking any time slice at $\tau \in (-\tauinf, +\tauinf)$, there are three corresponding momentums $P_u$, \ie three extremal surfaces anchoring at this boundary time.  Among these candidates, we should pick up the one with the maximal volume for holographic complexity $\cv$. We will prove in the following that the maximal one is always given by the extremal surface with larger momentum $|P_u|$, which is also the one smoothly related to the critical null surface $u=0$. 

Let us first take the boundary time $\tau=0$ as an example. The corresponding momentums are denoted by $P_+, P_0=0, P_-$, as shown in figure.~\ref{fig:max}. Assuming the complexity $\cv$ at $P_0=0$ is given by $\cv(P_0)$ and considering the evolution of the extremal surface from $P_0$ to $P_+$, we rewrite the holographic complexity $\cv(P_+)$ as 
\begin{equation}
\begin{split}
\cv(P_+) = \cv (P_-) &= \cv (P_0) + \int_0^{-\tau_{\rm max}} d\tau \frac{d \cv}{d\tau} + \int^0_{-\tau_{\rm max}} d\tau \frac{d \cv}{d\tau} \\
&=  \cv (P_0) + 8 N \int_{-\tau_{\rm max}}^0  d\tau  \( -P_u(-) + P_u (+)  \)   \,,\\
\end{split}
\end{equation}
where contributions from $P_u(-)$, $P_u(+)$ are negative and positive, respectively. As illustrated in figure.~\ref{fig:max}, the negative and positive contributions are represented by the area of the gray region and yellow region, respectively. Obviously, we have $P_u(+) > P_u(-)$ as well as 
\begin{equation}\label{eq:CVinequality01}
\cv(P_+) = \cv (P_-) >  \cv (P_0)  \,,\quad  \text {with} \qquad \tau =0 \,.
\end{equation}
We can then move to an arbitrary boundary time $\tau \in (-\tau_{\infty}, \tauinf)$ with three extremal surfaces labeled by conserved momentums $P_0', P_+', P_-'$. Without loss of generality, we consider an example with $\tau >0$ as indicated by the purple line in figure.~\ref{fig:max}. Following the method introduced above, one can find 
\begin{equation}\label{eq:CVinequality02}
\cv(P_+') >  \cv(P_+)  \,,\quad \cv(P_0') <  \cv(P_0)\,, \quad   \cv(P_-') <  \cv(P_-) \,, \quad \cv(P_-') >  \cv(P_0')\,,
\end{equation}
Combing the these inequalities with eq.~\eqref{eq:CVinequality01}, we finally conclude that 
\begin{equation}
\cv(P_+')  >  \cv(P_-') > \cv(P_0')\,, \qquad \text{with} \qquad \tau \in [0, +\tauinf]\,. 
\end{equation}
This inequality indicates that the maximization over all extremal surfaces at a fixed boundary time always arrives at the extremal surface with a larger $|P_u|$, which belongs to the branch with the critical null surface at $u=0$.

\bibliographystyle{jhep}
\bibliography{biographyCVdS}

\end{document}